\documentclass[12pt]{article}
\usepackage[dvips]{graphics}

\begin{document}

\begin{center}

\Large{The  rest masses of the electron and muon\\
and of the stable mesons and baryons}

\vspace{0.75cm}

\large{E.L. Koschmieder}

\vspace{0.5cm}

\small{Center for Statistical Mechanics\\
The University of Texas at Austin, Austin TX 78712, USA\\
e-mail: koschmieder@mail.utexas.edu}

\end{center}
\bigskip

\noindent
\small
{The rest masses of the electron, the muon and of the stable 
mesons and baryons can be explained, within 1\% accuracy, with the
 standing wave model, which uses only photons, neutrinos, charge and  
the weak nuclear force.  We do not need hypothetical particles for the
explanation of the masses of the electron, muon,  mesons and baryons.
 
We have determined the rest masses of the electron-, muon- and tau
neutrinos and found that the mass of the electron neutrino is equal to
the fine structure constant times the mass of the muon neutrino.


 }

\normalsize

\section*{Introduction}

 The so-called ``Standard Model" of the elementary particles 
has, until now, not come up with a precise theoretical determination  
of the masses of either the mesons and baryons or of the leptons, 
which means that neither the mass of the fundamental electron 
nor the mass of the fundamental 
proton have been explained. The quarks, which have been  
introduced by Gell-Mann [1] forty years ago, are said to explain the 
mesons and baryons. But the standard model does not explain 
neither the mass, nor the charge, nor the spin of the particles, the 
three fundamental time-independent properties of the particles.  
The fractional electric charges imposed on the quarks do not 
explain the charge e$^\pm$, neither does the spin imposed
on the quarks explain the spin. The measured values of the 
properties of the particles are given in the Review of Particle 
Physics [2]. There are many other attempts to explain the 
elementary particles or only one of the particles, too many 
to list them here.  For example El Naschie 
has, in the last years, proposed a topological theory for high 
energy particles and the spectrum of the quarks [3-6].

   The need for the present investigation has been expressed by  
Feynman [7] as follows: ``There remains one especially unsatisfactory 
feature: the observed masses of the particles, m. There is no theory 
that adequately explains these numbers. We use the numbers in all 
our theories, but we do not understand them - what they are, or where 
they come from. I believe that from a fundamental point of view, this 
is a very interesting and serious problem". Today, twenty
 years later, we still stand in front of the same problem.   

\section {The spectrum of the masses of the particles}

As we have done before [8] we will focus attention on the so-called 
``stable" mesons and baryons 
whose masses are reproduced with other data in Tables 1 and 2.
It is obvious that any attempt to explain the masses of the mesons and 
baryons should begin with the particles that are affected by the fewest 
parameters. These are certainly the particles without isospin (I = 0) and 
without spin (J = 0), but also with strangeness S = 0, and charm C = 0. 
Looking at the particles with I,J,S,C = 0 it is startling to find that 
their masses are quite close to integer multiples of the mass of the 
$\pi^0$\,meson. It is m($\eta) = (1.0140 \pm 0.0003)\,\cdot$\,4m($\pi^0$), 
and the mass of the resonance $\eta^\prime$ is m($\eta^\prime$) = (1.0137 
$\pm$ 0.00015)\,$\cdot$\,7m($\pi^0$). Three particles seem hardly to be 
sufficient to establish a rule. However, if we look a little further we 
find that m($\Lambda$) = 1.0332\,$\cdot$\,8m($\pi^0$) or m($\Lambda$) = 
1.0190\,$\cdot$\,2m($\eta$). We note that the $\Lambda$ particle has spin 
1/2, not spin 0 as the $\pi^0$,\,$\eta$ mesons. Nevertheless, the mass of 
$\Lambda$ is close to 8m($\pi^0$). Furthermore we have m($\Sigma^0$) = 
0.9817\,$\cdot\,$9m($\pi^0$), m($\Xi^0) = 0.9742\,\cdot\,$10m($\pi^0$), 
m$(\Omega^-)$ = 
1.0325\,$\cdot\,$12m($\pi^0)$ = 1.0183\,$\cdot$\,3m($\eta$), ($\Omega^-$ 
is charged and has spin 3/2). Finally the masses of the charmed baryons 
are m($\Lambda_c^+$) = 0.9958\,$\cdot$\,17m($\pi^0$) = 
1.024\,$\cdot$\,2m($\Lambda$), m($\Sigma_c^0$) = 
1.0093\,$\cdot$\,18m($\pi^0$), 
m($\Xi_c^0$) = 1.0167\,$\cdot$\,18m($\pi^0$), and m($\Omega_c^0$) = 
1.0017\,$\cdot$\,20m($\pi^0$).
 
	\begin{table}\caption{The ratios m/m($\pi^0$) of the $\pi^0$ and
$\eta$ mesons
\newline
\hspace*{1.5cm} 
 and of the baryons of the $\gamma$-branch.} 
	\begin{tabular}{lllllcl}\\
\hline\hline\\
 & m/m($\pi^0$) & multiples & decays & fraction & spin & 
mode
 \\
 & & & & (\%) & &\\
[0.5ex]\hline
\\
$\pi^0$ & 1.0000 & 1.0000\,\,$\cdot$\,\,$\pi^0$ & $\gamma\gamma$ & 98.798 
& 0 & (1.)\\
 & & & $e^+e^-\gamma$ & \,\,\,1.198 & &\\
\\
$\eta$ & 4.0559 & 1.0140\,$\cdot$\,\,4$\pi^0$ & $\gamma\gamma$ & 39.43 & 0 
& 
(2.)\\
 & & & 3$\pi^0$ & 32.51 & &\\
 & & & $\pi^+\pi^-\pi^0$ & 22.6 & &\\
 & & & $\pi^+\pi^-\gamma$ & \,\,\,4.68 & &\\
\\
$\Lambda$ & 8.26577 & 1.0332\,$\cdot$\,\,8$\pi^0$ & p$\pi^-$ & 63.9 & 
$\frac{1}{2}$ 
& 2$\ast$(2.)\\
 & & 1.0190\,$\cdot$\,\,2$\eta$ & n$\pi^0$ & 35.8 & &\\
\\
$\Sigma^0$ & 8.8352 & 0.9817\,$\cdot$\,\,9$\pi^0$ & $\Lambda \gamma$ & 100 
& 
$\frac{1}{2}$ & 2$\ast (2.) + (1.)$\\
\\
$\Xi^0$ & 9.7417 & 0.9742\,$\cdot$\,10$\pi^0$ & $\Lambda\pi^0$ & 99.52 & 
$\frac{1}{2}$ & 
2$\ast(2.) + 2(1.)$\\
\\
$\Omega^-$ & 12.390 & 1.0326\,$\cdot$\,12$\pi^0$ & $\Lambda$K$^-$ & 67.8 
& 
$\frac{3}{2}$ & 3$\ast(2.)$\\
 & & 1.0183\,$\cdot$\,\,3$\eta$ & $\Xi^0\pi^-$ & 23.6 & &\\
 & & &  $\Xi^-\pi^0$ & \,\,\,8.6 & &\\
\\
$\Lambda_c^+$ & 16.928 & 0.9958\,$\cdot$\,17$\pi^0$ & many & & 
$\frac{1}{2}$ & 
2$\ast(2.) + (3.)$\\
 & & 0.9630\,$\cdot$\,17$\pi^\pm$\\
\\
$\Sigma_c^0$ & 18.167 & 1.0093\,$\cdot$\,18$\pi^0$ & $\Lambda_c^+\pi^-$ & 
$\approx$100 & $\frac{1}{2}$ & $\Lambda_c^+ + \pi^-$\\
\\
$\Xi_c^0$ & 18.302 & 1.0167\,$\cdot$\,18$\pi^0$ & eleven & (seen) & 
$\frac{1}{2}$   & 
2$\ast(3.)$\\
\\
$\Omega_c^0$ & 20.033 & 1.0017\,$\cdot$\,20$\pi^0$ & seven & (seen) & 
$\frac{1}{2}$ & 
2$\ast(3.) + 2(1.)$\\
& & 0.9857\,$\cdot$\,\,5$\eta$ & & &\\
[0.2cm]\hline\hline
\vspace{0.1cm}
\end{tabular}
\footnotemark{\footnotesize The modes apply to neutral particles only. The 
$\ast$
 marks coupled modes.}
	
	\end{table}

   Now we have enough material to formulate the 
\emph{integer multiple rule} of the particle masses, 
according to which the masses of the 
$\eta$,\,$\Lambda$,\,$\Sigma^0$,\,$\Xi^0$,\,
$\Omega^-$,\,$\Lambda_c^+$,\,$\Sigma_c^0$,\,
$\Xi_c^0$, and $\Omega_c^0$ 
particles are, in a first approximation, integer multiples of the mass 
of the $\pi^0$\,meson, although some of the particles have spin, 
and may also have charge as well as strangeness and charm. A 
consequence of the integer multiple rule must be that the ratio of the 
mass of any meson or baryon listed above divided by the mass of another 
meson or baryon listed above is equal to the ratio of two integer numbers. 
And indeed, for example m($\eta$)/m($\pi^0$) is practically two times 
(exactly 0.9950$\,\cdot$\,2) the ratio m($\Lambda$)/m($\eta$). There is 
also the ratio m($\Omega^-$)/m($\Lambda$) = 0.9993\,$\cdot$\,3/2. 
We have furthermore e.g. the ratios m($\Lambda$)/m($\eta$) = 
1.019\,$\cdot$\,2, m($\Omega^-$)/m($\eta$) = 1.018\,$\cdot$\,3, 
m($\Lambda_c^+$)/m($\Lambda$) 
= 1.02399\,$\cdot$\,2, m($\Sigma_c^0$)/m($\Sigma^0$) = 
1.0281\,$\cdot$\,2,  m($\Omega_c^0$)/m($\Xi^0$) = 1.0282\,$\cdot$\,2, 
and m($\Omega_c^0$)/m($\eta$) = 0.9857\,$\cdot$\,5.

   We will call, for reasons to be explained soon, the particles 
discussed above, which follow in a first approximation the integer 
multiple rule, the \emph{$\gamma$-branch} of the particle spectrum.  
The mass 
ratios of these particles are in Table 1. The deviation of the mass ratios 
from exact integer multiples of m($\pi^0$) is at most 3.3\%, the average 
of the factors before the integer multiples of m($\pi^0$) of the nine 
$\gamma$-branch particles in Table 1 is 1.0066 $\pm$ 0.0184. From a least 
square analysis follows that the masses of the ten particles on Table 1 
lie on a straight line given by the formula

\begin{equation} \mathrm{m}(N)/\mathrm{m}(\pi^0) = 1.0065\,N - 0.0043 
\qquad  N\,>\,1, 
\end{equation}     
where N is the integer number nearest to the actual ratio of the particle 
mass divided by m($\pi^0$). The correlation coefficient in Eq.(1) 
has the nearly perfect value R$^2$ = 0.999.
 
   The integer multiple rule applies to more than just the stable mesons 
and baryons. The integer multiple rule applies also to the $\gamma$-branch 
baryon resonances which have spin J = 1/2 and the meson resonances with 
I,J  $\leq$ 1, listed in [2] or in Table 3 of the appendix. The 
$\Omega^-$\,baryon will not be considered because it has 
spin 3/2 but would not change the following equation significantly.
 If we consider all mesons and baryons 
of the $\gamma$-branch in Table 3, ``stable" or unstable,
then we obtain from a least square analysis the formula 

\begin{equation}    \mathrm{m}(N)/\mathrm{m}(\pi^0) = 0.999\,N + 
0.0867\qquad    N\,>\, 1,  
\end{equation}
with the correlation coefficient  0.9999. The line through the points is 
shown in Fig.\,1 which tells that 41 particles of the $\gamma$-branch  
of different spin and isospin, strangeness and charm; five I,J = 0,0 
$\eta$\,mesons, fifteen J = 1/2 baryons, ten I = 0, J = 0,1 
c$\bar{c}$ mesons, ten I = 0, J = 0,1 b$\bar{b}$ mesons and the
 $\pi^0$\,meson with I,J = 1,0, lie on a straight line with slope 
0.999. In other words they approximate the integer multiple rule 
very well. Spin 1/2 and spin 1 does not seem to affect the integer
 multiple rule, i.e. the ratios of the particle masses, neither
does strangeness S $\neq$ 0 and charm C $\neq$ 0.

\begin{figure}
	\includegraphics{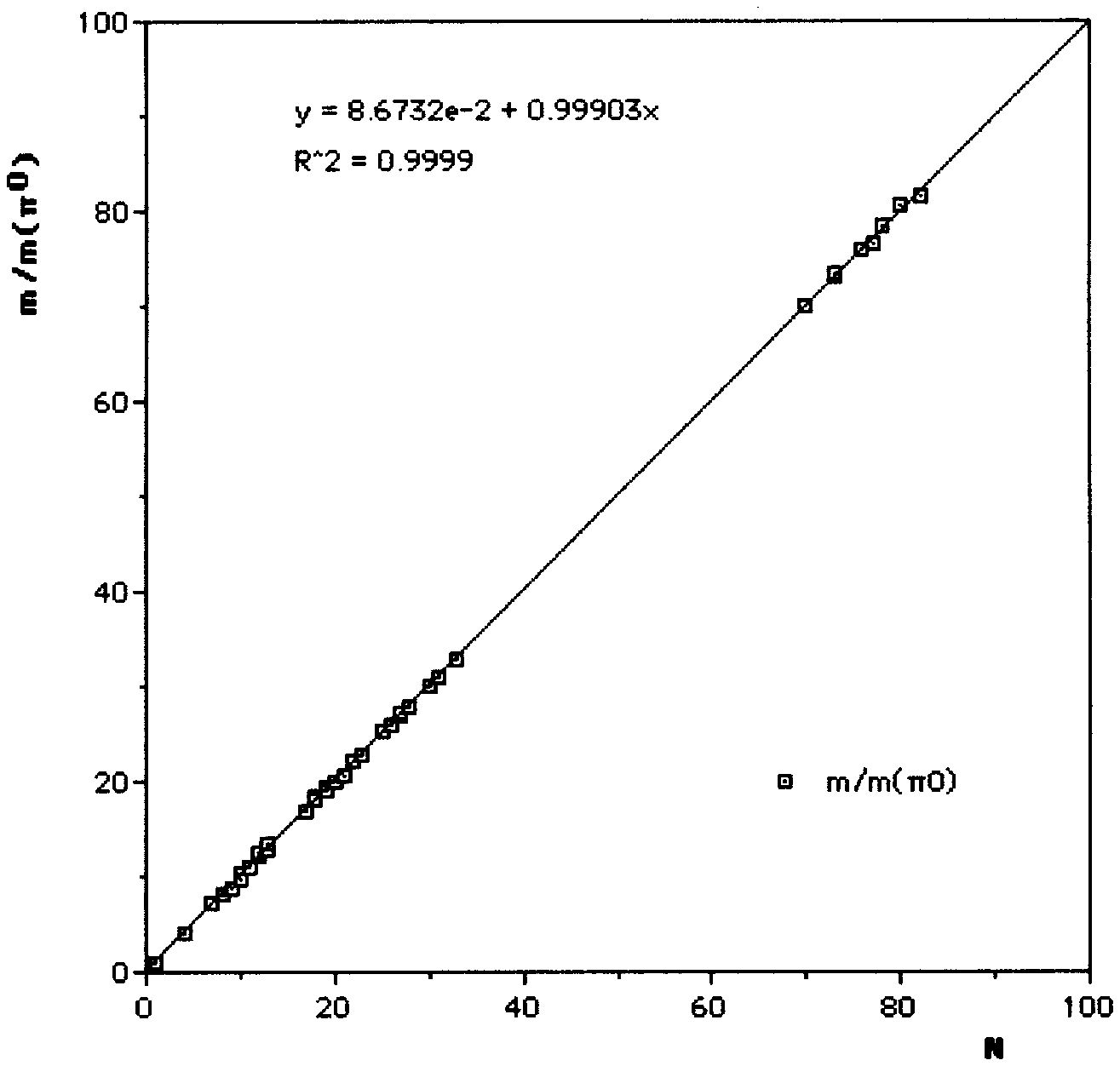}
	\begin{quote}
Fig.\,1: The mass of the mesons and baryons of the $\gamma$-branch,
  stable or unstable, with I\,$\leq$ 1,\,J\,$\leq$ 1 in units of m($\pi^0$) 
 as a function of the integer  N, demonstrating the integer multiple rule.
	\end{quote}
\end{figure}

   Searching for what else the  
$\pi^0$,\,$\eta$,\,$\Lambda$,\,$\Sigma^0$,\,$\Xi^0$,\,$\Omega^-$ 
particles have in common, we find that the principal decays (decays 
with a fraction\,$>$\,1\%) of these particles, 
 as listed in Table 1, involve primarily $\gamma$-rays, the 
 characteristic case is $\pi^0 \rightarrow \gamma\gamma$  (98.8\%). We 
will later on discuss a possible explanation for the 1.198\% of the decays 
of $\pi^0$ which do not follow the $\gamma\gamma$ route but decay via
$\pi^0 \rightarrow$ e$^+$ + e$^-$ + $\gamma$. After the 
$\gamma$-rays the next most 
frequent decay product of the heavier particles of the $\gamma$-branch are 
$\pi^0$\,mesons which again decay into $\gamma\gamma$. To describe the 
decays in another way, the principal decays of the particles listed above 
take place $\emph{always without the emission of neutrinos}$\,; see Table 
1. There the decays and the fractions of the principal decay modes are given, 
taken from the Review of Particle Physics. We cannot consider decays 
with fractions $<$\,1\%. We will refer to the particles whose masses are 
approximately integer multiples of the mass of the $\pi^0$\,meson, and 
which decay without the emission of neutrinos, as the 
$\gamma$-$\emph{branch}$ of the particle spectrum.

   To summarize the facts concerning the $\gamma$-branch. Within 
0.66\% on the average the masses of the stable particles of the 
$\gamma$-branch  in Table\,1 are integer multiples 
(namely 4,\,8,\,9,\,10,\,12, and even 17,\,18,\,20) of the mass of the 
$\pi^0$\,meson. 
It is improbable that nine particles have masses so close to integer 
multiples of m($\pi^0$) if there is no correlation between them and the 
$\pi^0$\,meson. It has, on the other hand, been argued that the integer 
multiple rule is a numerical coincidence. But the probability that the 
mass ratios of nine particles of the $\gamma$-branch fall by coincidence 
on integer numbers between 1 and 20 instead on all possible numbers 
between 1 and 20 with two decimals  
after the period is smaller than 10$^{-20}$, i.e.\,nonexistent. The 
integer multiple rule is not affected by more than 3\% by the spin, the 
isospin, the strangeness, and by charm. The integer multiple rule seems 
even to apply to the $\Omega^-$ and $\Lambda_c^+$ particles, although  
they are charged. In order for the integer multiple rule to be valid the 
deviation of the ratio m/m($\pi^0$) from an integer number must be smaller 
than 1/2N, where N is the integer number closest to the actual ratio 
m/m($\pi^0$). That means that the permissible deviation decreases rapidly 
with increased N. All particles of the $\gamma$-branch have deviations 
smaller than 1/2N.
 
   The remainder of the stable mesons and baryons are the  
$\pi^\pm$, K$^{\pm,0}$,\,p,\,\,n,\quad D$^{\pm,0}$, and D$_s^\pm$ 
particles which make up the neutrino-branch ({$\nu$-branch}) of the 
particle spectrum. The ratios of their masses are given in Table 2.

	\begin{table}\caption{The $\nu$-branch of the particle 
spectrum}
	\begin{tabular}{llllrcl}\\

\hline\hline\\
 & m/m($\pi^\pm$) & multiples & decays
 & fraction & spin & mode 
 \\
 & & & & (\%) & &\\
[0.5ex]\hline
\\
$\pi^\pm$ & 1.0000 & 1.0000\,\,$\cdot$\,\,$\pi^\pm$ & $\mu^+\nu_\mu$ & 
\,\,99.9877 & 0 & (1.)\\
\\
K$^{\pm,0}$ & 3.53713 & 0.8843\,$\cdot$\,\,4$\pi^\pm$ & $\mu^+\nu_\mu$ & 
\,\,63.43 & 0 & 
(2.) + $\pi^0$\\
 & & & $\pi^\pm\pi^0$ & \,\,21.13 && \,\,\,\,(K$^\pm$)\\
 & & & $\pi^+\pi^-\pi^+$ & \,\,\,\,\,5.58 && (2.) + $\pi^\mp$\\
 & & & $\pi^0 e^+ \nu_e$ (K$_{e3}^+$)& \,\,\,\,\,4.87 & 
&\,\,\,\,(K$^0$,$\overline{{\mathrm{K}}^0}$)\\
 & & & $\pi^0\mu^+\nu_\mu$ (K$_{\mu3}^+$) & \,\,\,\,\,3.27 & &\\
\\
n & 6.73186 & 0.8415\,$\cdot$\,\,8$\pi^\pm$ & $p\,e^-\overline{\nu}_e$ & 
100.\,\,\,\,\,\,\,\,  & $\frac{1}{2}$ & 2$\ast$(2.)\\
 & & 0.9516\,$\cdot$\,\,2K$^\pm$  & & & & \,\, + $2\pi^\pm$\\
 & &  0.9439\,$\cdot$\,\,(K$^0$ + $\overline{\mathrm{K}}^0$) & & & &\\
\\
D$^{\pm,0}$ & 13.393 & 0.8370\,$\cdot$\,16$\pi^\pm$ & $e^+$ anything & 
\,\,17.2 & 
0 & 2(2$\ast$(2.)\\
 & & 0.9466\,$\cdot$\,\,4K$^\pm$ & K$^-$ anything & \,\,24.2 & & \,\, + 
$2\pi^\pm$)\\
 & & 0.9954\,$\cdot$\,(p + $\bar{\mathrm{n}}$) & $\overline{\mathrm{K}}^0$ 
anything\\	
 & &                       & \,\,+\,K$^0$ anything & \,\,59\\
 & &                       & $\eta$ anything & $<$\,13\\
&&                          &K$^+$ anything &5.8\\
\\
D$^\pm_s$ & 14.104 & 0.8296\,$\cdot$\,17$\pi^\pm$ & K$^-$ anything & 
\,\,13 & 0 & 
body\\
 & & 0.9968\,$\cdot$\,\,4K$^\pm$ & $\overline{\mathrm{K}}^0$ 
 anything & & &  centered\\
 & &                            & \,\,+\,K$^0$ anything & \,\,39 & & 
cubic\\
 & &                            & K$^+$ anything & \,\,20\\
 & &                            & $e^+$ anything & 8\\
[0.2cm]\hline\hline

\vspace{0.1cm}
	\end{tabular}

	\footnotemark{\footnotesize The particles with negative 
	charges have conjugate charges of the listed  decays. Only the 
	decays of K$^\pm$ and D$^\pm$ are listed. The oscillation modes carry 
            one electric charge. The $\ast$ marks coupled modes.}

	\end{table}

These particles are in general charged, exempting the K$^0$ and 
D$^0$  mesons and the neutron n, in contrast to the particles of the 
$\gamma$-branch, which are in general neutral. It does not make a 
significant difference whether one considers the mass of a particular 
charged or neutral particle. After the $\pi$\,mesons, the largest mass 
difference between charged and neutral particles is that of the K\,mesons 
(0.81\%), and thereafter all mass differences between charged and neutral 
particles are $<0.5\%$. The integer multiple rule does not immediately 
apply to the masses of the $\nu$-branch particles if m($\pi^\pm$) (or 
m($\pi^0$)) is used as reference, because m(K$^\pm$) = 
0.8843\,$\cdot$\,4m($\pi^\pm$). 0.8843\,$\cdot$\,4 = 3.537 is far from 
integer. Since the masses 
 of the $\pi^0$\,meson and the $\pi^\pm$\,mesons differ by only 
3.4\% it has been argued that the $\pi^\pm$\,mesons are, but for the 
isospin, the same particles as the $\pi^0$\,meson, and that therefore  
the $\pi^\pm$\,mesons cannot start another particle branch. However, 
this argument is not supported 
by the completely different decays of the $\pi^0$\,mesons 
and the $\pi^\pm$\,mesons. The $\pi^0$\,meson decays almost 
exclusively into $\gamma\gamma$ (98.8\%),  whereas the 
$\pi^\pm$\,mesons decay practically exclusively into $\mu$\,mesons 
and neutrinos, as in $\pi^+$ $\rightarrow$  
$\mu^+$  + $\nu_\mu$  (99.9877\%). Furthermore, the lifetimes of the 
$\pi^0$  and the $\pi^\pm$ mesons differ by nine orders of magnitude, 
being $\tau$($\pi^0$) = 8.4\,$\cdot$\,10$^{-17}$ sec versus 
$\tau$($\pi^\pm$) 
= 2.6\,$\cdot$\,10$^{-8}$ sec.

   If we make the $\pi^\pm$\,mesons the reference particles of the 
$\nu$-branch, then we must multiply the mass ratios m/m($\pi^\pm$) of the 
above listed particles with an average factor 0.848 $\pm$ 0.025, as 
follows from the mass ratios on Table 2. 
The integer multiple rule may, however, apply 
directly if one makes m(K$^\pm$) the reference for masses larger than 
m(K$^\pm$). The mass of the neutron is 0.9516\,$\cdot$\,2m(K$^\pm$),
 which is only a fair approximation to an integer multiple. There are, on 
the other 
hand, outright integer multiples in m(D$^\pm$) = 0.9954\,$\cdot$\,(m(p) + 
m($\bar\mathrm{n}$)), and in m(D$_s^\pm$) = 0.9968\,$\cdot$\,4m(K$^\pm$). 
A least square 
analysis of the masses of the $\nu$-branch in Table 2 yields the formula

\begin{equation} \mathrm{m}(N)/0.853\mathrm{m}(\pi^\pm) = 1.000\,N + 
0.00575\qquad  N\,>\, 1 ,
\end{equation}
\noindent
with R$^2$ = 0.998. This means that the particles of the $\nu$-branch are 
integer multiples of m($\pi^\pm$) times the factor 0.853. One must, 
however, consider that the $\pi^\pm$\,mesons are not necessarily the 
perfect reference for all $\nu$-branch particles, because $\pi^\pm$ has  
I = 1, whereas for example K$^\pm$ has I = 1/2 and S = $\pm$1 and  
the neutron has also I = 1/2. Actually the 
factor 0.853 in Eq.(3) is only an average. The mass ratios indicate that 
this factor  decreases slowly with increased m(N).
 The existence of the factor and its decrease will be explained later.

   Contrary to the particles of the $\gamma$-branch, the $\nu$-branch 
particles decay preferentially with the emission of neutrinos, the 
foremost example is $\pi^\pm \rightarrow \mu^\pm$ + 
$\nu_\mu(\bar{\nu}_\mu)$ with a 
fraction of 99.9877\%. Neutrinos characterize the weak interaction. We 
will refer to the particles in Table 2 as the $\emph{neutrino branch}$ 
($\nu$-branch) of the particle spectrum. We emphasize that a weak decay of 
the particles of the $\nu$-branch is by no means guaranteed. Although the 
neutron decays via n $\rightarrow$ p + e$^-$ + $\bar{\nu}_e$ in 887 sec 
(100\%), the proton is stable. There are, on the other hand, weak decays 
such as e.g. K$^+ 
\rightarrow \pi^+\pi^-\pi^+$ (5.59\%), but the subsequent decays of the 
$\pi^\pm$\,mesons lead to neutrinos and e$^\pm$. 

   To summarize the facts concerning the $\nu$-branch of the mesons 
and bary-ons. The masses of these particles seem to follow the integer 
multiple rule 
 if one uses the $\pi^\pm$\,mesons as reference, however the mass 
ratios share a common factor 0.85 $\pm$ 0.025.

   To summarize what we have learned about the \emph{integer 
multiple rule}: In spite of differences in charge, spin, strangeness, and 
charm the masses of the ``stable" mesons and baryons of the 
$\gamma$-branch  are integer multiples of the mass 
 of the $\pi^0$\,meson within at most 3.3\% and on the average 
within 0.66\%. Correspondingly, the masses of the ``stable" 
particles of the $\nu$-branch are, after multiplication with a factor 0.85 
$\pm$ 0.025, integer multiples of the mass of the $\pi^\pm$\,mesons. 
The integer multiple rule has been anticipated 
much earlier by Nambu [9], who wrote in 1952 that ``some regularity 
[in the masses of the particles] might be found if the masses were 
measured in a unit of the order of the $\pi$-meson mass". A similar 
suggestion has been made by Fr\"ohlich [10]. The integer multiple rule 
suggests that the particles are the result of superpositions of modes 
and higher modes of a wave equation.

\section {Standing waves in a cubic lattice} 

 We will now study, as we 
have done in [11], whether the so-called ``stable" particles of the 
$\gamma$-branch cannot be described by the frequency spectrum of 
standing waves in a cubic lattice, which can accommodate automatically 
the Fourier frequency spectrum of an extreme short-time collision by 
which the particles are created. 
The investigation of the consequences of lattices for particle theory 
was initiated by Wilson [12] who studied a cubic fermion lattice. His 
study has developed over time into lattice QCD. 

   It will be necessary for the following to outline the most elementary 
aspects of the theory of lattice oscillations. The classic paper describing
 lattice oscillations is from Born 
 and v.\,Karman [13], henceforth referred to as B\&K. They looked at first 
at the oscillations of a one-dimensional chain of points with mass m, 
separated by a constant distance $\emph{a}$. This is the 
$\emph{monatomic}$ case, all lattice points have the same mass. B\&K 
assume 
that the forces exerted on each point of the chain originate only from the 
two neighboring points. These forces are opposed to and proportional to 
the displacements, as with elastic springs (Hooke's law). The equation of 
motion is in this case

\begin{equation} \mathrm{m}\ddot{u}_{n} = \alpha(u_{n+1} - u_n)
 - \alpha(u_n - u_{n-1})\,. \end{equation}
\noindent
The $u_n$ are the displacements of the mass points from their 
equilibrium position 
which are apart by the distance \emph{a}. The dots signify, as usual, 
differentiation with respect 
to time, $\alpha$ is a constant characterizing the force between the 
lattice points, and n is an integer number. For \emph{a} $\rightarrow$ 0 
Eq.(4) becomes the wave equation c$^2\partial^2$u/$\partial$x$^2$ = 
$\partial^2$u/$\partial$t$^2$ (B\&K).
 
   In order to solve Eq.(4) B\&K set 

\begin{equation} u_n = Ae^{i(\omega\,t\, +\, n\phi)}\,,
\end{equation}

\noindent
which is obviously a temporally and spatially periodic solution or 
describes \emph{standing waves}. n is an 
integer, with n  $<$ N, where N is the number of points in the chain. 
$\phi$ = 0 is the monochromatic case. We also consider higher modes,
 by replacing n$\phi$ in Eq.(5) by
$\ell$n$\phi$, with integer $\ell$\,$>$\,1. The wavelengths are then 
shorter by 1/$\ell$. At n$\phi$ = $\pi$/2 are 
nodes, where for all times $\emph{t}$ the displacements are zero, as 
with standing waves f(x,t) = A\,cos($\omega$t)\,cos(n$\phi$) = 
A\,cos($\omega$t)\,cos(kx). 
If a displacement is repeated after n points we have n$\emph{a}$ = 
$\lambda$, where $\lambda$ is the wavelength, $\emph{a}$ the lattice 
constant, and it must be n$\phi$ = 2$\pi$ according to (5). It follows that

\begin{equation} \lambda = 2\pi\emph{a}/\phi\,.
\end{equation}
 
Inserting (5) into (4) one obtains a continuous frequency spectrum of  
the standing waves as given by Eq.(5) of B\&K

\begin{equation}  \omega = \pm\,2\sqrt{\alpha/\mathrm{m}}
\,\mathrm{sin}(\phi/2)\,. \end{equation}
\noindent
 B\&K point out that there is not only a continuum 
of frequencies, but also a  \emph{maximal frequency} which 
is reached when $\phi$ = $\pi$, or at the 
minimum of the possible wavelengths $\lambda$ = 2$\emph{a}$. The 
boundary conditions are periodic, that means that $u_n$ = $u_{n + N}$, 
where N is the number of points in the chain. Born referred to the 
periodic boundary condition as a ``mathematical convenience". The number 
of normal modes must be equal to the number of particles in the lattice.
 
    Born's model of the crystals has been verified in great detail by X-ray
 scattering and even in much more complicated cases by neutron scattering. 
 The theory of lattice oscillations has been pursued in particular by 
Blackman [14], a summary of his and other studies is in [15]. 
Comprehensive reviews of the results of linear studies of lattice dynamics 
have been written by Born and Huang [16], by Maradudin et al. [17], and by 
Ghatak and Kothari [18].

\section {The masses of the $\gamma$-branch particles}

   We will now assume, as seems to be quite natural, that the particles
 \emph{consist of the same particles into which they 
decay}, directly or ultimately. We know this from atoms, which 
consist of nuclei and electrons, and from nuclei, which consist of 
protons and neutrons. Quarks 
have never been observed among the decay products of 
elementary particles. For the $\gamma$-branch particles 
our assumption means that they consist of photons. Photons 
and $\pi^0$\,mesons are the principal decay products of the 
$\gamma$-branch particles, the characteristic example 
 is $\pi^0$ $\rightarrow$ $\gamma\gamma$  (98.8\%). Table 1 
shows that there are decays of the $\gamma$-branch particles which 
lead to particles of the $\nu$-branch, in particular to pairs of $\pi^+$ 
and $\pi^-$ mesons.  It appears that this has to do with pair 
production in the $\gamma$-branch particles. Pair production 
is evident in the decay $\pi^0 \rightarrow$ e$^+$ + e$^- + 
\gamma$ (1.198\%) or in the $\pi^0$\,meson's third most frequent 
decay $\pi^0 \rightarrow$ e$^+$e$^-$e$^+$e$^-$ 
(3.14$\cdot10^{-3}$\%). Pair production requires the
presence of electromagnetic waves of high energy. Anyway, 
the explanation of the $\gamma$-branch particles must begin 
with the explanation of the most simple example of its kind, the 
$\pi^0$\,meson, which by all means seems to consist of photons. 
The composition of the particles of the $\gamma$-branch suggested here 
offers a direct route from the formation of a $\gamma$-branch particle, 
through its lifetime, to its decay products. Particles that are made of 
photons are necessarily neutral, as the majority of the particles of the 
$\gamma$-branch are.
 
   We also base our assumption that the particles of the $\gamma$-branch 
are made of photons on the circumstances of the formation of the 
$\gamma$-branch particles. The most simple and straightforward creation of 
a $\gamma$-branch particle are the reactions $\gamma$ + p $\rightarrow$ 
$\pi^0$ + p, or in the case that the spins of $\gamma$ and p are parallel 
\,\,$\gamma$ + p $\rightarrow$ 
$\pi^0$ + p + $\gamma^\prime$. A photon impinges on a proton and creates  
a $\pi^0$\,meson. The considerations which follow apply as well for other
 photoproductions such as $\gamma$ + p $\rightarrow \eta$ + p or $\gamma$ 
+ d  $\rightarrow \pi^0$ + d and to the photoproduction of $\Lambda$  in
$\gamma$ + p $\rightarrow \Lambda$ + K$^+$, but also for the 
electroproductions e$^-$ + p $\rightarrow 
\pi^0$ + e$^-$ + p or e$^-$ + d $\rightarrow$   $\pi^0$  + e$^-$ + d,  
see Rekalo et al. [19]. The most simple example of the creation of a 
$\gamma$-branch particle by a strong interaction is the reaction 
p + p $\rightarrow$ p + p + $\pi^0$. The electromagnetic energy 
accumulated in a proton during its acceleration reappears as the 
$\pi^0$\,meson. 

   In $\gamma$ + p $\rightarrow \pi^0$ + p   the pulse of the 
incoming electromagnetic wave is in 10$^{-23}$\,sec  converted 
into a continuum of electromagnetic waves with frequencies ranging  
from 10$^{23}$ sec$^{-1}$ to $\nu$ $\rightarrow$ $\infty$ according  
to Fourier analysis. There must be a cutoff frequency, 
 otherwise the energy in the sum of the frequencies would exceed the 
 energy of the incoming electromagnetic wave. The wave packet so 
created decays, according to experience, after 
8.4\,$\cdot$\,10$^{-17}$\,sec into two 
electromagnetic waves or $\gamma$-rays. It seems to be very 
unlikely that Fourier analysis does not hold for the case of an 
electromagnetic wave impinging on a proton. The question then arises of 
what happens to the electromagnetic waves in the timespan of 10$^{-16}$ 
seconds between the creation of the wave packet and its decay into two 
$\gamma$-rays\,? We will show that the electromagnetic waves 
can continue to exist for the 10$^{-16}$ seconds until the wave packet 
decays.

   If the wave packet created by the collision of a $\gamma$-ray with a 
proton
consists of electromagnetic waves, then the waves cannot be progressive
because the wave packet must have a \emph{rest mass}. The rest mass is 
the mass of a particle whose center of mass does not move. However 
\emph{standing electromagnetic waves}  have a rest mass. Standing 
electromagnetic waves are equivalent to a lattice, because in standing 
waves the waves travel back and forth between the nodes, just as lattice 
points oscillate between the nodes of the lattice oscillations. 
The oscillations in the lattice take care of the continuum of
 frequencies of the Fourier spectrum of the collision. So we assume that 
the very many photons in the wave packet are held together in a cubic
 lattice. It is not unprecedented that photons have been 
considered to be building blocks of the elementary particles. Schwinger 
[20] has once studied an exact one-dimensional quantum electrodynamical 
model in which the photon acquired a mass $\sim$ $e^2$. On the other 
hand,  it has been suggested by Sidharth [21] that the  $\pi^0$ meson 
consists of an electron and a positron which circle their center of mass.

   We will now investigate the standing waves in a cubic photon 
lattice. We assume that the lattice is held together by a weak force 
acting 
from one lattice point to its nearest neighbors. We assume that the range  
of this force is 10$^{-16}$\,cm, because the range of the weak nuclear  
force is on the order of 10$^{-16}$\,cm, as stated e.g. on p.25 of
 Perkins [22].  For the sake of simplicity we set the 
sidelength of the lattice at 10$^{-13}$\,cm, the exact size of the nucleon 
is given in [23] and will be used later. With \emph{a} = $10^{-16}$\,cm 
there are then 10$^9$ lattice 
points. As we will see the ratios of the masses of the particles are 
independent of the sidelength
 of the lattice. Because it is the most simple case, we assume 
that a central force acts between the lattice points. We cannot consider 
spin, isospin, strangeness or charm of the particles. The frequency 
equation for the oscillations in an isotropic monatomic cubic lattice with 
central forces is, in the one-dimensional case, given by Eq.(7). The 
direction of the oscillation is determined by the direction of the 
incoming wave.

   According to Eq.(13) of B\&K the force constant $\alpha$ is

\begin{equation}   \alpha = \emph{a}\,(c_{11} - c_{12} - c_{44})\,,
\end{equation}

\noindent
where c$_{11}$, c$_{12}$ and c$_{44}$ are the elastic constants in 
continuum mechanics which applies in the limit $\emph{a}$ $\rightarrow$  
0. If we consider central forces then c$_{12}$ = c$_{44}$ which is the 
classical Cauchy relation. Isotropy requires that c$_{44}$ = 
(c$_{11}\,-\,$  c$_{12}$)/2.  
The waves are longitudinal. Transverse waves in a cubic lattice 
with concentric forces are not possible according to [18]. All frequencies 
that solve Eq.(7) come with either a plus or a minus sign which is, as we 
will see, important. The reference frequency in Eq.(7) is

\begin{equation}  \nu_0 = \sqrt{\alpha/4\pi^2\mathrm{m}}\,,
\end{equation}

\noindent
or as we will see, using Eq.(11),  $\nu_0$  =  c$_\star/2\pi\emph{a}$.

   The \emph{limitation of the group velocity} has 
now to be considered. The group velocity is given by

\begin{equation}  c_g = \frac{d\omega}{dk} = 
\emph{a}\sqrt{\frac{\alpha}{\mathrm{m}}}\cdot\frac{df(\phi)}{d\phi}\,.
\end{equation}
\noindent
The group velocity of the photons has to be equal to the 
\emph{velocity of  light} 
c$_\star$ throughout the entire frequency spectrum, because 
photons move with the velocity of light. In order to learn how this 
requirement affects the frequency distribution we have to know the value 
of $\sqrt{\alpha/\mathrm{m}}$ in a photon lattice. But we do not have 
information 
about what either $\alpha$ or m might be in this case. In the following 
we set $\emph{a}\sqrt{\alpha/\mathrm{m}}$ = c$_\star$, which means, 
since $\emph{a} = 10^{-16}$\,cm, that $\sqrt{\alpha/\mathrm{m}}$ = 
3\,$\cdot$\,10$^{26}$ sec$^{-1}$, or that the corresponding period is 
$\tau$ = 1/3\,$\cdot$\,10$^{-26}$ sec, which is the time it takes for a 
wave 
to travel with the velocity of light over one lattice distance. With 

\begin{equation} c_\star = \emph{a}\sqrt{\alpha/\mathrm{m}}\,
\end{equation}

\noindent
the equation for the group velocity is

\begin{equation} c_g = c_\star\cdot df/d\phi\,.
\end{equation} 

  For photons that means, since c$_g$ 
must then always be equal to the velocity of light 
 c$_\star$, that df/d$\phi$ = 1. This requirement determines the form of 
the frequency distribution regardless of the 
order of the mode of oscillation or it means that instead of the sine 
function in Eq.(7) the frequency is given by

\begin{equation}\nu = \pm\,\nu_0(\,\phi + \phi_0\,)\,. \end{equation} 

   For the time being we will disregard $\phi_0$ in Eq.(13). The 
frequencies
 of the corrected spectrum in Eq.(13) 
must increase from $\nu$ = 0 at the origin $\phi$ = 0 with slope 
1 (in units of $\nu_0$) until the maximum is reached at $\phi = \pi$. 
The energy  contained in the oscillations  must be proportional to
 the sum of all frequencies (Eq.14). The
 \emph{second mode}  of the lattice oscillations contains 4 times as much 
energy as  the basic mode, because the frequencies are twice the 
frequencies of the basic mode, and there are twice as many oscillations. 
 Adding, by superposition, to the second mode different 
numbers of basic modes or of second modes will give exact integer 
multiples of the  
energy of the basic mode. Now we understand the integer multiple 
rule of the particles of the $\gamma$-branch. There is, in the framework 
of this theory, on account of Eq.(13), no alternative but $\emph{integer 
multiples}$ of the basic mode for the energy contained in the frequencies 
of the different modes or for superpositions of different modes. In other 
words, the masses of the different particles are integer multiples of the 
mass of the $\pi^0$\,meson, if there is no spin, isospin, 
strangeness or charm.

   We remember that the measured masses in Table 1, which 
incorporate different spins, isospins, strangeness and charm, spell 
out the integer multiple rule within on the average 0.65\% accuracy. It is 
worth noting that $\emph{there is no free parameter}$ if one takes
 the ratio of the energies contained in the frequency distributions of 
the different modes, because the factor $\sqrt{\alpha/\mathrm{m}}$ 
in Eq.(7) or $\nu_0$ in Eq.(13) cancels. This means, 
in particular, that the ratios of the frequency distributions, or the mass 
ratios, are independent of the mass of the photons at the lattice points, 
as well as of the magnitude of the force between the lattice points.

   It is obvious that the integer multiples of the frequencies are only a 
first approximation of the theory of lattice oscillations and of the mass 
ratios of the particles. The equation of motion in the lattice Eq.(4) does 
not apply in the eight corners of the cube, nor does it 
apply to the twelve edges nor, in particular, to the six sides of the 
cube. A cube with 10$^9$ lattice points is not correctly described by  
the periodic boundary condition we have used to derive Eq.(7), but  
is what is referred to as a microcrystal. 
 A phenomenological theory of the frequency distributions in 
microcrystals, considering in particular the surface energy, can be found in 
Chapter 6 of Ghatak and Kothari [18]. The surface energy may account for the 
small deviations of the mass ratios of the mesons and baryons from the 
integer multiple rule of the oscillations in a cube. However, it seems to 
be futile to pursue 
a more accurate determination of the oscillation frequencies before 
we know what the interaction of the electron with mass is. The mass 
of the electron is 0.378\% of the mass of the $\pi^0$\,meson and 
hence is a substantial part of the deviation of the mass ratios from  
the integer multiple rule.
 
   Let us summarize our findings concerning the $\gamma$-branch. 
The particles of the $\gamma$-branch consist of standing 
electromagnetic waves. The $\pi^0$\,meson is the basic mode.  
The $\eta$ meson corresponds to the 
second mode, as is suggested by m($\eta$) $\approx$ 
4m($\pi^0$). The $\Lambda$ particle corresponds to the superposition
 of two second modes, as is suggested by m($\Lambda) \approx$ 
2m($\eta$). This superposition apparently results in the creation of 
spin 1/2. The two modes would then have to be coupled. The
 $\Sigma^0$ and $\Xi^0$ baryons are superpositions 
 of one or two basic modes on the $\Lambda$  particle. The 
$\Omega^-$ particle corresponds to the superposition of three coupled 
second modes  as is suggested by m($\Omega^-$) $\approx$ 
3m($\eta$). This procedure apparently causes spin 3/2. The charmed 
$\Lambda_c^+$ particle seems to be the first particle incorporating a 
third mode. $\Sigma_c^0$ is apparently the superposition of  
a negatively charged basic mode on  $\Lambda_c^+$, as is 
suggested by the decay of $\Sigma_c^0$. The easiest explanation 
of $\Xi_c^0$ is that it is the superposition of two coupled 
third modes. The superposition of two modes of the same type is, 
as in the case of $\Lambda$, accompanied by spin 1/2. The $\Omega_c^0$ 
baryon is apparently the superposition of two basic modes on the $\Xi_c^0$ 
particle. All neutral particles of the $\gamma$-branch are thus accounted 
for. The explanation of the charged $\gamma$-branch particles 
$\Sigma^\pm$  and $\Xi^-$ has been discussed in [56]. 
The modes of the particles are listed in Table 1.

   We have also found the $\gamma$-branch $\emph{antiparticles}$,  
which follow from the negative frequencies which solve Eq.(7) or Eq.(13). 
Antiparticles have always been associated with negative energies. 
Following Dirac's argument for electrons and positrons, we associate  
the masses with the negative frequency distributions with 
antiparticles. We emphasize that the existence of antiparticles is an 
automatic consequence of our theory.

   All particles of the $\gamma$-branch are unstable with lifetimes on the 
order of 10$^{-10}$ sec or shorter. Born [24] has shown that the 
oscillations in cubic lattices held together by central forces are 
unstable. It seems, however, to be possible that the particles can be 
unstable for reasons other than the instability of the lattice which 
apparently 
causes the most frequent decay of the $\pi^0$\,meson $\pi^0 \rightarrow 
\gamma\gamma$ (98.798\%), or the most frequent decay of the 
$\eta$ meson $\eta \rightarrow \gamma\gamma$ (39.43\%).
 Pair production seems to make it possible to understand the decay 
of the $\pi^0$\,meson $\pi^0 \rightarrow$ e$^-$ + e$^+ + \gamma$  
(1.198\%).  Since in our model the $\pi^0$\,meson consists of a 
multitude of electromagnetic waves it seems that pair 
production takes place within the $\pi^0$\,meson, and even more so 
in the higher modes of the $\gamma$-branch where the electrons and 
positrons created by pair production tend to settle on mesons, as e.g. 
in $\eta \rightarrow \pi^+ +  \pi^- + \pi^0$ (22.6\%) or in the decay 
$\eta \rightarrow \pi^+ + \pi^- + \gamma$ (4.68\%), where the origin 
of the pair of charges is more apparent. Pair production is also evident 
in the decays $\eta \rightarrow$ e$^+$e$^-\gamma$ (0.6\%)  or 
$\eta \rightarrow$ e$^+$e$^-$e$^+$e$^-$ (6.9$\cdot10^{-3}$\%).

   Finally we must explain the reason for which the photon lattice or the 
$\gamma$-branch particles are limited in size to a particular value of 
about $10^{-13}$ cm, as 
the experiments tell. Conventional lattice theory using the periodic 
boundary condition does not limit the size of a crystal, and in fact very 
large crystals exist. If, however, the lattice consists of standing 
electromagnetic waves the size of the lattice is limited by the radiation 
pressure. The lattice will necessarily break up at the latest when the 
outward directed radiation pressure is equal to the inward directed 
elastic force which holds the lattice together. For details we refer to 
[25].

\section {The rest mass of the $\pi^0$\,meson}

   So far we have studied the ratios of the masses of the particles. We 
will now determine the mass of the $\pi^0$\,meson in order to validate 
that the mass ratios link with the actual masses of the particles. The 
energy of the $\pi^0$\,meson is \vspace{0.5cm}

\centerline{E(m($\pi^0$)) = 134.9766\,MeV =
 2.16258\,$\cdot\,10^{-4}$\,erg.}
\vspace{0.5cm}
\noindent
The sum of the energies  E = h$\nu$ of the frequencies of all standing 
one-dimensional waves in $\pi^0$ seems to be given by the equation

\begin{equation}\mathrm{E}_\nu = 
\frac{\mathrm{Nh}\nu_0}{2\pi}\,\,\int\limits_{-\pi}^{\pi}\,f(\phi)d\phi\,.
\end{equation}

  This equation originates from B\&K. N is the number of all lattice 
points. The total energy of the frequencies in a cubic lattice is equal 
to the number N of the oscillations times the average of the energy of the 
individual frequencies. In order to arrive at an exact value of N in 
Eq.(14) 
we have to use the correct value of the radius of the proton, which is 
r$_p$ = (0.880 $\pm$ 0.015)\,$\cdot$\,10$^{-13}$\,cm according to [23] or
r$_p$ = (0.883 $\pm$  0.014)\,$\cdot$\,10$^{-13}$\,cm according to [26].  
With $\emph{a}$ = 10$^{-16}$\,cm it follows that the number of all lattice 
points in the cubic lattice is

\begin {equation} \mathrm{N} = 2.854\cdot10^9 \cong 1\,418^{\,3}\,.
 \end{equation}
\noindent
When the number of the grid points of a cubic lattice is derived from the 
volume of a sphere one cannot arrive at an integer number to the third.
The radius of the $\pi^\pm$\,mesons has also been measured [27] and after 
further analysis [28] was found to be 0.83\,$\cdot$\,10$^{-13}$\,cm, which 
means that within the uncertainty of the radii we have r$_p$ = r$_\pi$. 
And according to [29] the charge radius of $\Sigma^-$ is (0.78 $\pm$ 
0.10)\,$\cdot$\,$10^{-13}$\,cm.

  If the oscillations are parallel to an axis, the 
limitation of the group velocity is taken into account, that means if 
Eq.(13) applies and the absolute 
values of the frequencies are taken, then the value of the 
 integral in Eq.(14) is $\pi^2$. With N = 
2.854\,$\cdot$\,10$^9$ and $\nu_0$ = c$_\star/2\pi\emph{a}$  follows  
from Eq.(14) that the sum of the energy of the frequencies of the basic 
mode corrected for the group velocity limitation is 
E$_{corr}$ = 1.418\,$\cdot$\,10$^9$\,erg. That means  
that the energy is 6.56\,$\cdot\,10^{12}$ times larger than E(m($\pi^0$)). 
This discrepancy is inevitable, because the basic frequency of the 
Fourier spectrum after a collision on the order of 
10$^{-23}$ sec duration is $\nu$ = 10$^{23}$ sec$^{-1}$, which means,  
when E = h$\nu$, that one basic frequency alone contains an energy of  
about 9\,m($\pi^0$)c$_\star^2$.
 
   To eliminate this discrepancy we use, instead of the simple 
form E = h$\nu$, the complete quantum mechanical energy of a linear 
oscillator as given by Planck

\begin{equation} E = \frac{h\nu}{e^{h\nu/kT} -\,1}\,\,.
\end{equation}
\noindent
This equation was already used by B\&K for the determination of the 
specific heat of cubic crystals or solids. Equation (16) calls into 
question the value of the temperature T in the interior of a particle. We 
determine T empirically with the formula for the internal energy of solids

\begin{equation} u = \frac{R\Theta}{e^{\Theta/T} -1}\,\,,
\end{equation}
\noindent
which is from Sommerfeld [30]. In this equation R = 
2.854\,$\cdot\,10^9$\,k, 
where k is Boltzmann's constant, and $\Theta$ is the 
characteristic temperature introduced by Debye [31] for the explanation of 
the specific heat of solids. It is $\Theta = h\nu_m$/k, where $\nu_m$ is a 
maximal frequency. In the case of the oscillations making up the 
$\pi^0$\,meson the maximal frequency is $\nu_m = \pi\nu_0$,  therefore 
$\nu_m = 1.5\cdot10^{26}$ sec$^{-1}$, and we find that $\Theta = 
7.2\cdot10^{15}$\,K.
 
  In order to determine T we set the internal energy u equal to 
m$(\pi^0)$c$_\star^2$. It then follows from Eq.(17) that $\Theta$/T = 
30.20, or 
T = 2.38\,$\cdot$\,10$^{14}$\,K. That means that Planck's formula (16) 
introduces a 
factor $1/(e^{\Theta/T} - 1\,) \cong 1/e^{30.2}$ = 1/(1.305$\cdot10^{13}$) 
into Eq.(14). In other words, if we determine the temperature T of the 
particle through Eq.(17), and correct Eq.(14) accordingly then we arrive 
at a sum of the oscillation energies in the $\pi^0$\,meson which is 
\begin{equation} \sum_1^N\mathrm{E}_\nu = 1.0866\cdot10^{-4}\,
\mathrm{erg} = 67.82\,\mathrm{MeV}\,.
\end{equation}
 That  means that the sum of the energies of the one-dimensional 
oscillations  consisting of N waves  is 0.502\,E(m($\pi^0$)). We 
have to double this amount because standing waves consist  of 
two waves traveling in opposite direction with 
the same absolute value of the frequency. The sum of the 
energy of the oscillations in the $\pi^0$\,meson is therefore 
\begin{equation} \mathrm{E}_\nu(\pi^0)(theor) = 
2.1732\cdot10^{-4}\,\mathrm{erg} = 135.64\,\mathrm{MeV} = 
1.005\,\mathrm{E(m}(\pi^0))(exp)\,,\end{equation}
if the oscillations are parallel to the $\phi$ axis. The energy in the 
measured mass of the $\pi^0$\,meson and the energy 
in the sum of the oscillations agree fairly well, considering the 
uncertainties of the parameters involved. The theoretical mass of the 
$\eta$ meson is then m($\eta$) = 4\,$\cdot$\,m($\pi^0$)(\emph{theor}) =
542.56\,MeV = 0.991\,m($\eta$)(\emph{exp}), and the theoretical mass 
of the $\Lambda$\,baryon, the superposition of two $\eta$ mesons, is 
then m($\Lambda$)(\emph{theor}) = 8\,$\cdot$\,m($\pi^0$)(\emph{theor})
 = 1085.1\,MeV = 0.9726\,m($\Lambda$)(\emph{exp}).  

   To sum up: The $\pi^0$\,meson is formed when a $\gamma$-ray
collides with a proton, $\gamma$ + p $\rightarrow \pi^0$ + p. By the 
collision the incoming $\gamma$-ray is converted into a packet of standing 
electromagnetic waves, the $\pi^0$\,meson. After $10^{-16}$ seconds
the wave packet decays into two electromagnetic waves, 
$\pi^0 \rightarrow \gamma\gamma$. Electromagnetic waves prevail 
throughout the entire process. The energy in the rest mass 
of the $\pi^0$\,meson
 and the other particles of the $\gamma$-branch is correctly given 
by the sum of the energy of  standing electromagnetic waves in a cube,  
if the energy of the oscillations is determined by Planck's formula for 
the energy of a linear oscillator. \emph{The $\pi^0$\,meson is like an 
adiabatic, cubic black body filled with standing electromagnetic waves}. 
A black body of a given size and temperature can certainly contain the 
energy in the rest mass of the $\pi^0$\,meson, which is O($10^{-4})$
erg, if only the frequencies are sufficiently high. We know from 
Bose's work [32] that Planck's formula applies to a photon gas as well.
For all $\gamma$-branch particles we have found a simple mode of 
standing electomagnetic waves. Since the equation determining 
 the frequency of the standing waves is quadratic it follows 
\emph{automatically} that for each positive frequency there  
is also a negative 
frequency of the same absolute value, that means that for each particle 
there exists also an antiparticle. For the explanation of the stable mesons 
and baryons of the $\gamma$-branch $\emph{we use only photons,
 nothing else}$. This is a rather conservative
 explanation of the $\pi^0$\,meson and the $\gamma$-branch 
particles.  \emph{We do not use hypothetical particles}.

   From the frequency distributions of the standing waves follow 
the ratios of the masses of the particles which obey the integer 
multiple rule. It is important to note that in this theory the ratios of 
the masses of the $\gamma$-branch particles to the mass of the 
$\pi^0$\,meson $\emph{do not depend}$ on the sidelength of 
the lattice, and the distance between 
the lattice points, neither do they depend on the strength 
of the force between the lattice points nor on the mass of the 
lattice points. The mass ratios are determined only by the spectra  
of the frequencies of the standing electromagnetic waves. 

\section {The neutrino branch particles}

The masses of the neutrino branch, the $\pi^\pm$, K$^{\pm,0}$, n, 
D$^{\pm,0}$ and D$^\pm_s$ particles, are integer multiples of the 
mass of the $\pi^\pm$\,mesons times a factor $0.85\,\pm\,0.02$ as 
we stated before. We assume, 
as appears to be quite natural, that the $\pi^\pm$\,mesons and the 
 other particles of the $\nu$-branch $\emph{consist of the 
same particles into}\\ 
\emph{which they decay}$, that means in the case of the 
$\pi^\pm$\,mesons of muon neutrinos, anti-muon neutrinos, 
electron neutrinos, anti-electron neutrinos
 and of an electron or positron,  as exemplified by the 
decay sequence 
$\pi^\pm$  $\rightarrow$  $\mu^\pm$ + $\nu_\mu$($\bar{\nu}_\mu$),
$\mu^\pm$ $\rightarrow$ e$^\pm$ + $\bar{\nu}_\mu$($\nu_\mu$) 
+ $\nu_e$($\bar{\nu}_e$). The absence of an electron neutrino $\nu_e$
in the decay branches of $\pi^-$ or of an anti-electron neutrino 
$\bar{\nu}_e$ in the decay
branches of $\pi^+$ can be explained with the composition of 
the electron or positron, which will be discussed in Section 9. The   
existence of neutrinos and antineutrinos is unquestionable.
 Since the particles of the $\nu$-branch decay through weak  
decays, we assume, as appears likewise to be natural, that 
\emph{the weak nuclear force holds the particles of the $\nu$-branch 
together}. This assumption has far reaching consequences, it is not
only fundamental for the explanation of the $\pi^\pm$\,mesons, but
 leads also to the explanation of the $\mu^\pm$\,mesons and 
ultimately to the explanation of the mass of the electron.  
The existence of the weak nuclear force is unquestionable. Since 
the range of the weak interaction, which is about 10$^{-16}$ cm [22],
 is only about a thousandth of the diameter of the particles, which is 
about
 10$^{-13}$ cm, the weak force can hold particles together only 
if the particles have a lattice structure, just as macroscopic crystals 
are  held together by microscopic forces between atoms. In the 
absence of a force which originates in the center of the particle and 
affects all neutrinos of the particle the configuration of the particle is 
not spherical but cubic,
 reflecting the very short range of the weak nuclear force.  We will  
show that the energy in the rest mass of the $\nu$-branch particles is 
the energy in the oscillations of a cubic lattice consisting of electron 
and muon neutrinos and their antiparticles, plus the energy  in the rest 
masses of the neutrinos.

   First it will be necessary to outline the basic aspects of diatomic 
lattice 
oscillations. In $\emph{diatomic}$ lattices the lattice points have 
alternately the masses m and M, as with the masses of the electron
 neutrinos m($\nu_e$) and muon neutrinos m($\nu_\mu$).
 The classic example of a diatomic lattice is the salt 
crystal with the masses of the Na and Cl atoms in the lattice points. 
The theory of diatomic harmonic lattice oscillations was started by 
Born and v.\,Karman [13]. They first discussed a diatomic chain. The 
equation of motions in the chain are according to Eq.(22) of B\&K

\begin{equation} \mathrm{m}\ddot{u}_{2n} = \alpha(u_{2n+1} +
 u_{2n-1} - 2u_{2n})\, ,\end{equation}

\begin{equation} \mathrm{M}\ddot{u}_{2n+1} = \alpha(u_{2n+2} +
 u_{2n} - 2u_{2n+1}) \,, \end{equation}
\noindent    
where the u$_n$ are the displacements, n an integer number and $\alpha$ a 
constant characterizing the force between the particles. Eqs.(20,21) are 
solved with

\begin{equation}u_{2n} = Ae^{i(\omega\,t\,+\,2n\phi)} ,\end{equation} 

\begin{equation}u_{2n+1} = 
Be^{i(\omega\,t\,+\,(2n+1)\phi)}\,,\end{equation} 

\noindent
where A and B are constants and $\phi$ is given by $\phi = 
2\pi\emph{a}/\lambda$ as in Eq.(6). $\emph{a}$ is the lattice constant as 
before and $\lambda$ the wavelength, $\lambda$ = n$\emph{a}$. The 
solutions of Eqs.(22,23) are obviously periodic in time and space and 
describe again standing waves. Using (22,23) to solve (20,21) leads to a 
secular equation from which according to Eq.(24) of B\&K the frequencies 
of the oscillations of the chain follow from 

\begin{equation}4\pi^2\nu^2_\pm  = \alpha/\mathrm{Mm}\cdot((\mathrm{M+m}) 
\pm\sqrt{(\mathrm{M-m})^2 + 
4\mathrm{mM}\mathrm{cos}^2\phi}\,)\,.\end{equation}    

Longitudinal and transverse waves are distinguished by the minus or plus 
sign in front of the square root in (24).

\section{The masses of the $\nu$-branch particles}

The characteristic case of the neutrino branch particles are the
 $\pi^\pm$\,mesons which can be created in the process $\gamma$ + p
$\rightarrow \pi^- +  \pi^+$ + p. A photon impinges on a proton and is
converted in $10^{-23}$ sec into a pair of particles of opposite charge. 
A simple example of the creation of a $\nu$-branch particle by strong
interaction is the case p + p $\rightarrow$  p + p + $\pi^-$  + $\pi^+$. 
Fourier
analysis dictates that a continuum of frequencies must be in the collision
products. The waves must be standing waves in order to be part of the
rest mass of a particle. The $\pi^\pm$\,mesons decay via 
 $\pi^\pm$ $\rightarrow$  $\mu^\pm$ + $\nu_\mu(\bar{\nu}_\mu)$ 
(99.98770\%) followed by e.g. 
$\mu^+$ $\rightarrow$ e$^+$ + $\bar{\nu}_\mu$ + $\nu_e$
 ($\approx$ 100\%). Only $\mu$\,mesons,
 which decay into charge and neutrinos, and neutrinos 
result from the decay of the $\pi^\pm$\,mesons.  If the 
particles consist of the particles into which they decay, then 
 the $\pi^\pm$\,mesons  are made of neutrinos, antineutrinos
 and e$^\pm$. Since neutrinos interact through the weak 
force which has a range of 10$^{-16}$\,cm according to p.25 of [22], 
and since the size of the $\pi^\pm$\,mesons [28] is on the order of 
10$^{-13}$\,cm, \emph{the $\nu$-branch particles must have a lattice
 with  N neutrinos}, N being the same as in Eq.(15). It is not known 
with certainty that neutrinos actually have a 
rest mass as was originally suggested by Bethe [33] and Bahcall [34]
and what the values of m($\nu_e$) and m($\nu_\mu$) are. However, the 
results of the Super-Kamiokande [35] and the Sudbury [36] experiments 
indicate that the neutrinos have rest masses. The neutrino lattice must be 
diatomic, meaning that the lattice points have alternately larger 
(m($\nu_\mu$)) and smaller (m($\nu_e$)) masses. We will retain the 
traditional term diatomic. \emph{The term neutrino lattice will refer to a 
lattice consisting of neutrinos and antineutrinos}. The lattice we 
consider  
is shown in Fig.\,2. Since the neutrinos have spin 1/2 this is a 
four-Fermion lattice which is required for the explanation of the weak 
decays. The first investigation of cubic Fermion lattices in context with 
the elementary particles was made by Wilson [12]. A neutrino lattice is 
electrically neutral. Since we do not know the interaction of the electron
with a neutrino lattice we cannot consider lattices with a charge.

	\begin{figure}[h]
	\hspace{3cm}
	\includegraphics{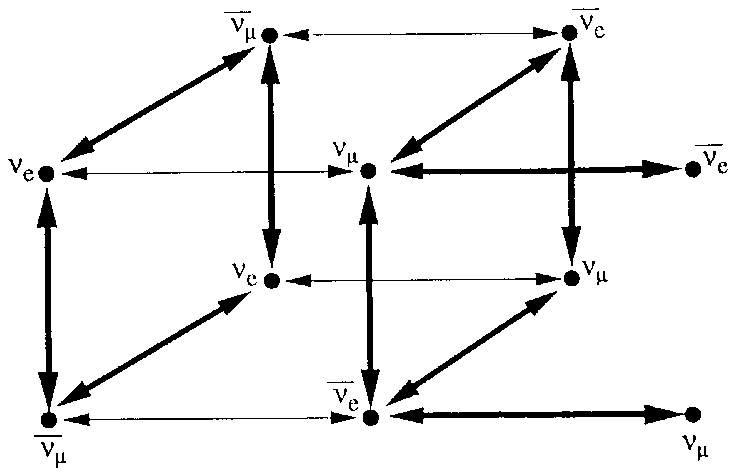}
	\vspace{-0.2cm}
	\begin{quote}
Fig.\,2: A cell in the neutrino lattice. 
Bold lines mark the forces between 
neutrinos and antineutrinos. Thin lines mark the forces between either 
neutrinos only, or antineutrinos only.
	\end{quote}
	\end{figure}

A neutrino lattice takes care of the continuum of frequencies which must, 
according to Fourier analysis, be present after the high energy collision 
which created the particle. We will, for the sake of simplicity, first set 
the sidelength of the lattice at 10$^{-13}$\,cm that means approximately 
equal to the size of the nucleon. The lattice then contains about 10$^9$ 
lattice points. The sidelength of the lattice does not enter Eq.(24) 
for the frequencies of diatomic oscillations. The calculation of the 
ratios of the masses  is consequently
 independent of the size of the lattice, as was the case with 
the $\gamma$-branch. The size of the lattice can be explained with 
the pressure which the lattice oscillations exert on a crossection of the 
lattice. The pressure cannot exceed Young's modulus of the lattice. We 
require that the lattice is isotropic.

	\begin{figure}[h]
	\hspace{1.5cm}
	\includegraphics{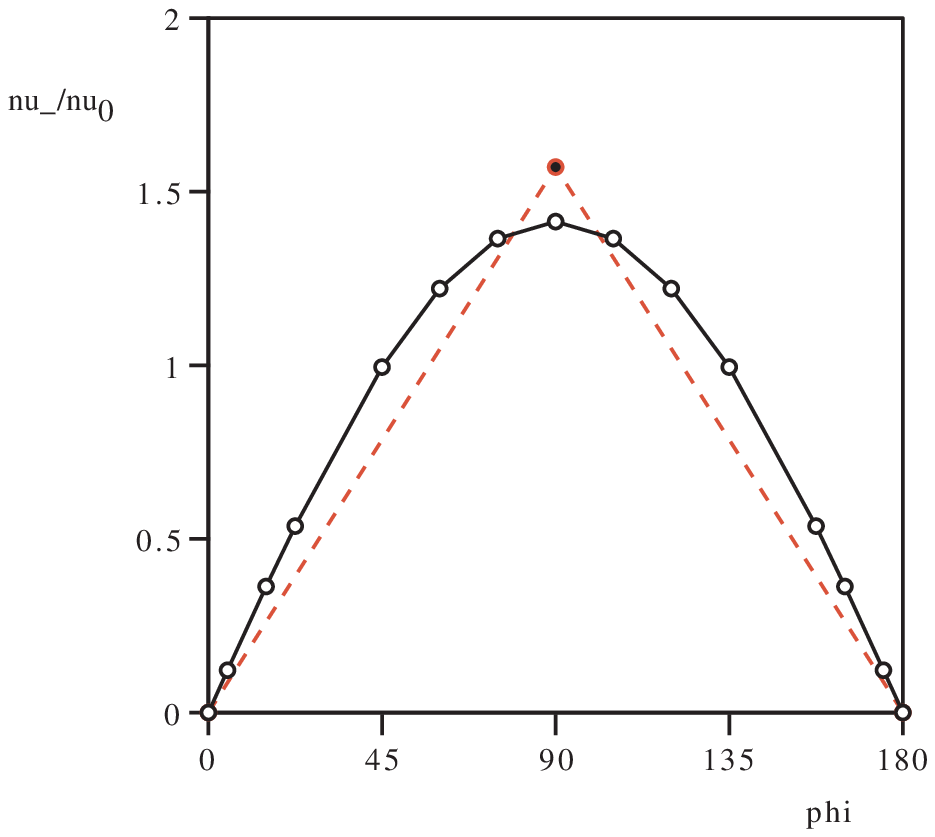}
	\vspace{-.5cm}
	\begin{quote}
Fig.\,3: The frequency distribution $\nu_-/\nu_0$ of the basic diatomic 
mode 
according to Eq.(24) with M/m = 100. The dashed line shows the 
distribution of the frequencies corrected for the group velocity 
limitation.
	\end{quote}
	\end{figure}

   From the frequency distribution of the axial diatomic oscillations 
   (Eq.24), shown in Fig.\,3, follows the group velocity 
   $\mathrm{d}\omega/\mathrm{dk} 
= 2\pi\emph{a}\,\,d\nu/d\phi$\, at each point $\phi$. With $\nu = 
\nu_0f(\phi)$ and $\nu_0$ = $\sqrt{\alpha/4\pi^2\mathrm{M}}$ = 
c$_\star/2\pi\emph{a}$
 as in Eq.(9) we find

\begin{equation} c_g = d\omega/dk = 
\emph{a}\sqrt{\alpha/\mathrm{M}}\cdot df(\phi)/d\phi\,.
\end{equation}
\noindent	
In order to determine the value of d$\omega/dk$ we have to know the value 
of $\sqrt{\alpha/\mathrm{M}}$. From Eq.(8) for $\alpha$ follows that 
$\alpha = 
\emph{a}\,c_{44}$ in the isotropic case with central forces. The group 
velocity is therefore

\begin{equation} c_g = \sqrt{a^3c_{44}/\mathrm{M}}\cdot df/d\phi\,.
\end{equation}

\noindent
In Eq.(25) we now set $\emph{a}\sqrt{\alpha/\mathrm{M}}$ = c$_\star$,
as in Eq.(11), where c$_\star$ is the velocity of light. It follows that

\begin{equation} c_g = c_\star\cdot df/d\phi\,,
\end{equation}

\noindent
as it was with the $\gamma$-branch, only that now on account of 
the rest masses of the neutrinos the group velocity must be smaller than 
c$_\star$, so the value of df/d$\phi$ is limited to $<$\,1, but c$_g 
\cong$\, c$_\star$, 
which is a necessity because the neutrinos in the lattice soon approach 
the velocity of light as we will see. Equation (27) applies regardless 
whether we consider $\nu_+$ or $\nu_-$ in  
Eq.(24). That means that there are no separate transverse 
oscillations with their theoretically higher frequencies.

  The rest mass M of the heavy neutrino can be determined with lattice 
theory from Eq.(26) as we have shown in [11]. This involves the
 inaccurately known compression modulus of the proton. We will, 
therefore, rather determine the rest mass  
of the muon neutrino with Eq.(29), which leads to m($\nu_\mu$) 
$\approx$  50\,milli-eV/c$_\star^2$. It can be verified easily that
 m($\nu_\mu$) = 50\,milli-eV/c$_\star^2$ makes sense. The energy 
of the rest mass of the $\pi^\pm$ mesons is 139\,MeV, and we
 have N/4 = 0.7135$\cdot10^9$ 
muon neutrinos and the same number of anti-muon neutrinos with 
an energy of about 50\,milli-eV. It follows that the energy in the 
rest masses of all muon and anti-muon neutrinos is 
71.35\,MeV, that is 51\% of the energy of the rest mass of the 
$\pi^\pm$ mesons, m($\pi^\pm$)c$_\star^2$ = 139.57\,MeV. A very 
small part of m($\pi^\pm$)c$_\star^2$ goes, as we will see, into the 
electron neutrino masses, the rest of the energy in $\pi^\pm$ is
in the lattice oscillations.

   The energy in the rest mass of the $\pi^\pm$\,mesons is the sum 
 of the oscillation energies plus the sum of the energy in the rest masses
 of the neutrinos. \emph{The $\pi^\pm$\,mesons are like cubic black bodies 
filled with oscillating neutrinos}. For the sum of the energies of the 
frequencies we use Eq.(14) modified with Eq.(16), with the same
 N and $\nu_0$ we used 
for the $\gamma$-branch. For the  integral in Eq.(14) of the axial 
diatomic frequencies corrected for the group velocity limitation we find   
the value $\pi^2$/2 as can be easily derived from the plot of the 
corrected frequencies in Fig.\,3. The value of the integral in 
Eq.(14) for the axial diatomic frequencies $\nu = \nu_0\phi$ is 1/2 of 
the value $\pi^2$ of the same integral in the case of axial monatomic 
frequencies, because in the latter case the increase of the corrected 
frequencies continues to $\phi$ = $\pi$, whereas in the diatomic case the 
increase of the corrected frequencies ends at $\pi$/2, see Fig.\,3.  We 
consider c$_g$ to be so close to c$_\star$ that it does not 
change the value of the integral in Eq.(14) significantly. It can be 
calculated that the time average of 
the velocity of the electron neutrinos in the $\pi^\pm$\,mesons is 
$\bar{v}$ = 0.99994c$_\star$ if m($\nu_e$) = 0.365\,milli-eV/c$_\star^2$
as will be shown in Eq.(37). Consequently we find that the sum of the
 energies of the corrected diatomic neutrino 
frequencies is 0.5433$\cdot$10$^{-4}$\,erg = 33.91\,MeV. We double 
 this amount because we deal with standing waves or the superposition 
of two waves of the  same energy and find with Eq.(18) that the 
energy of the neutrino oscillations in $\pi^\pm$ is

\begin{equation} \mathrm{E}_\nu(\pi^\pm)  \cong   
1/2\cdot\mathrm{E}_\nu(\pi^0)  = 67.82\,\mathrm{MeV} = 
0.486\,\mathrm{m}(\pi^\pm)\mathrm{c}_\star^2\,,
\end{equation}
\noindent
or that \emph{ $\approx$ 1/2 of the energy of $\pi^\pm$  is in the 
oscillation energy 
E$_\nu(\pi^\pm)$}.

   In order to determine the sum of the rest masses of the neutrinos we
make use of E$_\nu(\pi^\pm)$ and obtain an approximate value of the
 rest mass of the muon neutrino from 
\begin{equation} \mathrm{m}(\pi^\pm)\mathrm{c}_\star^2 
- \mathrm{E}_\nu(\pi^\pm) = \sum\,[m(\nu_\mu) +
 m(\bar{\nu}_\mu) + m(\nu_e) + m(\bar{\nu}_e)]\mathrm{c}_\star^2 = 
71.75\,\mathrm{MeV}\,,
\end{equation}
\noindent
that means that \emph{$\approx$ 1/2 of the energy of $\pi^\pm$  is in 
the neutrino rest masses}. Since nothing else but the electric charge
contributes to the rest mass of $\pi^\pm$ it appears that in a good 
approximation the oscillation energy in $\pi^\pm$ is equal to the
energy in the sum of the neutrino rest masses in $\pi^\pm$, i.e. 

\begin{displaymath} \mathrm{E}_\nu(\pi^\pm) \cong 
\Sigma\,\mathrm{m(neutrinos)c}^2 = 
\mathrm{N}/2\cdot(\mathrm{m}(\nu_\mu) + 
\mathrm{m}(\nu_e))\mathrm{c}^2 
\cong 1/2\cdot\mathrm{m(\pi^\pm)c}^2. (29a) \end{displaymath}

   If m($\nu_e$) $\ll$ m($\nu_\mu$) and m($\nu_\mu$)
 = m($\bar{\nu}_\mu$), as we will justify later, we arrive 
with N/2 = 1.427$\cdot10^9$ at

\hspace{4.5cm} m($\nu_\mu$) $\approx$ 50\,milli-eV/$\mathrm{c}_\star^2$.

\noindent The sum of the energy of the rest masses of all neutrinos Eq.(29)
plus the oscillation energy Eq.(28) gives
 the theoretical rest mass of the $\pi^\pm$\,mesons which is, since we 
used m($\pi^\pm)$  in the determination of the neutrino rest masses with
Eq.(29), equal to the experimental rest mass of 139.57\,MeV/c$^2_\star$.

   A cubic lattice and conservation of neutrino numbers during the reaction
 $\gamma$  + p $\rightarrow \pi^+ + \pi^- +$ p  \emph{necessitates} 
that the $\pi^+$ and $\pi^-$  lattices contain just as many  electron
 neutrinos as anti-electron neutrinos. If the lattice is cubic it
 must have a center neutrino (Fig.\,4). Conservation of neutrino numbers
 requires furthermore that the center neutrino of  $\pi^+$ is matched  
by an antineutrino in $\pi^-$. In the decay sequence of (say)
 the $\pi^-$\,meson $\pi^- \rightarrow  \mu^- + \,\bar{\nu}_\mu$ and 
$\mu^- \rightarrow$ e$^-$ +\,$\nu_\mu\,+\,\bar{\nu}_e $ an electron
 neutrino $\nu_e$ does not appear. But since (N\,-\,1)/4 electron 
neutrinos  $\nu_e$ must be in the $\pi^-$ lattice it follows that 
(N\,-\,1)/4 electron neutrinos must go with the electron emitted 
in the $\mu^-$ decay.

\begin{figure}[h] 
	\hspace{2.2cm}
	\includegraphics{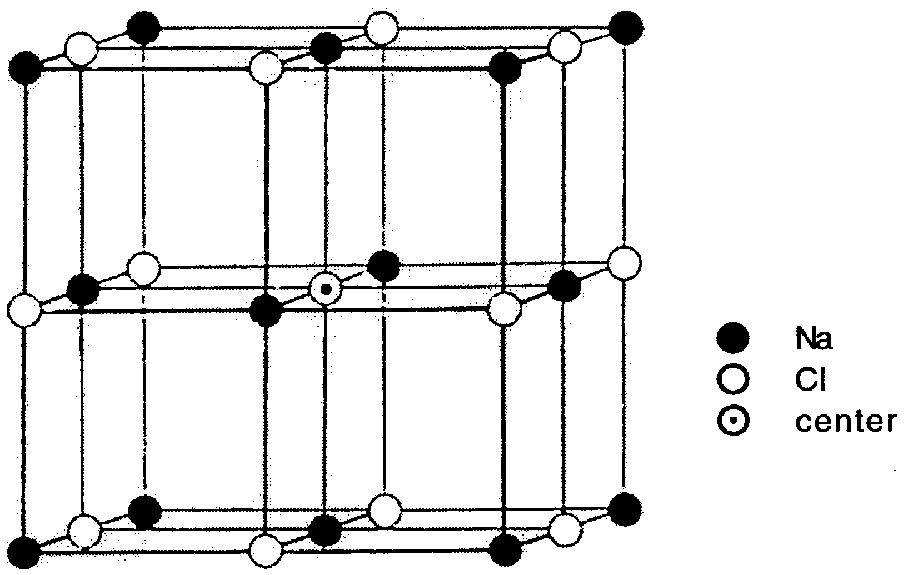}
	\begin{quote}
Fig.\,4: The center of a NaCl lattice. (After Born and Huang).
	\end{quote}
\end{figure}

    We must now be more specific about N,
 which is an odd number, because a cubic lattice (Fig.\,4) 
 has a center particle, just as the NaCl lattice.
In the $\pi^\pm$\,mesons there are then (N\,-\,1)/4 muon 
neutrinos $\nu_\mu$ and the same number of 
anti-muon neutrinos $\bar{\nu}_\mu$, as well as 
 (N\,-\,1)/4 electron neutrinos $\nu_e$ and the same number  
of anti-electron neutrinos $\bar{\nu}_e$, \emph{plus} a center  
neutrino or antineutrino. We replace
 N\,-\,1 by N$^\prime$. Since N$^\prime$ differs from N by 
 only one in $10^9$ we have N$^\prime$ $\cong$ N. Although the 
numerical difference between N and N$^\prime$ is negligible we 
cannot consider any integer number N because that would mean 
that there could be fractions  of a neutrino. N$^\prime$ is an even 
number, because each cell (Fig.\,2) of the lattice (Fig.\,4) consists
 of four pairs of neutrinos. 
  
   The \emph{antiparticle} of the $\pi^+$\,meson is the particle in   
which all frequencies of the 
neutrino lattice oscillations have been replaced by frequencies with 
the opposite sign, all neutrinos replaced by their antiparticles and the 
positive charge replaced by the negative charge. If, as we will show, the 
antineutrinos have the same rest mass as the neutrinos it follows that the 
antiparticle of the $\pi^+$\,meson has the same mass as $\pi^+$ but 
opposite charge, i.e. is the $\pi^-$\,meson. As we will see, the 
explanation of the mass of the $\pi^\pm$\,mesons opens the door to 
the explanation of the mass of the muon and of the electron.  
 
   Now we turn to the K\,mesons whose mass is m(K$^\pm$) = 
0.8843\,$\cdot$\,4m($\pi^\pm$). The primary decay of the K$^\pm$\,mesons 
K$^\pm \rightarrow \mu^\pm + \nu_\mu(\bar{\nu}_\mu)$ (63.5\%)  
 leads to the same end products as the $\pi^\pm$\,meson decay 
$\pi^\pm \rightarrow \mu^\pm + \nu_\mu(\bar{\nu}_\mu)$ (99.98\%). 
From this and the decay of the $\mu$\,mesons we learn 
that the K\,mesons must, at least partially, be made of the same four 
neutrino types as in the $\pi^\pm$\,mesons namely of muon neutrinos, 
anti-muon neutrinos, electron neutrinos and anti-electron neutrinos and 
their oscillation energies. However the K$^\pm$\,mesons cannot be 
solely the second mode of the lattice oscillations of the 
$\pi^\pm$\,mesons, because the second mode of the 
$\pi^\pm$\,mesons has an energy of 
\begin{eqnarray}
{ \mathrm{E}((2.)\pi^\pm) = 4\mathrm{E}_{\nu}(\pi^\pm) + 
\mathrm{N}/2\cdot(\mathrm{m}(\nu_\mu) + 
\mathrm{m}(\nu_e))\,c_\star^2\nonumber }\\
\cong 2\mathrm{m(\pi^\pm)c_\star^2} + 1/2\cdot \mathrm{m(\pi^\pm)c_\star^2} = 348.92\,\mathrm{MeV}\,,\end{eqnarray}
with 2E$_\nu(\pi^\pm$) $\cong$ m($\pi^\pm$)c$_\star^2$ and 
N/2\,$\cdot$\,(m($\nu_\mu$) + m($\nu_e$)) $\cong$ m($\pi^\pm$)/2 from 
Eqs.\, (29,29a). The 348.9\,MeV characterize 
the second or (2.) mode of the $\pi^\pm$ mesons, which fails 
m(K$^\pm$)c$_\star^2$ = 493.7\,MeV by a wide margin.

   The concept that the K$^\pm$\,mesons are alone a higher mode of  
the $\pi^\pm$\,mesons also contradicts our point that the particles 
consist of  
the particles into which they decay. The decays K$^\pm$ $\rightarrow$ 
$\pi^\pm \,+ \,\pi^0$ (21.13\%), as well as K$^+$ $\rightarrow$ $\pi^0$ + 
e$^+ + \nu_e$ (4.87\%), called K$^+_{e3}$, and 
K$^+$ $\rightarrow$ $\pi^0 + \mu^+ + \nu_\mu$ (3.27\%), called 
K$^+_{\mu3}$, make up 29.27\% of the K$^\pm$\,meson
 decays. A $\pi^0$\,meson figures 
in each of these decays. If we add the energy in the rest mass of a 
$\pi^0$\,meson m($\pi^0$)c$_\star^2$ = 134.97\,MeV to the 
348.9\,MeV in the second mode of the $\pi^\pm$\,mesons then 
we arrive at an energy of 483.9\,MeV, which is 98.0\% of 
m(K$^\pm$)c$_\star^2$. Therefore we conclude that the 
K$^\pm$\,mesons consist of the second 
 mode of the $\pi^\pm$\,mesons $\emph{plus}$ a $\pi^0$\,meson 
or are the state (2.)$\pi^\pm$ + $\pi^0$. Then 
it is natural that  $\pi^0$\,mesons from the $\pi^0$ component 
in the K$^\pm$\,mesons are among the decay products of the 
K$^\pm$\,mesons.

   The average factor 0.85 $\pm$ 0.025 in the ratios of the masses  
of the particles of the $\nu$-branch to the mass of the 
$\pi^\pm$\,mesons is a consequence of the neutrino rest masses. 
They make it impossible that the ratios of the particle masses are  
outright integer multiples because the 
particles consist of the energy in the neutrino oscillations plus the 
neutrino rest masses which are independent of the order of the lattice 
oscillations. Since the contribution in percent of the neutrino rest 
masses to the $\nu$-branch particle masses decreases with 
increased particle mass 
 the factor in front of the mass ratios of the $\nu$-branch particles 
must decrease with increased particle mass.

The K$^0$,$\overline{{\mathrm{K}}^0}$\,mesons have a rest 
mass m(K$^0$,$\overline{{\mathrm{K}}^0}$) = 1.00809\,m(K$^\pm$),
or it is m(K$^0$,$\overline{{\mathrm{K}}^0}$) = 
0.99984\,(m(K$^\pm$) + $\alpha_f\cdot4$m$(\pi^\pm$)).
 We obtain the K$^0$\,meson if we superpose onto the 
second mode of the $\pi^\pm$\,mesons instead of 
a $\pi^0$\,meson a basic mode of the $\pi^\pm$\,mesons 
with a charge opposite 
to the charge of the second mode of the $\pi^\pm$\,meson. 
The K$^0$ and $\overline{{\mathrm{K}}^0}$\,mesons, 
or the state (2.)$\pi^\pm$ + $\pi^\mp$, is made  
of neutrinos and antineutrinos only, without a photon component,
 because the second mode of $\pi^\pm$ as well as the basic mode 
$\pi^\mp$ consist of neutrinos and antineutrinos only.  The 
K$^0$\,meson has a measured mean square charge radius
 $\langle$r$^2$$\rangle$ = -\,0.077 $\pm$ 0.010\,fm$^2$ according
 to [2], which can only be if there are two charges of opposite sign 
within K$^0$, as this model implies. Since the mass of a 
$\pi^\pm$\,meson is by 4.59\,MeV/c$_{\star}^2$ larger than
the mass of a $\pi^0$\,meson the mass of K$^0$ should be larger  
than m(K$^\pm$), and indeed m(K$^0$)\,$-$\,m(K$^\pm$) = 
3.972\,MeV/c$_\star^2$ according to [2]. Similar differences occur 
with m(D$^\pm$)\,$-$\,m(D$^0$) and 
m($\Xi_c^0$)\,$-$\,m($\Xi_c^+$). The decay 
K$^0_S \rightarrow \pi^++ \pi^-$ (68.6\%) creates directly the 
$\pi^+$ and $\pi^-$ mesons which are part of the (2.)$\pi^\pm$ + 
$\pi^\mp$ structure of K$^0$ we have suggested.
 The decay K$^0_S \rightarrow \pi^0 + \pi^0$ (31.4\%) apparently 
originates from the 2$\gamma$ branch of electron positron annihilation. 
Both decays account for 100\% of the decays of K$^0_S$. The 
decay K$^0_L \rightarrow 3\pi^0$ (21.1\%) apparently comes from the 
3$\gamma$ branch of electron positron annihilation. The two decays of 
K$^0_ L$ called K$^0_{\mu3}$ into $\pi^\pm\,\mu^\mp\,\nu_\mu$ 
(27.18\%) and K$^0_{e3}$ into $\pi^\pm$\,e$^\mp\,\nu_e$ 
(38.79\%) which together make up 65.95\% of 
the K$^0_{L}$ decays apparently originate from the decay of  
the second mode of the $\pi^\pm$\,mesons in the 
K$^0$ structure, either into $\mu^\mp$ + $\nu_\mu$  
 or into e$^\mp$ + $\nu_e$. The same types of decay, 
 apparently tied to the (2.)$\pi^\pm$ mode, accompany also the 
K$^\pm$  decays K$^\pm$ $\rightarrow \pi^\pm\pi^0$ (20.92\%)
in which, however, a $\pi^0$\,meson replaces the 
$\pi^\pm$\,mesons in  the K$^0_L$ decay products. Our rule 
that the particles consist of the particles into which they decay also 
holds for the K$^0$ and $\overline{{\mathrm{K}}^0}$\,mesons. The
explanation of the K$^0$,$\overline{{\mathrm{K}}^0}$\,mesons with the
state (2.)$\pi^\pm$ + $\pi^\mp$ confirms that the state (2.)$\pi^\pm$ +
$\pi^0$ was the correct choice for the explanation of the K$^\pm$\,mesons.
The state (2.)$\pi^\pm$ + $\pi^\mp$ is also crucial for the explanation of 
the absence of spin of the K$^0$,$\overline{{\mathrm{K}}^0}$\,mesons, 
as we will see later.   
   
   The neutron whose mass is m(n) = 0.95156\,$\cdot$\,2m(K$^\pm$)  
is either the superposition of a K$^+$ and a K$^-$\,meson or of   
a K$^0$\,meson and a $\overline{{\mathrm{K}}^0}$\,meson. As
has been shown in [56], the spin rules out
  a neutron consisting of a K$^+$ and a K$^-$\,meson. On the 
other hand, the neutron can be the superposition of a K$^0$   
and a $\overline{{\mathrm{K}}^0}$\,meson which guarantees that 
the neutron consists of neutrinos without a photon component.
 In this case the neutron lattice contains at each lattice point a 
$\nu_\mu,\bar{\nu}_\mu,\nu_e,\bar{\nu}_e$ neutrino quadrupole 
 and there is a single quadrupole of positive and negative electric 
charges because in each K$^0$  and $\overline{{\mathrm{K}}^0}$\,meson
are neutrino pairs at the lattice points and each K$^0$  and 
$\overline{{\mathrm{K}}^0}$\,meson carries a pair of opposite 
elementary electric charges. There must be opposite charges in 
the neutron because it has a mean square charge radius
$\langle$r$^2\rangle$ = - 0.1161\,fm$^2$ [2]. 
The lattice oscillations in the neutron must be a coupled  pair in 
order for the neutron to have spin 1/2, just as the $\Lambda$ 
baryon with spin 1/2 is a superposition of two $\eta$\,mesons. With 
m(K$^0$)(\emph{theor}) = m(K$^\pm$) + 4\,MeV/c$^2$ =
 487.9\,MeV/c$^2_\star$
 from above it follows that m(n)(\emph{theor}) $\approx$ 
2m(K$^0)$(\emph{theor}) $\approx$ 975.8\,MeV/c$^2_\star$ 
= 1.04\,m(n)(\emph{exp}). 

   The proton, whose mass is m(p) = 0.99862\,m(n), does not decay 
and does not tell which particles it is made of. 
 However, we learn about the structure of the proton through the 
decay of the neutron n $\rightarrow$ p + e$^- + \bar{\nu}_e$ (100\%). 
An electron and one single anti-electron neutrino is emitted when the 
neutron decays and 1.29333 MeV are released. But there is no place 
for a permanent vacancy of a single missing neutrino and for a small 
amount of permanently missing oscillation energy in a nuclear lattice. As 
it appears all anti-electron neutrinos are removed from the structure of 
the neutron in the neutron decay and converted into
 the kinetic energy of the decay products. This type of 
process will be explained in the following 
Section. On the other hand, it is certain that the proton consists 
 of a neutrino lattice because the neutron has a neutrino lattice. The  
proton carries a net positive elementary electric charge because 
the neutron carries an e$^+$e$^-$e$^+$e$^-$ quadrupole, of 
which one e$^-$ is lost  in the $\beta$-decay.
The concept that the proton carries just one elementary electric charge 
has been abandoned a long time ago when it was said that the 
proton consists of three quarks carrying fractional electric charges. 
Each elementary charge in the proton has a magnetic moment,
 all of them point in the same direction because the spin
 of the one e$^-$ must be opposite to the spin of the two e$^+$.
Each magnetic moment of the elementary charges has a g-factor $\cong$\,\,2.
All three electric charges in the proton must then have a g-factor 
$\approx$\,\,6, whereas the measured g-factor of the magnetic
moment of the proton is g(p) = 5.585 = 0.93\,$\cdot$\,6.   

  The D$^\pm$\,mesons with m(D$^\pm$) = 0.9954$\cdot$(m(p) + 
m($\bar{\mathrm{n}}$))
are the superposition of a proton and an antineutron of opposite spin or 
of their antiparticles, whereas the superposition 
of a proton and a neutron with the same spin creates the deuteron 
 with spin 1 and a mass m(d) = 0.9988\,(m(p) + m(n)). In this 
case the proton and neutron interact with the strong force, nevertheless 
the deuteron consists of a neutrino lattice with standing waves. The 
D$_s^\pm$\,mesons seem to be made of a body centered cubic 
lattice (Fig.\,5) as discussed in [38].

\begin{figure}[h] 
	\hspace{4cm}
	\includegraphics{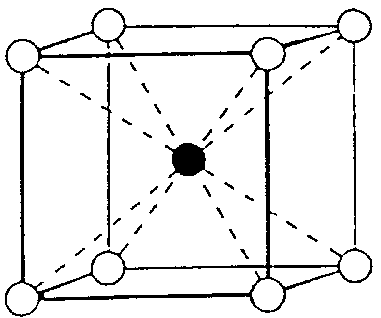}
	\begin{quote}
Fig.\,5: A body-centered cell. (After Born and Huang). 
\indent\qquad\qquad\,\,\,\,In the center of a D$^\pm_s$ cell is a 
$\tau$\,neutrino
 in the corners are  
\indent \qquad\,\,\,\,$\nu_{\mu}$,$\bar{\nu}_\mu$,$\nu_e$,$\bar{\nu}_e$ 
neutrinos.\\ 
	\end{quote}
\end{figure}

   Summing up: The particles of the $\nu$-branch consist of oscillating 
neutrinos and one or more positive and/or negative elementary electric 
charges. The characteristic feature of the $\nu$-branch particles is 
the cubic lattice consisting of $\nu_\mu,\bar{\nu}_\mu,\nu_e,\bar{\nu}_e$ 
neutrinos. The rest masses of the $\nu$-branch particles is the sum of 
the rest masses of the neutrinos and antineutrinos in the lattice plus the 
mass in 
the energy of the  lattice oscillations. The existence of the neutrino
 lattice is a necessity if one wants to explain the spin, or the absence
 of spin, of the $\nu$-branch particles. \emph{We do not use hypothetical 
particles} for the explanation of the $\nu$-branch particles, just as we 
did not use hypothetical particles for the explanation of the 
$\gamma$-branch particles.   

\section{The rest mass of the muon}

Surprisingly one can also explain the mass of the $\mu^\pm$\,mesons 
with the standing wave model. The $\mu$\,mesons belong to the lepton 
 family which is distinguished from the mesons and baryons by the 
absence of strong interaction with the mesons and baryons. The charged 
leptons make up 1/2 of the number of stable elementary particles. The 
standard model of the particles does not deal with the lepton masses. 
Barut [39] has given a simple and quite 
accurate empirical formula relating the masses of the electron, 
$\mu$\,meson and $\tau$\,meson, which formula has been 
extended by Gsponer and Hurni [40] to the quark masses.

    The origin of most of the $\mu^\pm$\,mesons is the decay
 of the $\pi^\pm$\,mesons, $\pi^\pm$ $\rightarrow \mu^\pm
+ \nu_\mu (\bar{\nu}_\mu$), or the decay of the K$^\pm$\,mesons, 
K$^\pm \rightarrow \mu^\pm + \nu_\mu(\bar{\nu}_\mu)$. 
The mass of the muons is 
m($\mu^\pm$) = 105.658\,369 $\pm$ 
9$\cdot$$10^{-6}$\,MeV/c$_\star^2$, according to the Review of 
Particle Physics [2]. The mass of the muons is usually compared to  
the mass of the electron and is very often said to be 
m($\mu^\pm$) = m(e$^\pm$)\,$\cdot$\,(1 + 3/2$\alpha_f$) =
 206.554\,m(e$^\pm$), ($\alpha_f$ being the fine structure 
constant), whereas the experimental value is 206.768\,m(e$^\pm$).  
This formula for m($\mu^\pm$) was given by Barut [41] following 
an earlier suggestion by Nambu [9] that m($\mu^\pm$) $\approx$
 m(e$^\pm$)\,$\cdot$\,3/2$\alpha_f$. The muons are ``stable", their 
lifetime  
$\tau(\mu^\pm) = 2.19703\cdot10^{-6} \pm 4\cdot10^{-11}$\,sec is
about a hundred times longer than the lifetime of the $\pi^\pm$\,mesons,  
that means longer than the lifetime of any other elementary particle, but 
for the electrons, protons and neutrons.

   Comparing the mass of the $\mu^\pm$\,mesons to the mass of the 
$\pi^\pm$\,mesons  from which the $\mu^\pm$\,mesons emerge
we find, with m($\pi^\pm$) = 139.570\,18\,MeV/c$_\star^2$, that\\
 
m($\mu^\pm$)/m($\pi^\pm$) = 0.757027 = 1.00937\,$\cdot$\,3/4

$\cong$ 3/4 + $\alpha_f$ = 0.757297 = 
1.00036\,$\cdot$\,m($\mu^\pm$)/m($\pi^\pm)$(\emph{exp}).\\

 \noindent
 The term  +\,$\alpha_f$ in the preceding equation will be explained 
later, Eq.(71). 
The mass of the $\mu^\pm$ mesons is in a good approximation 3/4 
of the mass of the $\pi^\pm$\,mesons. We have also m($\pi^\pm)\,-$ 
m($\mu^\pm$) = 33.9118\,MeV/c$_\star^2$ = 0.24297m($\pi^\pm$)  
or approximately 1/4\,$\cdot$\,m$(\pi^\pm$). The
 mass of the electron is approximately 1/207 of the mass of the muon,
 the contribution of m(e$^\pm$) to m($\mu^\pm$) will therefore 
be neglected in the following. We assume, as we have done before   
and as appears to be natural, that the particles, including the muons, 
$\emph{consist of the particles into}$ \emph{which they decay}. The 
$\mu^\pm$\,mesons decay via 
$\mu^\pm \rightarrow$ e$^\pm$ + $\bar{\nu}_\mu(\nu_\mu)$ 
+ $\nu_e(\bar{\nu}_e)$ ($\approx$ 100\%). The muons are 
apparently composed of an elementary electric charge and 
some of the neutrinos,
antineutrinos and their oscillations which are, according to the 
standing wave model,  present in the cubic neutrino lattice of  
the $\pi^\pm$\,mesons from which the $\mu^\pm$\,mesons 
come. The $\mu^\pm$\,mesons with a mass 
m($\mu^\pm$) $\cong$ 3/4\,$\cdot$\,m($\pi^\pm$) seem to 
be related to the $\pi^\pm$\,mesons
rather than to the electron with which the $\mu^\pm$\,mesons have
been compared traditionally, although m($\mu^\pm$) is separated 
from m(e$^\pm$) by a factor $\cong$ 207. 

   From Eq.(29) followed that the rest mass of a muon neutrino   
should be about 50\,milli-eV/c$_\star^2$. Provided that 
the mass of an electron neutrino m($\nu_e)$ is small as compared  
to m($\nu_\mu$), as will be shown by Eq.(42), we find,
 with N = 2.854$\cdot10^9$, that:
\bigskip

(a) The difference of the rest masses of the $\mu^\pm$ and $\pi^\pm$ 
mesons is nearly equal to the sum of the rest masses of all muon, 
respectively anti-muon, neutrinos in the $\pi^\pm$\,mesons.
\bigskip

\noindent
    m($\pi^\pm)\,-\,$m$(\mu^\pm$) = 33.912\,MeV/$\mathrm{c}_\star^2$ 
\quad versus \quad $\mathrm{N^\prime}/4\cdot\mathrm{m}(\nu_\mu)$ 
$\approx$ 35.68\,$\mathrm{MeV}/\mathrm{c}_\star^2$\,.

\bigskip

(b) The energy in the oscillations of all 
$\nu_\mu,\bar{\nu}_\mu,\nu_e,\bar{\nu}_e$
 neutrinos in the $\pi^\pm$\,mesons
 is nearly the same as the energy in the oscillations of 
all $\bar{\nu}_\mu,\nu_e,\bar{\nu}_e$, respectively 
$\nu_\mu,\bar{\nu}_e,\nu_e$, neutrinos in the $\mu^\pm$\,mesons.
The oscillation energy is the rest mass of a particle minus the sum of 
the rest masses of all neutrinos in the particle as in Eq.(29). With 
m($\nu_\mu$) = m($\bar{\nu}_\mu$) and m($\nu_e$) = 
m($\bar{\nu}_e)$ from Eqs.(38,41) we have
\begin{equation} \mathrm{E}_{\nu}(\pi^\pm) =
 \mathrm{m}(\pi^\pm)\mathrm{c}_\star^2 - 
\mathrm{N^\prime}/2\cdot[\mathrm{m}(\nu_\mu) + 
\mathrm{m}(\nu_e)]\mathrm{c}_\star^2 = 68.22\,\mathrm{MeV}
 \end{equation}
\quad versus 
\begin{equation} \mathrm{E}_{\nu}(\mu^\pm) =
 \mathrm{m}(\mu^\pm)\mathrm{c}_\star^2 - 
\mathrm{N^\prime/4}\cdot \mathrm{m}(\nu_\mu)\mathrm{c}_\star^2 - 
\mathrm{N^\prime}/2\cdot\mathrm{m}(\nu_e)\mathrm{c}_\star^2 = 
69.98\,\mathrm{MeV}\,.\end{equation}
\noindent
Equation (32) means that either all N$^\prime$/4 muon neutrinos or  
all N$^\prime$/4 anti-muon neutrinos have been removed from the 
$\pi^\pm$  lattice in its decay.  If, e.g.,  any $\nu_\mu$ neutrinos   
were to remain in the $\mu^+$\,meson after the decay of the 
$\pi^+$\,meson they ought to appear in the decay of  $\mu^+$, 
but they do not.

   We attribute the  1.768\,MeV difference 
 between the left and right side of (a) to the second order effects
which cause the deviations of the masses of the particles from
the integer multiple rule. There is also the difference that the left 
side of (a) deals with two charged particles, whereas the right side 
deals with neutral particles. (b) seems to say that the
oscillation energy of all neutrinos in the $\pi^\pm$ lattice is conserved
in the $\pi^\pm$ decay, which seems to be necessary because the 
oscillation frequencies in $\pi^\pm$ and $\mu^\pm$ must follow Eq.(13)
as dictated by the group velocity limitation. If indeed 
\begin{equation} \mathrm{E}_\nu(\pi^\pm) = \mathrm{E}_\nu(\mu^\pm)
\end{equation} 
then it follows from the difference of Eqs.(31) and (32) that
\begin{equation}
 \mathrm{m}(\pi^\pm)\,-\,\mathrm{m}(\mu^\pm) = 
\mathrm{N}^\prime/4\cdot\mathrm{m}(\nu_\mu) =
\mathrm{N}^\prime/4\cdot \mathrm{m}(\bar{\nu}_\mu)\,. \end{equation}

The N$^\prime$/4 electron neutrinos, respectively anti-electron neutrinos, 
which come, as we will see in Section 9, into the 
$\mu^\pm$\,mesons with the elementary electric charge, 
are the recipients of 1/4 of the oscillation energy E$_\nu(\pi^\pm$) of 
the 
$\pi^\pm$\,mesons which becomes available when the muon neutrinos or 
anti-muon neutrinos leave the $\pi^\pm$ lattice in the $\pi^\pm$ decay.  
The neutrinos coming with e$^\pm$ make it possible that   
$\mathrm{E}_\nu(\pi^\pm) = \mathrm{E}_\nu(\mu^\pm)$. After the $\pi^\pm$
decay the muon neutrinos in $\mu^\pm$ retain their original oscillation 
energy E$_\nu(\pi^\pm$)/4, the electron neutrinos and anti-electron 
neutrinos 
in $\mu^\pm$ retain their original oscillation energies E$_\nu(\pi^\pm$)/4 
as well. The remaining oscillation energy E$_\nu(\pi^\pm$)/4 of the $\pi^\pm$
lattice so far not accounted for is picked up by the electron neutrinos,
 respectively anti-electron neutrinos, brought into $\mu^\pm$ by the  
elementary electric charge. Without a recipient for this oscillation energy
E$_\nu(\pi^\pm$)/4 a stable new particle can apparently not be formed
in the $\pi^\pm$ decay, that means there is no $\mu^0$\,meson.

   We should note that in the $\pi^\pm$ decays only \emph{one single} 
 muon neutrino is emitted, not N$^\prime$/4 of them, but that in the 
 $\pi^\pm$ decay 33.912\,MeV are released. Since according to (b) the 
oscillation energy of the neutrinos in the $\pi^\pm$ mesons is conserved 
in their decay the 33.912\,MeV released in the $\pi^\pm$ decay can come 
from \emph{no other source} then from the rest masses of all muon or
 all anti-muon neutrinos in the $\pi^\pm$\,mesons. The average kinetic energy 
of the neutrinos in the $\pi^\pm$ lattice is about 50\,milli-eV, so it is not  
possible for a single neutrino in $\pi^\pm$ to possess an energy
 of 33.9\,MeV. The 33.9\,MeV can come only from the sum of the
muon neutrino rest masses. However, what happens then to the neutrino
 numbers\,? Either conservation of neutrino numbers is violated or the 
decay
 energy comes from equal numbers  of muon and anti-muon neutrinos. Equal
numbers N$^\prime$/8  muon and anti-muon neutrinos would then be in
the $\mu^\pm$\,mesons. This would not make a difference in either the 
oscillation energy or in the sum of the rest masses of the neutrinos or in 
the spin of the $\mu^\pm$\,mesons. The 
sum of the spin vectors of the N$^\prime$/4 muon or anti-muon neutrinos
 converted into kinetic energy is zero, as will become clear in Section 12.

   Inserting m($\pi^\pm$)\,$\mathrm{-}$\,m($\mu^\pm$) = 
N$^\prime$/4\,$\cdot$\,m($\nu_\mu$)
from Eq.(34) into Eq.(31) we arrive at an equation for \emph{the 
theoretical value of the mass of the $\mu^\pm$\,mesons}. It is
\begin{equation} \mathrm{m}(\mu^\pm)\mathrm{c}_\star^2 = 
1/2\cdot[\,\mathrm{E}_\nu(\pi^\pm) + 
\mathrm{m}(\pi^\pm)\mathrm{c}_\star^2 +
\mathrm{N^\prime\,m}(\nu_e)\mathrm{c}_\star^2/2\,] = 
103.95\,\mathrm{MeV}\,\,,\end{equation}
which is 0.9838\,m($\mu^\pm$)c$_\star^2$(\emph{exp}) and  expresses
 m($\mu^\pm$) through the well-known mass of $\pi^\pm$,
the calculated oscillation energy of $\pi^\pm$, and a small contribution 
(0.4\%) of the electron neutrino and anti-electron neutrino masses. 
Eq.(35) shows that 
our explanation of the mass of the $\mu^\pm$\,mesons comes close to 
the experimental value m($\mu^\pm$) = 105.658\,MeV/c$_\star^2$.
With E$_\nu(\pi^\pm$) = E$_\nu(\mu^\pm$) and with m($\pi^\pm)$ from
Eq.(31) we find a different form of  Eq.(35) which is, in the case of  
the $\mu^+$\,meson, 
\begin{equation}\mathrm{m}(\mu^+) =
 \mathrm{E}_\nu(\mu^\pm)/\mathrm{c}_\star^2 + 
\mathrm{N^\prime m}(\bar{\nu}_\mu)/4 + 
\mathrm{N^\prime}\mathrm{m}(\nu_e)/2  \,\,. \end{equation}
As Eq.(36) tells, the rest mass of the muons is the sum of the rest masses
of the muon neutrinos, respectively anti-muon neutrinos, and of the masses 
of the electron- and anti-electron neutrinos which are in the muon 
lattice, plus
the oscillation energy of these neutrinos, neglecting the mass of e$^\pm$. 
 The ratio  m($\mu^\pm$)/m($\pi^\pm$) is 3/4, as it must be, if we 
divide Eq.(36) by m($\pi^\pm$) which follows from Eq.(29) and if we 
neglect the small masses of the electron and anti-electron neutrinos. 

   \emph{The  $\mu$\,mesons cannot be point particles} because  
they have a neutrino lattice.  The commonly held belief that the 
$\mu$\,mesons are point particles  is based on the results of scattering
 experiments. But at a true point the density of a ``point particle" would
be infinite, which poses a problem. Since, on the other hand, neutrinos
 do not interact, in a 
very good approximation, with charge or mass it will not be possible to 
determine the size of the $\mu$\,meson lattice through conventional 
scattering experiments. The $\mu$\,mesons  \emph{appear} to be 
point particles because only the electric charge of the muons 
participates in the scattering process and the elementary electric
charge scatters like a point particle. \emph{99.5\% of the muon mass 
consists of non-interacting neutrinos}. There is therefore no measurable
difference in e$^-$-p and $\mu^-$-p scattering.

   Finally we must address the question for what reason do the 
muons or leptons not interact strongly with the mesons 
and baryons\,? We have shown in [8] that a strong force emanates 
from the sides of a cubic lattice caused by the unsaturated weak 
forces of about $10^6$  lattice points at the surface of the lattice of 
the mesons and baryons. This follows from the study of Born and Stern [42] 
which dealt with the forces between two parts of a cubic lattice cleaved 
in vacuum. The strong force between two particles is an automatic 
consequence of the weak internal force which holds the particles together.
If the muons have a lattice consisting of one type of muon 
neutrinos, say, $\bar{\nu}_\mu$ and of $\nu_e$ and $\bar{\nu}_e$ 
neutrinos their lattice surface is not the same as the surface of the 
cubic 
$\nu_\mu,\,\bar{\nu}_\mu,\,\nu_e,\,\bar{\nu}_e$ lattice of the mesons and 
baryons in the standing wave model. Therefore  the muon lattice does  
not bond with the cubic lattice of the mesons and baryons.

   To summarize what we have learned about the $\mu^\pm$\,mesons.
Eq.(36) says that the energy in m($\mu^\pm$)c$_\star^2$ is the sum of 
the oscillation energies plus the sum of the energy of the rest masses of 
the neutrinos and antineutrinos in m($\mu^\pm$), neglecting the energy in 
e$^\pm$. The three neutrino types in the $\mu^\pm$\,mesons (Fig.\,6) 
are the remains 
of the cubic neutrino lattice in the $\pi^\pm$\,mesons. Since all 
$\nu_\mu$ respectively all $\bar{\nu}_\mu$ neutrinos have been
 removed from the $\pi^\pm$ lattice in the $\pi^\pm$ decay the rest 
mass of the $\mu^\pm$\,mesons must be
  $\cong$ 3/4$\cdot$m($\pi^\pm$),  in agreement with the experimental 
results. The $\mu^\pm$\,mesons are not point particles.

\begin{figure}[h] 
	\vspace{0.5cm}
	\hspace{0.2cm}
	\includegraphics{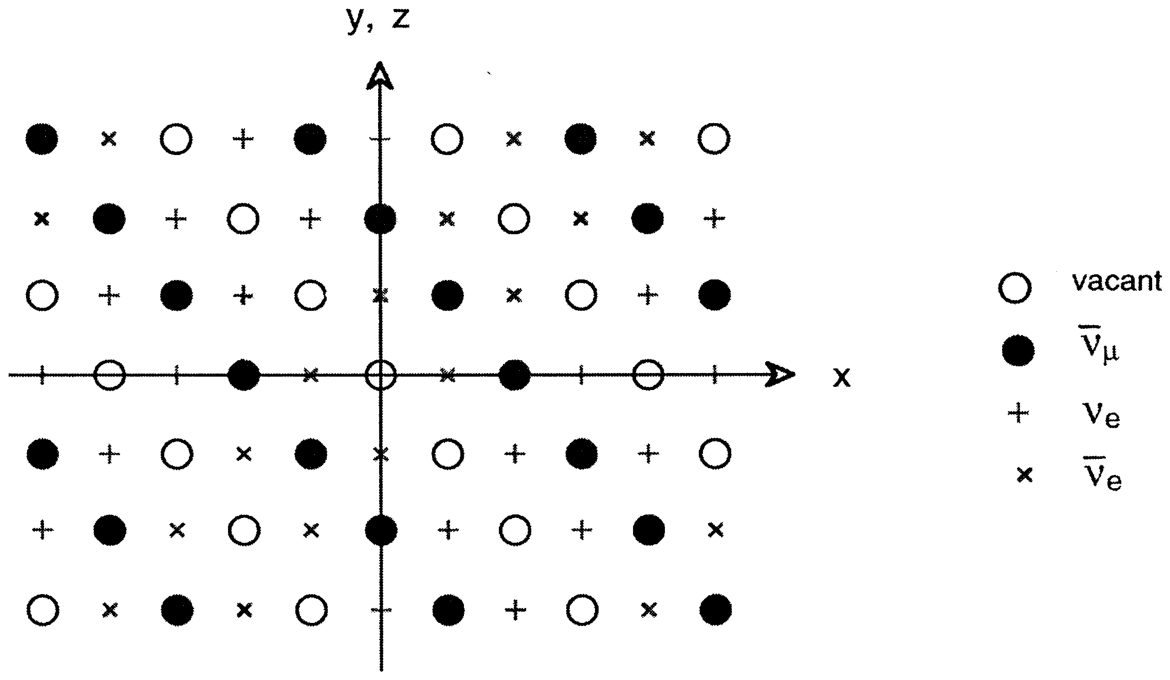}
	\vspace{-0.3cm}
	\begin{quote}
Fig.\,6: A section through the central part of the neutrino lattice  \\
\indent \hspace{1.1cm} of the $\mu^+$ meson without its charge.
	\end{quote}
\end{figure}

   The mass of the $\tau^\pm$\,mesons follows from the most frequent  
decay of the D$^\pm_s$\,mesons, D$^\pm_s \rightarrow \tau^\pm +
 \nu_\tau(\bar{\nu}_\tau)$ (6.4\%). It can be shown readily that the 
oscillation 
energies of the lattices in D$^\pm_s$ and in $\tau^\pm$ are the same. 
From that follows that the energy in the rest mass
 of the $\tau^\pm$\,mesons is the sum of the oscillation energy in the 
$\tau$\,meson lattice plus the sum of the energy of the rest masses of all 
neutrinos and antineutrinos in the $\tau$\,meson lattice, just as with the 
$\mu^\pm$\,mesons. We will skip the details. The tau mesons are not 
point particles either.

\section{The neutrino masses}

   Now we come to the neutrino masses. There is no certain knowledge
what the neutrino masses are. Numerous values for m($\nu_e$) and
m($\nu_\mu$) have been proposed and upper limits for them have
been established experimentally which have, with time, decreased 
steadily. The Review of Particle Physics [2] gives 
for the mass of the electron neutrino the value $<$\,2\,eV/c$^2$. Neither  
the Superkamiokande [35] nor the Sudbury [36] experiments determine 
a neutrino mass, however, both experiments make it very likely that the 
neutrinos have rest masses. We will now determine the neutrino masses 
from the composition of the $\pi^\pm$\,mesons and from the 
$\beta$-decay of the neutron.
 
   If the same principle that applies to the decay of the 
 $\pi^\pm$\,mesons, namely that in the decay the oscillation energy
 of the decaying particle is conserved and that an entire neutrino type 
supplies the energy released in the decay, also applies to the decay of  
the neutron n $\rightarrow$ p + e$^-\,+\,\bar{\nu}_e$, then the mass of  
the anti-electron neutrino can be determined from the known difference 
$\Delta$ =  m(n)\,$-$\,m(p) = 1.293\,332\,MeV/c$_\star^2$ [2]. Nearly
 one half of $\Delta$ comes 
from the energy lost by the emission of the electron, whose mass is 
0.510\,9989\,MeV/c$_\star^2$. N anti-electron neutrinos are in the neutrino
 quadrupoles in the neutron, one-fourth of them is carried away by the 
emitted electron. We have seen in the paragraph below Eq.(29) that 
the decay sequence of the $\pi^\pm$\,mesons requires
that the electron carries with it N$^\prime$/4 electron neutrinos, if the
 $\pi^\pm$\,mesons consist of a lattice with a center neutrino or 
antineutrino  
and equal numbers of $\nu_e,\bar{\nu}_e,\nu_\mu,\bar{\nu}_\mu$ 
neutrinos as required by conservation of 
neutrino number during the creation of $\pi^\pm$.  The electron can carry
 N$^\prime$/4 anti-electron neutrinos as well as N$^\prime$/4 electron 
neutrinos. Since, as we will see shortly, m$(\nu_e$) = m($\bar{\nu}_e$)
 this does not make a difference energetically but is relevant
 for the orientation of the spin vector of the emitted electron. After the
neutron has lost N$^\prime$/4 anti-electron neutrinos to the electron 
emitted in the $\beta$-decay the other 3/4$\cdot$N$^\prime$
 anti-electron neutrinos in the neutron provide
 the energy $\Delta$ $-$ m(e$^-$)c$_\star^2$ = 
0.782\,333\,MeV released in the decay of the neutron. After division by
 3/4$\cdot$N$^\prime$ the rest mass of the anti-electron neutrino is
\begin {equation} \mathrm{m}(\bar{\nu}_e) = 
0.365\,\mathrm{milli\,eV/c}_\star^2\,.\end{equation}

\noindent
 Since theoretically the antineutron decays as 
$\bar{\mathrm{n}} \rightarrow\bar{\mathrm{p}}$ + e$^+$ + $\nu_e$ it 
follows from the same considerations as with the decay of the neutron that 
\begin{equation} \mathrm{m}(\nu_e) = \mathrm{m}(\bar{\nu}_e)\,. 
\end{equation}
We note that 
\begin{equation}\mathrm{N^\prime}/4\cdot \mathrm{m}(\nu_e) =
 \mathrm{N^\prime}/4\cdot \mathrm{m}(\bar{\nu}_e) =
 0.51\,\mathrm{m(e}^\pm)\,. \end{equation}
This equation is, as we will see, fundamental for the explanation of  
the mass of the electron.
 
   Inserting Eq.(37) into Eq.(29) we find that 
\begin{equation} \mathrm{m}(\nu_\mu) = 
49.91\,\mathrm{milli\,eV}/\mathrm{c}_\star^2\,.
\end{equation}
Since the same considerations apply for either the $\pi^+$ or 
the $\pi^-$ meson it follows that  
\begin{equation}\mathrm{m}(\nu_\mu) = \mathrm{m}(\bar{\nu}_\mu)\,.
\end{equation}
Experimental values for the rest masses of the different neutrino types are
not available. However, it appears that for the $\nu_\mu \leftrightarrow 
\nu_e$
oscillation the value for $\Delta$m$^2$ = m$^2_2$ $\mathrm{-}$ m$^2_1$  =
3.2$\times10^{-3}$\,eV$^2$ given on p.1565 of [35] can be used to
 determine m$_2$ = m($\nu_\mu$) if m$_1$ = m$(\nu_e$) is much smaller than
 m$_2$. We have then m($\nu_\mu$) $\approx$ 56.56\,milli-eV/c$^2_\star$,
which is compatible with the value of m($\nu_\mu$) given in Eq.(40).

From Eqs.(37,40) follows that 

\begin{equation}\mathrm{m}(\nu_e) = 1/136.74\cdot \mathrm{m}(\nu_\mu)
 \cong \alpha_f\mathrm{m}(\nu_\mu)\,. \end{equation}
\noindent
1/136.74 is 1.0021 times the fine structure constant $\alpha_f$ = 
e$^2/\hbar c$ = 1/137.036. It does not seem likely that Eq.(42) is just 
a coincidence. The probability for this being a coincidence is zero 
considering the infinite pool of numbers on which the 
ratio m($\nu_e$)/m($\nu_\mu)$ could settle.

   The mass of the $\tau$\,neutrino $\nu_\tau$ can be determined from 
the decay D$_s^\pm$  $\rightarrow$ 
$\tau^\pm$ + $\nu_\tau\,(\bar{\nu}_\tau)$,
 and  the subsequent decay $\tau^\pm \rightarrow \pi^\pm + \bar{\nu}_\tau
(\nu_\tau)$, which are stated in [2]. The appearance of $\nu_\tau$
in the decay of D$^\pm_s$ 
and the presence of $\nu_\mu,\bar{\nu}_\mu,\nu_e,\bar{\nu}_e$ neutrinos 
in the $\pi^\pm$ decay product of the $\tau^\pm$\,mesons means that 
 $\nu_\tau,\bar{\nu}_\tau,\nu_\mu,\bar{\nu}_\mu,\nu_e,\bar{\nu}_e$ 
neutrinos must be  in the D$^\pm_s$ lattice. 
The additional $\nu_\tau$ and $\bar{\nu}_\tau$ neutrinos can be
accomodated in the D$^\pm_s$ lattice by a body-centered cubic
lattice, in which there is in the center of each cubic cell one particle 
different
from the particles in the eight cell corners (Fig.\,5). In a body-centered 
cubic lattice are N$^\prime$/4  cell centers, if N$^\prime$ is the number 
of lattice points without the cell centers. If the particles in the cell 
centers are tau neutrinos then N$^\prime$/8 \,$\nu_\tau$ and
N$^\prime$/8 \,$\bar{\nu}_\tau$ neutrinos must be present, because  
of conservation of neutrino numbers. From m(D$^\pm_s$) = 
1968.5\,MeV/c$_\star^2$ and m($\tau^\pm$) = 
1777\,MeV/c$_\star^2$  follows that

\begin{equation} \mathrm{m}(\mathrm{D}^\pm_s) - \mathrm{m}(\tau^\pm) = 
191.5\,\mathrm{MeV/c_\star^2} = \mathrm{N^\prime}/8
\cdot\mathrm{m}(\nu_\tau)\,.\end{equation}
\noindent
The rest mass of the $\tau$\,neutrinos is therefore
\begin{equation} \mathrm{m}(\nu_\tau) = \mathrm{m}(\bar{\nu}_\tau) = 
0.537\,\mathrm{eV/c}_\star^2\,.
\end{equation}
\noindent
From the neutrino masses given by Eq.(44) and Eq.(40) follows that 
\begin{equation}
\mathrm{m}(\nu_\tau) = 10.76\,\mathrm{m}(\nu_\mu) = 
1.048\,(\alpha_w/\alpha_f)\mathrm{m}(\nu_\mu)\,,
\end{equation}
 where $\alpha_w$  is the weak coupling constant $\alpha_w$  = 
$g_w^2/4\pi\hbar c$ = 1.02$\cdot10^{-5}(m_W/m_p)^2$ [22] 
and $\alpha_f$  is  the fine  structure constant. We keep in mind that 
g$_w^2$ in $\alpha_w$ is not nearly as accurately known as 
e$^2$  in $\alpha_f$. With Eq.(42) we find that 
\begin{equation}
\mathrm{m}(\nu_\tau) = 1.048\cdot\alpha_w/\alpha_f^2\cdot
\mathrm{m}(\nu_e) = 1474\,\mathrm{m}(\nu_e)\,. \end{equation} 

   To summarize what we have learned about the masses of the leptons: 
We have found an explanation for the mass of the $\mu^\pm$\,mesons 
and $\tau^\pm$\,mesons. We have also determined the rest masses of 
the \emph{e},\,$\mu,\tau$ neutrinos and antineutrinos. In other words,
 we have found the masses of all leptons, exempting the electron, 
which will be dealt with in the next Section.

\section{The electron}

The electron differs from the other particles we have considered in so far 
as it appears that its electric charge cannot be separated from its mass, 
whereas 
in the other charged particles the mass of the electric charge is, in a 
first 
approximation, unimportant for the mass of the particles. Even in the 
muons the mass of the electron contributes only five thousandth of the 
muon mass. On the other hand, the electron is fundamental for the 
stability 
of the charged 
particles whose lifetime is sometimes orders of magnitude larger than the 
lifetime of their neutral counterparts. For example the lifetime of the 
$\pi^\pm$\,mesons is eight orders of magnitude larger than the lifetime of 
$\pi^0$, the lifetime of the proton is infinite, whereas the neutron 
decays 
in about 900 seconds and, as a startling example, the lifetime of 
$\Sigma^\pm$ is O($10^{-10}$) seconds, whereas the lifetime of 
$\Sigma^0$ is O($10^{-20}$) seconds. There is something particular 
to the interaction of electric charge with the particle masses.  

   After J.J. Thomson [43] discovered the small corpuscle which soon 
 became known as the electron an enormous amount of theoretical work
 has been done to explain the existence of the electron. Some of the most
 distinguished physicists have participated in this effort. Lorentz [44], 
Poincar\'{e} [45], Ehrenfest [46], Einstein [47], Pauli [48], and others 
showed 
that it is fairly certain that the electron cannot be explained as a 
purely electromagnetic particle. In particular it was not clear 
 how the elementary electric charge could be held together in its  
small volume because the internal parts of the charge repel each other. 
Poincar\'{e} [49] did not leave it at showing that such an electron
 could not be stable, but suggested a solution for the problem by 
introducing what has become known as the Poincar\'{e} stresses whose origin
however remained unexplained. These studies were concerned with the
 static properties of the electron, its mass m(e$^\pm$) and its electric 
charge e. In order to explain the electron with its existing mass and 
charge it appears to be necessary to add to Maxwell's equations a 
non-electromagnetic mass and a non-electromagnetic force which could 
hold the electric charge together. We shall see what this mass and force 
is. 

   The discovery of the spin of the electron by Uhlenbeck and 
Goudsmit [50] increased the difficulties of the problem in so far as it now
 had also to be explained how the angular momentum $\hbar$/2 and the 
magnetic moment $\mu_e$  come about.
The spin of a point-like electron seemed to be explained by Dirac's [51] 
equation, however it turned out later [52] that Dirac type equations can  
be constructed for any value of the spin. Afterwards Schr\"{o}dinger [53] 
tried to explain the spin and the magnetic moment of the electron with 
 the so-called Zitterbewegung. Later on many other models of the electron
 were proposed. On p.74 of his book ``The Enigmatic Electron" Mac 
Gregor [54] lists more than thirty such models.
At the end none of these models has been completely successful 
because the problem developed a seemingly insurmountable difficulty
 when it was shown through electron-electron scattering experiments 
that the radius of the electron must be
 smaller than $10^{-16}$\,cm, in other words that the electron appears 
to be a point particle, at least by three orders of magnitude smaller than 
the classical electron radius r$_e$  =  e$^2$/mc$^2$  = 
2.8179$\cdot10^{-13}$\,cm. This, of 
course, makes it very difficult to explain how a particle can have
a finite angular momentum when the radius goes to zero, and how an 
electric charge can be confined in an infinitesimally small volume. If the 
elementary electric
charge would be in a volume with a radius of O($10^{-16}$)\,cm the
 Coulomb self-energy would be orders of magnitude larger than the 
rest mass of the electron, which is not realistic. The choice is between 
a massless point charge and a finite size particle with a non-interacting 
mass to which an elementary electric charge is attached. It seems fair to 
say that at present, more than 100 years after the discovery of the 
electron,
we do not have an accepted theoretical explanation of the electron.

   We propose in the following, as in [55], that the non-electromagnetic 
mass which seems  to be necessary in order to explain the electron 
consists of neutrinos. This is actually a necessary consequence of our 
standing wave model  of the masses of the mesons and baryons. 
And we propose that the non-electromagnetic force required to 
hold the  electric charge and the neutrinos in the electron together is  
the weak nuclear force which, as we
 have suggested, holds together the masses of the mesons and
baryons and also the mass of the muons. Since the range of the weak 
nuclear force is on the order of $10^{-16}$\,cm the neutrinos must be 
arranged in a lattice with the weak force extending from each lattice  
point only to the nearest neighbors.  The O($10^{-13}$)\,cm size  
of the neutrino lattice in the electron does 
not at all contradict the results of the scattering experiments, just as 
the  
explanation of the mass of the  muons with the standing wave model 
 does not contradict the apparent point particle characteristics of the 
muon, 
 because neutrinos are in a very good approximation non-interacting and 
therefore are not noticed in scattering experiments with electrons. 

   The rest mass of the electron is m(e) = 0.510\,998\,92 $\pm$ 
4$\cdot10^{-8}$\,MeV/c$^2_\star$ and the electrostatic charge
 of the electron is e = 4.803\,2044\,$\cdot\,10^{-10}$\,esu, as stated 
in the Review  of Particle Physics [2].  \emph{ The objective of a 
theory  of the electron must, first of all, be the explanation of 
m(e$^\pm$) and e}, but also of s(e$^\pm$) and of the magnetic 
moment $\mu_e$. We will first 
explain the rest mass of the electron making use of what we
 have learned from the standing wave model, in particular of what
 we have learned about the explanation of the mass of the 
$\mu^\pm$\,mesons in Section 7. The muons are leptons, just as
 the electrons, that 
means that they interact  with other particles exclusively through 
the electric force. The muons have a mass which is 206.768 
times larger than the mass of the electron, but they have the same 
elementary electric charge as the electron or positron and the same 
spin. Scattering experiments tell that the $\mu^\pm$\,mesons are point 
particles  with a size $<$\,$10^{-16}$\,cm, just as the electron. In other 
words, the muons have the same characteristics as the electrons and
 positrons but for a mass which is about 200 times larger. 
Consequently the muon is often referred to as a ``heavy"  electron. 
If a non-electromagnetic  mass is required to explain
 the mass of the electron then a non-electromagnetic mass 200 times
 as large as in the electron is required to explain the mass of the muons.
These non-electromagnetic masses must be \emph{non-interacting}, 
otherwise scattering experiments could not find the size of either the 
electron or the muon at 10$^{-16}$\,cm. 
 
   We have explained the mass of the muons with the standing wave 
model in Section 7. According to our model the muons consist of an 
elementary electric charge and an oscillating lattice of neutrinos. 
Neutrinos are the only non-interacting matter we know  of. 
In the muon lattice  are, according to our model,
 (N\,-\,1)/4 = N$^\prime$/4 muon neutrinos $\nu_\mu$  
(respectively anti-muon  neutrinos $\bar{\nu}_\mu$), 
N$^\prime$/4 electron neutrinos $\nu_e$ and the same 
 number of anti-electron neutrinos $\bar{\nu}_e$,  one 
elementary electric charge and the energy of the lattice oscillations. 
The letter N stands for the number of all neutrinos and antineutrinos in
 the cubic lattice of the $\pi^\pm$ mesons, N = 2.854$\cdot10^9$, 
Eq.(15). It is, as explained in Section 7, a necessary consequence 
of the decay of, say, the $\mu^-$ muon 
$\mu^-$  $\rightarrow$ e$^-$ + $\bar{\nu}_e$ + $\nu_\mu$ 
that there must be
N$^\prime$/4 electron neutrinos $\nu_e$ in the emitted electron. 
For the mass of the electron neutrinos and anti-electron neutrinos  
we found in Eq.(37)  that m($\nu_e$) = m($\bar{\nu}_e$) = 
0.365\,milli-eV/c$^2_\star$.
The sum of the energies in the rest masses of the N$^\prime$/4 neutrinos
or antineutrinos in the lattice of the electron or positron is then
\begin{equation} \sum{\,\mathrm{m(\nu_e)c_{\star}^2}} =
 \mathrm{N}^\prime/4\cdot\mathrm{m}(\nu_e)\mathrm{c}_{\star}^2 
= 0.260\,43\,\mathrm{MeV}
 = 0.5096\,\mathrm{m(e^\pm)}\mathrm{c}_{\star}^2\,. \end{equation}
\noindent
In other words, \emph{the sum of the rest masses of the neutrinos
in the electron is approximately equal to 1/2 of the rest mass of the
electron}.  The other half of the rest mass of the 
electron must originate from the energy in the electric charge 
carried by the electron. The explanation of the $\pi^\pm$\,mesons 
led to the  explanation of the $\mu^\pm$\,mesons and now leads to 
the explanation of the mass of e$^\pm$. 

From pair production 
$\gamma$ + M   $\rightarrow$  e$^-$ + e$^+$ + M, (M being any
 nucleus), and from conservation of neutrino numbers follows 
necessarily that there must also be a neutrino lattice 
 composed of N$^\prime$/4 anti-electron neutrinos, which make 
up the lattice of the positrons, which lattice has, since m($\nu_e$) = 
m($\bar{\nu}_e$), the same rest mass as the neutrino lattice of the 
electron, as it must be for the antiparticle of the electron. 
Conservation of charge requires conservation of neutrino numbers.     
   
   Fourier analysis dictates that a continuum of high frequencies must be 
in the electrons or positrons created by pair production in a timespan of 
$10^{-23}$ seconds. We will now determine the energy 
E$_\nu$(e$^\pm$) in the oscillations in the  interior of the electron. 
Since we want to explain the \emph{rest mass} of the electron we can 
only consider the frequencies of non-progressive waves, either standing 
waves or circular waves. The sum of the energies of the lattice 
oscillations 
is, in the case of the $\pi^\pm$\,mesons,  given by 

\begin{equation} \mathrm{E}_\nu(\pi^\pm) =  
\frac{\mathrm{h}\nu_0\mathrm{N}}{2\pi(\mathrm{e^{h\nu/kT}}\,\mathrm{-}\,1)}
\,\int\limits_{-\pi}^{\pi}\,\phi\,d\phi\,. \end{equation}
\noindent
This is Eq.(14) combined with Eq.(16) which were used to determine
 the oscillation energy in the $\pi^0$ and $\pi^\pm$ mesons. This equation 
was introduced by Born and v.\,Karman [13] in order to explain the 
internal energy of cubic crystals.  If we apply Eq.(48) to the  
 oscillations  in the electron which has N$^\prime$/4 electron neutrinos
 $\nu_e$ we arrive at  E$_\nu$(e$^\pm)$  = 1/4$\cdot$E$_\nu(\pi^\pm$), 
which is mistaken because E$_\nu(\pi^\pm$) $\approx$ 
m($\pi^\pm$)c$_{\star}^2$/2 and m($\pi^\pm$) = 273\,m(e$^\pm$). 
Eq.(48) must be modified in order to be suitable for the oscillations in 
the electron. It turns out that we must use
\begin{equation} \mathrm{E}_\nu(\mathrm{e}^\pm) =  
\frac{\mathrm{h}\nu_0\mathrm{N}\cdot\alpha_f}{2\pi(\mathrm{e^{h\nu/kT}}\,
\mathrm{-}\,1)}\,\int\limits_{-\pi}^{\pi}\,\phi\,d\phi\,,
\end{equation}
\noindent
where $\alpha_f$ is the fine structure constant. As is well-known the 
fine structure constant $\alpha_f$ characterizes the strength of the 
electromagnetic forces. The appearance of $\alpha_f$ in Eq.(49)
means that the nature of the oscillations in the electron is different 
from the oscillations in the $\pi^0$ or $\pi^\pm$ lattices. With 
$\alpha_f$  = e$^2/\hbar$c$_{\star}$ 
and $\nu_0$ = c$_{\star}/2\pi$\emph{a} we have
\begin{equation} \mathrm{h}\nu_0\alpha_f = \mathrm{e}^2/\emph{a}\,, 
\end{equation}
 which shows that the oscillations in the electron are \emph{electric 
oscillations}. The appearance of e$^2$  in Eq.(50) guarantees that
the oscillation energy of the electron and positron are the same, as it
must be. 

   There must be N$^\prime$/2 oscillations of the elements of the electric 
charge  in e$^\pm$, because we deal with non-progressive waves, which
are the superposition of two waves. As we will see later the spin requires 
that the oscillations are circular. That means that  2$\times$N$^\prime$/4 
$\cong$ N/2 oscillations are in Eq.(49). From Eqs.(48,49) then follows that

\begin{equation} \mathrm{E_\nu(e}^\pm) =
 \alpha_f/2\cdot\mathrm{E}_\nu(\pi^\pm)\,. \end{equation}
\noindent
E$_\nu(\pi^\pm)$  is the oscillation energy in the $\pi^\pm$\,mesons 
which  can be calculated with Eq.(48). According to Eq.(28) it is
\begin{equation} \mathrm{E}_\nu(\pi^\pm) = 67.82 \,\mathrm{MeV} =
0.486\,\mathrm{m}(\pi^\pm)\mathrm{c}_{\star}^2 \approx 
\mathrm{m}(\pi^\pm)\mathrm{c}_{\star}^2/2\,.
 \end{equation}
 With E$_\nu(\pi^\pm$) $\approx$ m($\pi^\pm$)c$_{\star}^2$/2 = 
139.57/2\,MeV and $\alpha_f$  = 1/137.036  follows from Eq.(51) 
that the oscillation energy of the electron or positron is 
\begin{equation} \mathrm{E_\nu(e}^\pm) = \frac{\alpha_f}{2}\cdot
\frac{\mathrm{m}(\pi^\pm)\mathrm{c}_{\star}^2}{2} = 
0.254\,623\,\mathrm{MeV} = 
0.996\,570\,\mathrm{m}(\mathrm{e}^\pm)\mathrm{c}_{\star}^2/2\,. 
\end{equation}
\noindent
If we replace in Eq.(53) the experimental value for m($\pi^\pm$) by the 
good
empirical approximation m($\pi^\pm$) $\cong$ m(e$^\pm$)(2/$\alpha_f$),
 Eq.(73), then it follows that 
\begin{equation}\mathrm{E}_\nu(\mathrm{e}^\pm) \cong 
1/2\cdot \mathrm{m(e}^\pm)c^2_\star \,.\end{equation} 
In other words,\,\,\emph{ 1/2 of the energy in the rest mass of the 
electron is made up by the oscillation energy in} e$^\pm$. This equation
corresponds to Eq.(29a) for the oscillation energy in the $\pi^\pm$\,mesons.

   In Eq.(53) we have determined the value of the oscillation energy in 
e$^\pm$  from the product of the very accurately known fine structure 
constant and the very accurately 
known rest mass of the $\pi^\pm$\,mesons which establishes 
a firm value of the oscillation energy of e$^\pm$.  The other half of  
the energy  in m(e$^\pm$) is in the rest masses
of the neutrinos in the electron.  We can confirm Eq.(53) without 
 using E$_\nu(\pi^\pm)$ with the formula for the oscillation energy  
 in the form of Eq.(58) with N/2 = 1.427$\cdot10^9$, 
e = 4.803$\cdot10^{-10}$\,esu,
 $\emph{a}$ = 1$\cdot10^{-16}$\,cm, f(T) = 1/1.305$\cdot10^{13}$ and, 
with the integral being $\pi^2$, we obtain E$_\nu$(e$^\pm$) = 
0.968\,m(e$^\pm$)c$_{\star}^2$/2. This calculation involves more 
parameters than in Eq.(53) and is consequently less accurate than Eq.(53).

   In a good approximation the oscillation energy of e$^\pm$ in Eq.(53) 
is equal to the sum of the energies in the rest masses
 of the electron neutrinos in the e$^\pm$ lattice in Eq.(47), or 
E$_\nu$(e$^\pm$) $\cong$ N$^\prime$/4\,$\cdot$\,m($\nu_e$)c$^2$. 
Since
 \begin{equation} \mathrm{m(e}^\pm)\mathrm{c}_{\star}^2
 =  \mathrm{E}_\nu(\mathrm{e}^\pm) + 
\sum{\,\mathrm{m}(\nu_e)\mathrm{c}_{\star}^2} 
= \mathrm{E_\nu(e^\pm)} + \mathrm{N^\prime/4\cdot m(\nu_e)c_{\star}^2}\,,
\end{equation}
\noindent 
it follows from Eqs.(47) and (53) that  
\begin{equation} \mathrm{m(e^\pm)c_{\star}^2}(theor) = 
0.5151\,\mathrm{MeV} = 
1.0079\,\mathrm{m(e^\pm)c_{\star}^2}(exp)\,. \end{equation}
The theoretical rest mass of the electron or positron agrees within the 
accuracy of the parameters N and m($\nu_e)$ with the measured rest mass.

   From Eq.(51) follows with E$_\nu(\pi^\pm)$ $\cong$ 
m($\pi^\pm$)c$_{\star}^2$/2  from Eq.(52) that

\vspace{0.5cm} 
\centerline{2E$_\nu$(e$^\pm) = \alpha_f$E$_\nu(\pi^\pm)$ $\cong$ 
$\alpha_f$m$(\pi^\pm)$c$_{\star}^2$/2  $\cong$ 
m(e$^\pm$)c$_{\star}^2$\,,}

\vspace{0.5cm}
\noindent
or that
\begin{equation} \mathrm{m(e^\pm)}\cdot2/\alpha_f  =
 274.072\,\mathrm{m(e^\pm)} 
 \cong \mathrm{m(\pi^\pm)}\,, \end{equation}
whereas the actual ratio of the mass of the $\pi^\pm$\,mesons  to 
the mass of the electron is m($\pi^\pm$)/m(e$^\pm$) = 
273.132 or 0.9965\,$\cdot$\,2/$\alpha_f$. We have
 here the same ratio m($\pi^\pm$)/m(e$^\pm$) which we find
with the standing wave model of the $\pi^\pm$\,mesons, Eq.(73).
This is a necessary condition for the validity of our model 
of the electron.

   We have thus shown that the \emph{rest mass of the electron or 
positron can be explained} by the sum  of the rest masses of the 
electron neutrinos or anti-electron neutrinos in a cubic  lattice with 
N$^\prime$/4 electron neutrinos 
$\nu_e$ plus the mass in the sum of the energy of 
N/2 standing electric oscillations in the lattice, Eq.(53). The one 
oscillation added to the even numbered N$^\prime$/4 oscillations   
is the oscillation at the center of the lattice, Fig.\,7. From this model 
follows, since it deals with a cubic neutrino lattice, that \emph{the 
electron is not a point particle}. However, since neutrinos are 
non-interacting their presence will not be detected in 
electron-electron scattering experiments. 

\begin{figure}[h]
\vspace{0.5cm}
\hspace{2.2cm}
\includegraphics{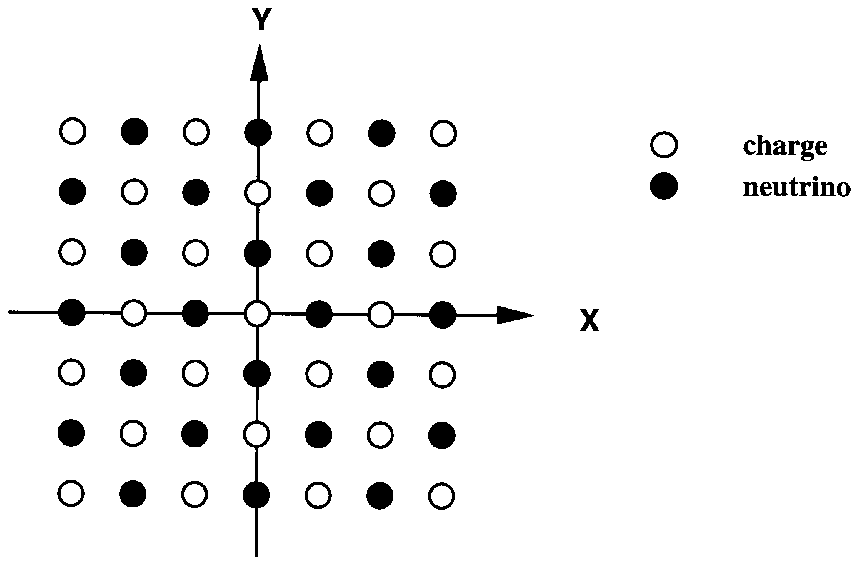}
\vspace{-0.2cm}
\begin{quote}
Fig.\,7: Horizontal or vertical section through the central part of\\ 
\indent\hspace{1.1cm} the electron lattice.
\end{quote}
\end{figure}

   The size of the electron determined by 
 electron-electron scattering at $<$\,$10^{-16}$\,cm seems
to contradict the model of the electron proposed here. However, this
experimental size does not apply to the circumstances considered here. 
One has to find the scattering formula for finite size charged cubic 
lattices  
and analyze the experimental data with such a scattering formula in order  
to see whether our model is in contradiction to the experiments. If the 
electron lattice behaves like a rigid body then electron-electron 
scattering 
should take place at the center of the lattice, which is a point [54].

   Let us compare the rest mass of the electron to the \emph{rest mass of  
the muon}, which was explained in Section 7 with an oscillating lattice of 
muon and electron neutrinos and an elementary electric charge. The 
electron 
has the most simple neutrino lattice, consisting of only one neutrino type,
either electron neutrinos or anti-electron neutrinos, and it has the 
smallest 
sum of the rest masses of the neutrinos in a particle. The heavy weight 
of the  muon m($\mu^\pm$) = 206.768\,m(e$^\pm$) is a 
consequence of the heavy weight of the
 N$^\prime$/4 muon neutrinos or N$^\prime$/4 anti-muon neutrinos in
the muon lattice. The mass of either a muon neutrino or an anti-muon 
neutrino is 137 times the mass of an electron or anti-electron neutrino,
Eq.(42). That makes the electron neutrinos and anti-electron neutrinos in 
$\mu^\pm$ as well as the mass of the 
electric charge in a first approximation negligible. It then follows from 
 Eq.(68) that m($\mu^\pm$)(\emph{theor}) $\cong$ 
3/2\,$\cdot$\,m($\nu_\mu$)/m($\nu_e$)\,$\cdot$\,m(e$^\pm$) = 
205.11\,m(e$^\pm$) = 0.99198\,m($\mu^\pm$)(\emph{exp}), 
which proves that the heavy mass of the muons is caused by the 
heavy muon neutrinos.  
 
   In order to confirm the \emph{validity} of our preceding explanation of 
the mass
of the electron we must show that the sum of the charges in the electric 
oscillations in the interior of the electron is equal to the elementary 
electric charge of the electron.   
We recall that Fourier analysis requires that, after pair production,
there must be a continuum of frequencies in the electron and positron.
With h$\nu_0\alpha_f$ = e$^2$/\emph{a} from Eq.(50) follows 
from Eq.(49) that the oscillation energy in e$^\pm$ is the sum of 
2$\times$(N$^\prime$/4 + 1) $\cong$ N/2  electric oscillations

\begin{equation} \mathrm{E}_\nu(\mathrm{e}^\pm) =  
\frac{\mathrm{N}}{2} \cdot \frac{\mathrm{e^2}}{\emph{a}}\cdot 
\frac{f(T)}{2\pi}\,
\int\limits_{-\pi}^{\pi}\,\phi\,\mathrm{d}\phi\,, \end{equation}
with f(T) = 1/(e$^{h\nu/kT} \mathrm{-}$ 1) = 1/1.305$\cdot10^{13}$
from p.17. Inserting the values for N, f(T) and \emph{a} we find
that E$_\nu$(e$^\pm$) = 0.968\,m(e$^\pm$)c$_\star^2$/2 $\cong$ 
m(e$^\pm$)c$_\star^2$/2 as in Eq.(53). The discrepancy
between m(e$^\pm$)c$_\star^2$/2 and E$_\nu$(e$^\pm$) so calculated
must originate from the uncertainty of the parameters N, f(T) and \emph{a}
in Eq.(58).

   In order to determine the charge e in the electric oscillations we 
replace the integral divided by 2$\pi$ in Eq.(58), which has the value 
$\pi$/2, by the sum $\Sigma\,\phi_k\Delta\,\phi$, where k is an integer 
number with the maximal value k$_m$ = (N/4)$^{1/3}$. $\phi_k$ is 
equal to k$\pi$/k$_m$  and we have

\begin{displaymath}  \Sigma\,\phi_k\,\Delta\phi = 
\sum_{k=1}^{k_m}\,\frac{k\pi}{k_m}\cdot\frac{1}{k_m}
= \frac{ k_m(k_m + 1)\pi}{2\,k_m^2} \cong \frac{\pi}{2}\,,
\end{displaymath}
\noindent as it must be. The energy in the individual electric oscillation 
with index k is then 
\begin{equation} \Delta\mathrm{E}_\nu(k) = \phi_k\,\Delta\phi = 
k\pi/k_m^2\,.
\end{equation}

      Suppose that the energy of the electric oscillations is correctly 
described by the self-energy of an electric charge 
\begin{equation}\mathrm{U} = 1/2\,\cdot\,\mathrm{Q}^2/\mathrm{r}\,.
\end{equation}
The self-energy of the elementary electric charge is normally used to 
determine the mass of the electron from its charge, here we use Eq.(60) 
the other way around, we determine the charge from the energy  
in the oscillations. 

   The charge of the electron is contained in the electric oscillations. 
That means that the electric charge of e$^\pm$ is not concentrated in a
point, but is distributed over N/4 = O($10^9)$ charge elements
 Q$_k$. \emph{The charge elements are distributed in a cubic lattice} and
 the resulting electric field is cubic, not spherical. In the absence of a 
central 
force which originates at the center of the particle and affects all parts 
of the
particle the configuration of the particle is not spherical, just as it 
was with
the shape of the $\pi^\pm$\,mesons. For distances large as compared 
to the sidelength of the cube, (which is O($10^{-13}$)\,cm), say at the 
first Bohr radius which is on the order of $10^{-8}$\,cm, the deviation 
of the cubic 
field from the spherical field will be reduced by about $10^{-10}$.   
 The charge in all electric oscillations in the electron is  
\begin{equation} \mathrm{Q} = \sum_{k}\,\mathrm{Q}_\mathrm{k}\,. 
\end{equation}

   Setting the radius r in the formula for the self-energy  equal to 
2\,\emph{a} we find, 
with Eqs.(58,59,60), that the charge in the individual electric 
oscillations is 

\begin{equation} \mathrm{Q_k} = 
\pm\,\sqrt{2\pi\,N\,e^2f(T)/k_m^2}\,\cdot\,\sqrt{k}\,.
\end{equation}  

\noindent
and with k$_m$ = 1/2\,$\cdot$\,(N/4)$^{1/3}$ = 447 and
\begin{equation}  \sum_{k=1}^{k_m}\,\sqrt{k} = 6310.8\, \end{equation}

\noindent follows, after we have doubled the sum over $\sqrt{k}$, because 
for each index k there is a second oscillation on the negative axis of 
$\phi$, 
that

\begin{equation} \mathrm{Q} = \Sigma\,\mathrm{Q_k} = 
\pm\,5.027\cdot10^{-10}\,\,\mathrm{esu}\,,\end{equation}

\noindent whereas the elementary electric charge is e = 
$\pm$\,4.803\,$\cdot\,10^{-10}$\,esu.  That means that our theoretical 
charge of the electron is 1.047 times the elementary electric charge. 
Within the uncertainty of the parameters the theoretical charge of the 
electron agrees with
the experimental charge e. We have confirmed that it follows from our 
explanation of the mass of the electron that the electron has, within a 
5\% error, the correct electric charge.

   Each element of the charge distribution is surrounded in the horizontal
plane by four electron neutrinos as in Fig.\,7, and in vertical direction by  
an electron neutrino above and also below the element. The electron 
neutrinos hold the charge elements in place.  We must assume that 
the charge elements are bound to the neutrinos by the weak nuclear 
force. The weak nuclear force plays here a role similar 
to its role in holding, for example,  the $\pi^\pm$ or $\mu^\pm$ lattices 
together. It is not possible, in the absence of a definitive explanation 
of the neutrinos, to give an explanation for the electro-weak 
interaction between the electric oscillations and the neutrinos. 
However, the presence of the range
 \emph{a} of the weak nuclear force in e$^2$/\emph{a} is a sign  
that the weak force is involved in the electric oscillations. The 
attraction of the charge elements by the neutrinos
overcomes the Coulomb repulsion of the charge elements.
 The weak nuclear force is the missing non-electromagnetic force   
or the Poincar\'{e} stress which holds the elementary electric charge 
together. The same considerations apply 
for the positive electric charge of the positron, only that then the 
electric oscillations are all of the positive sign and that they are 
bound to  anti-electron neutrinos.

   Finally we learn that  Eq.(58) precludes the possibility that the 
charge of the electron sits only on its surface. The number N in 
Eq.(58) would then be on the order 
of $10^6$, whereas N must be on the order of $10^9$ so that
E$_\nu$(e$^\pm$) can be m(e$^\pm$)c$_{\star}^2$/2 as is necessary. 
In other words, the charge of the electron must be distributed throughout 
the interior of the electron, as we postulated.  
 
   Summing up: The rest mass of the electron or positron originates from 
the sum of the rest masses of N$^\prime$/4 electron neutrinos or 
anti-electron neutrinos in cubic lattices plus the mass in the energy of 
the electric oscillations in their neutrino lattices. The neutrinos, as well 
as the electric oscillations, make
 up 1/2 of the rest mass of e$^\pm$ each.
 The  electric oscillations are bound to the 
neutrinos by the weak nuclear force. The sum of the charge elements
 of the electric oscillations accounts for the elementary charge of the 
electron, respectively positron. The electron or the positron are 
not point particles.

   One hundred years of sophisticated theoretical work have made it 
abundantly 
clear that the electron is not a purely electromagnetic particle. There 
must be 
something else in the electron but electric charge. It is equally clear 
from the 
most advanced scattering experiments that the  ``something else"  in the 
electron must be non-interacting, otherwise it could not  be that we find 
that 
the radius of the electron must be smaller than $10^{-16}$\,cm. The only 
non-interacting matter we know of with certainty are the neutrinos. So it 
seems 
to be natural to ask whether neutrinos are not part of the electron. 
Actually we have
not introduced the neutrinos in an axiomatic manner but rather as a 
consequence of our standing wave model of the stable mesons, baryons
and $\mu$\,mesons. It follows necessarily from this model that after the 
decay of, say, the $\mu^-$\,meson there must be electron neutrinos in the 
emitted electron, and that they make up one half of the rest mass of the 
electron. The other half of the energy in the electron originates from the 
energy of the electric oscillations. We have thus explained the rest 
masses of the electron or positron which agree, within 1\% accuracy, 
with the experimental
 value of m(e$^\pm$). We have learned that the charge of the electron
 is not concentrated in a single point, but rather is distributed over 
O(10$^9$) elements which are held together with the neutrinos by the 
weak nuclear force. The sum of the charges in the electric oscillations 
is, within the accuracy of
the parameters, equal to the elementary charge of the electron. From the 
explanation of the mass and charge of the electron follows, as we will 
show, the correct spin and 
magnetic moment of the electron, the other two fundamental features of the 
electron. With a cubic lattice of anti-electron neutrinos we also arrive 
with the same considerations as above at the correct 
mass, charge, spin and magnetic moment of the positron.

\section{The ratios m($\mu^\pm$)/m(e$^\pm$), m($\pi^\pm$)/m(e$^\pm$) \\ 
and  m(p)/m(e)}

   In order to check on the validity of our explanation of m($\pi^\pm$),
m($\mu^\pm$) and also of m(e$^\pm$) we will now look at the ratios
m($\mu^\pm$)/m(e$^\pm$), m($\pi^\pm$)/m(e$^\pm$) and
 m(p)/m(e). In order to determine m($\mu^\pm$)/m(e$^\pm$) we first
modify Eq.(39) by setting  N$^\prime$/4\,$\cdot$\,m($\nu_e$)
= 0.5\,m(e$^\pm$), not at 0.51\,m(e$^\pm$). In other words we say that 
1/2 of the mass of the electron is made of neutrinos. If the other half of
the mass of the electron originates from the electric charge of the 
electron,
 as we have shown in the preceding Section,  then the mass of the electron
 is twice the mass of the sum of the rest masses of the neutrinos in the 
electron (Eq.39) and we have 

\begin{equation}  \mathrm{m(e}^\pm) = \mathrm{N}^\prime/2\cdot
\mathrm{m}(\nu_e) \quad\mathrm{or}\quad 
\mathrm{N}^\prime/2\cdot\mathrm{m}(\bar{\nu}_e)\,.
\end{equation}
 We also set E$_\nu(\pi^\pm$) = 0.5\,m($\pi^\pm$)c$_\star^2$, not 
at 0.486\,m($\pi^\pm$)c$_\star^2$
as in Eq.(28). With E$_\nu(\pi^\pm$) = E$_\nu(\mu^\pm$) from Eq.(33)
follows with Eq.(31) and m($\pi^\pm$)c$_\star^2$ = 2\,E$_\nu(\pi^\pm$) 
that 
\begin{equation} \mathrm{E}_\nu(\mu^\pm) = 
\mathrm{N}^\prime/2\cdot[\mathrm{m}(\nu_\mu) + 
\mathrm{m}(\nu_e)]c_\star^2\,. \end{equation} 
From Eq.(36) and with Eqs.(38,41) then follows that 

\begin{equation} \mathrm{m}(\mu^\pm) = 3/4\cdot\mathrm{N^\prime}
\mathrm{m}(\nu_\mu) + \mathrm{N^\prime}\mathrm{m}(\nu_e)\,,
\end{equation}
considering only the neutrino masses, not the charge in $\mu^\pm$.
With m(e$^\pm$) = N$^\prime$/2\,$\cdot$\,m($\nu_e$) from
Eq.(65) we have
 \begin{equation} \frac{\mathrm{m}(\mu^\pm)}{\mathrm{m(e}^\pm)} = 
\frac{3}{2}\cdot\frac{\mathrm{m}(\nu_\mu)}{\mathrm{m}(\nu_e)} + 2\,,
\end{equation}
or with m($\nu_\mu$)/m($\nu_e$) $\cong$ 1/$\alpha_f$ from Eq.(42) 
it turns out that

\begin{equation} \frac{\mathrm{m}(\mu^\pm)}{\mathrm{m(e}^\pm)}
\cong \frac{3}{2}\cdot \frac{1}{\alpha_f} + 2 = 207.55\,, \end{equation}
\noindent
which is 1.0038 times the real ratio 206.768. In order to
arrive at the proper ratio of m($\mu^\pm$)/m(e$^\pm$) \emph{the
ratio of the neutrino masses m($\nu_e$)/m($\nu_\mu$) must be equal 
to $\alpha_f$}, as in Eq.(42).

   The mass of the muon is, according to Eq.(68), much larger than 
the mass of the electron because the mass
 of the muon neutrino is much larger, 
m($\nu_\mu)$ $\cong$ 137\,m($\nu_e$), than the mass of the electron
 neutrino. The ratio of the mass of the muon to the mass of the electron 
is independent of the number N$^\prime$ of the neutrinos in both  
lattices. 

   Barut's [41] empirical formula for the mass ratio is 
\begin{equation}\mathrm{m}(\mu^\pm)/\mathrm{m(e}^\pm) = 
3/2\alpha_f + 1 = 206.55\,, \end{equation}
\noindent 
whereas the actual ratio is 206.768 = 1.0010\,$\cdot$\,206.55.
 A much better approximation to  
the experimental mass ratio is obtained when the +\,1 in Barut's 
formula is replaced by +\,1.25. The thus calculated 
m($\mu^\pm$)/m(e$^\pm$) = 206.804
differs then from the measured m($\mu^\pm$)/m(e$^\pm$) =
206.7683 by the factor 1.00017. 

   A better relation between the mass of the muon and the mass of the 
pion can now be established. According to the third paragraph of 
Section 7 it is, empirically, 
\begin{equation} \mathrm{m}(\mu^\pm)/\mathrm{m}(\pi^\pm) =
3/4 + \alpha_f\,. \end{equation}
With $\alpha_f$E$_\nu(\pi^\pm$) = 2E$_\nu$(e$^\pm$), (Eq.51), 
and m($\pi^\pm)$c$^2_\star$ = 2E$_\nu(\pi^\pm$), as well as
m(e$^\pm$)c$^2_\star$ = 2E$_\nu$(e$^\pm$), follows that
\begin{equation}  \mathrm{m}(\mu^\pm) \cong 3/4\cdot\mathrm{m}(\pi^\pm) + 
2\,\mathrm{m}(\mathrm{e}^\pm)\,. \end{equation} 
The right hand side of Eq.(72) is 1.00039 times larger than the 
experimental value of m($\mu^\pm$). The reason for the additional 
term 2m(e$^\pm$) on the right hand side of Eq.(72) has still to
be found. 

   Similarly we obtain for the $\pi^\pm$\,mesons the ratio 
\begin{equation} \frac{\mathrm{m}(\pi^\pm)}{\mathrm{m(e^\pm)}}
= 2\,[\frac{\mathrm{m}(\nu_\mu)}{\mathrm{m}(\nu_e)} +1] 
 \cong \frac{2}{\alpha_f} + 2 = 276.07\,,
\end{equation}
which is 1.0108 times the real ratio 273.1321. We have, however, 
only considered the ratio of the rest masses of the neutrinos in 
$\pi^\pm$  and e$^\pm$, not the consequences of either the charge
or the spin in $\pi^\pm$  or e$^\pm$. The empirical formula for
the mass ratio m($\pi^\pm$)/m(e$^\pm$) is
\begin{equation} \mathrm{m}(\pi^\pm)/\mathrm{m}(\mathrm{e}^\pm) 
= 2/\alpha_f\,-\,1 = 273.07\,, \end{equation}
 which is 0.99977 times  the experimental ratio  273.1320. The rest mass 
of the $\pi^\pm$\,mesons is 273 $\cong$ $2\times1/\alpha_f$ times larger 
than the rest mass of the 
electron because the muon or antimuon neutrino masses in $\pi^\pm$ are 
$\cong\,$137 times larger than the electron or anti-electron neutrino 
masses in the electron or positron.

For the $\pi^0$\,meson we find empirically that
\begin{displaymath}\hspace{2.5cm} \mathrm{m}(\pi^0)/\mathrm{m(e}^\pm) =
1.000\,268\,(2/\alpha\, - \,10) = 
264.1426\,.\hspace{1.4cm}(74a)\end{displaymath}
The masses of the $\gamma$-branch particles are, according to Eq.(1), in a 
good approximation integer multiples of the mass of the $\pi^0$\,meson. 
Their masses are consequently multiples of the right hand side of Eq.(74a). 
For example we find that m($\eta$)/m(e$^\pm$)(\emph{exp}) 
= 1.0144\,$\cdot$\,4\,$\cdot$\,(2/$\alpha$\,$-$\,10) and 
m($\Lambda$)/m(e$^\pm$)(\emph{exp}) = 
1.0335\,$\cdot$\,8\,$\cdot$\,(2/$\alpha$\,$-$\,10).   

   In order to determine m(n)/m(e$^\pm$) we start with K$^0$ = 
(2.)$\pi^\pm$ + $\pi^\mp$ and E((2.)$\pi^\pm$) = 4E$_\nu(\pi^\pm$) + 
N$^\prime$/2\,$\cdot$\,[m($\nu_\mu)$ + m$(\nu_e)$]c$_\star^2$, Eq.(30).
Then m(K$^0$) =  7N$^\prime$/2\,$\cdot$\,[m($\nu_\mu)$ + m($\nu_e$)],
 and with m(n) $\cong$ m(K$^0$ +  $\overline{{\mathrm{K}}^0}$) = 2m(K$^0$) 
 follows that

\begin{equation} \frac{\mathrm{m(n)}}{\mathrm{m(e^\pm)}} = 
14\,[\frac{\mathrm{m}(\nu_\mu)}{\mathrm{m}(\nu_e)} + 1] = 1932.5\,,
\end{equation}
that is 105.2\% of the experimental value 1836.15. The 5.2\% excess
of the calculated m(n)/m(e) is the consequence of the 2.6\%
excess of the theoretical m(K$^0$) over the
experimental m(K$^0$). With m(p) = 0.9986\,m(n) we have 
\begin{equation} \frac{\mathrm{m(p)}}{\mathrm{m(e)}} =
 0.9986\cdot14\,[\frac{\mathrm{m}(\nu_\mu)}{\mathrm{m}(\nu_e)} + 1] 
\cong 0.9986\,[\frac{14}{\alpha_f} + 14] = 1929.79\,, \end{equation}
that is 104.98\% of the experimental ratio 1836.15. The empirical 
formula for m(p)/m(e) is 
\begin{equation}
\mathrm{m(p)}/\mathrm{m(e)} = 
14\,[1/\alpha_f - 6] = 0.9980\,\mathrm{m(p)}/\mathrm{m(e)}(exp)\,.
\end{equation}

   We arrive at a better agreement between the theoretical values  
of m(e$^\pm$), m($\mu^\pm$)/m(e$^\pm$) and 
m($\pi^\pm$)/m(e$^\pm$) and their actual values if 
we introduce a modified oscillation energy of the electron 
E$^\prime_\nu$(e$^\pm$) given by 
\begin{equation} \mathrm{E}^\prime_\nu(\mathrm{e}^\pm) = 
\mathrm{E}_\nu(\mathrm{e}^\pm)(1 + \alpha_f/2)\,.
\end{equation}
 With Eq.(53) we then have
\begin{equation} \mathrm{E}^\prime_\nu(\mathrm{e}^\pm) = 
1.000\,206\,\mathrm{m(e^\pm)c}_{\star}^2/2\,. \end{equation}
The agreement of E$^\prime_\nu$(e$^\pm$) given by Eq.(79) with 
the actual m(e$^\pm$)c$_\star^2$/2 is an order of magnitude better 
than it was with Eq.(53).

   The modification of the oscillation energy applies only to the electron,
not to either $\mu^\pm$ or $\pi^\pm$, because according to Eq.(53)
E$_\nu$(e$^\pm$) is proportional to m($\pi^\pm$) and if m($\pi^\pm$)
were also modified by (1 + $\alpha_f$/2) then E$^\prime_\nu$($\pi^\pm$)
would be proportional to (1 + $\alpha_f$/2)$^2$ and would be a 
worse approximation then Eq.(28).

   With m(e$^\pm$)$^\prime$ standing for the mass of the electron 
calculated with the modified oscillation energy E$_\nu^\prime$(e$^\pm$), 
and assuming that m(e$^\pm)^\prime$c$_{\star}^2$ = 
2E$_\nu^\prime$(e$^\pm$), as seems to be the case according to 
Eq.(79), we obtain

\begin{equation} \frac{\mathrm{m}(\mu^\pm)}{\mathrm{m(e}^\pm)^\prime} = 
\frac{\mathrm{m}(\mu^\pm)}{\mathrm{m(e}^\pm)(1 + \alpha_f/2)} \cong 
\frac{\mathrm{m}(\mu^\pm)}{\mathrm{m(e}^\pm)}(1 
\mathrm\,{-}\,\alpha_f/2)\,,
\end{equation}
and with Eq.(69) we have 
\begin{equation} \frac{\mathrm{m}(\mu^\pm)}{\mathrm{m(e}^\pm)^\prime} = 
( \frac{3}{2\alpha_f} + 2)(1\mathrm\,{-}\,\alpha_f/2) =
\frac{3}{2\alpha_f} + 1.25 \,\mathrm\,{-}\,\alpha_f\,,\end{equation}
which agrees remarkably well with the slightly modified form of Barut's 
Eq.(70)
in which the +\,1 is replaced by +\,1.25. So we have
\begin{equation} \frac{\mathrm{m}(\mu^\pm)}{\mathrm{m(e}^\pm)^\prime} = 
206.7967 = 
1.000\,139\,\frac{\mathrm{m}(\mu^\pm)}{\mathrm{m(e}^\pm)}(exp)\,.
\end{equation}

   Similarly we obtain for m($\pi^\pm$)/m(e$^\pm$)$^\prime$ with Eq.(51)
\begin{equation} \frac{\mathrm{m}(\pi^\pm)}{\mathrm{m(e}^\pm)^\prime} =
\frac{2\mathrm{E}_\nu(\pi^\pm)}{2\mathrm{E}_\nu^\prime(\mathrm{e}^\pm)}
\cong \frac{2\mathrm{E}_\nu(\pi^\pm)(1\,\mathrm{-}\, \alpha_f/2)}
{\alpha_f/2\cdot2\mathrm{E}_\nu(\pi^\pm)}\,,
\end{equation}
or \begin{equation} \frac{\mathrm{m}(\pi^\pm)}{\mathrm{m(e}^\pm)^\prime} =
\frac{2}{\alpha_f}(1\,\mathrm{-}\,\alpha_f/2) = 
\frac{2}{\alpha_f}\,\mathrm{-}\,1
= 273.072\,,\end{equation}
whereas the experimental m($\pi^\pm$)/m(e$^\pm$) is 273.132 =
1.000\,2129\,(2/$\alpha_f$\,$\mathrm{-}$\,1).

   Our theoretical calculations of  m($\pi^\pm$)/m(e$^\pm$) and of
m($\mu^\pm$)/m(e$^\pm$) agree within the percent range with their
experimental values. This can only be if our explanation of m($\pi^\pm$),
m($\mu^\pm$) 
and m(e$^\pm$) are correct in the same approximation, and if the ratio 
m($\nu_e$)/m($\nu_\mu$)  in Eq.(42) is valid. In other words, our 
theoretical 
values of m($\pi^\pm$)/m(e$^\pm$) and of m($\mu^\pm$)/m(e$^\pm$)
confirm the \emph{validity} of our explanation of m($\pi^\pm$), 
m($\mu^\pm$) and of m(e$^\pm$), as well as the validity of the relation
m($\nu_e$) = $\alpha_f$m($\nu_\mu$). We have found 
that the leading term in the mass ratios, namely the term given by the 
ratio 
$\alpha_f$ of the mass of the electron neutrino to the mass of the muon 
neutrino is, in our model, the same as in the empirical formulas for the 
mass ratios.  On the other hand, there are differences
 on the right hand side of Eqs.(69,71) in the small second
term of the mass ratios as compared to the same term in the empirical
formulas for the mass ratios. The second term deals with the ratios of the 
number of the electron neutrinos in the mesons and baryons to the number 
of the electron neutrinos in the electron, without considering charge and 
spin.  We have remedied these differences with the small 
modification of the oscillation energy of the electron. The mass 
ratios m($\mu^\pm$)/m(e$^\pm$) and m($\pi^\pm$)/m(e$^\pm$) then 
agree with the experimental mass ratios to the fourth decimal. The mass
 ratio of the proton to the electron cannot be remedied by the 
modification 
of the oscillation energy of the electron, because the difference between 
calculated and experimental mass ratios is primarily caused by the 2.6 \%
excess of the theoretical rest mass of the K$^0$\,mesons.

\section{The spin of the $\gamma$-branch particles}

  It appears to be crucial for the validity of a model of the elementary 
particles that the model can also explain the spin of the particles 
without additional assumptions. The spin or the intrinsic angular 
momentum is, after the mass, the second most important property 
of the elementary particles. The standard model does not explain 
the spin, the spin is imposed on the quarks. 
As is well-known the spin of the electron 
was discovered by Uhlenbeck and Goudsmit [50] more than 80 
years ago. Later on it was established that the baryons have spin as 
well, but not the mesons. We have proposed an explanation of the spin 
of the particles in [56].  For current efforts to 
understand the spin of the nucleon see Jaffe [57] and of the 
spin structure of the $\Lambda$ baryon see G\"ockeler et al.\,[58]. 
Rivas has described his own model of the spin and other spin models
in his book [59]. The explanation of the spin requires an unambiguous 
answer, the spin must be 0 or 1/2 or integer multiples thereof, nothing 
else. 

    For the explanation of the spin of the particles it seems to be 
necessary 
to have an explanation of the structure of the particles. The spin of  
a particle is, of course, the sum of the angular momentum 
 vectors of the oscillations in the particle, plus the sum of the spin 
vectors of its  neutrinos and antineutrinos, plus the spin of the 
electric charges which the particle carries. It is
 striking that the particles which, according to the standing wave model, 
consist of a single oscillation mode do not have spin, as the $\pi^0, 
\pi^\pm$ and $\eta$ mesons do, see Tables 1 and 2. 
 It is also striking that particles whose mass is approximately 
twice the mass of a smaller particle have spin 1/2 as is the 
case with the $\Lambda$ baryon, 
m($\Lambda$) $\approx$ 2m($\eta$),  and with the nucleon 
 m(n) $\approx$ 2m(K$^\pm$) $\approx$ 2m(K$^0$). The $\Xi^0_c$ 
baryon which is a doublet of one mode has also spin 1/2. Composite 
particles which consist of a doublet of one mode plus one or two other
single modes have spin 1/2, as the  $\Sigma^0$, $\Xi^0$\, 
and $\Lambda_c^+$, $\Sigma_c^0$, $\Omega_c^0$ baryons do. 
 The only particle which seems to be the triplet of a single mode,
 the $\Omega^-$ baryon with m($\Omega^-$) $\approx$ 3m($\eta$),
 has spin 3/2. It appears that the relation between
the spin and the oscillation modes of the particles is straightforward.
 
  In the standing wave model  the $\pi^0$ and $\eta$ mesons consist 
of N = 2.85$\cdot10^9$ standing electromagnetic waves, each with its
 own frequency.
Their oscillations are longitudinal. The longitudinal oscillations of
 frequency $\nu_i$ in the $\pi^0$ and 
$\eta$   mesons do not have angular momentum or 
 $\sum_{i}j(\nu_i)$ = 0, with the index running from 0 to N. 
 Longitudinal oscillations 
 cannot cause an intrinsic angular momentum because for longitudinal 
 oscillations $\vec{r}\,\times\,\vec{p}$ = 0.

   Each of the standing electromagnetic waves in the $\pi^0$ and 
$\eta$  mesons may, on the other hand, have spin s = 1 of its 
own, because circularly polarized electromagnetic waves have an angular 
momentum as was first suggested by Poynting [60] and verified by, 
among others, Allen [61]. The creation of the $\pi^0$\,meson in the 
reaction 
$\gamma$ \,+\, p $\rightarrow \pi^0$ + p and conservation of angular 
momentum dictates that the sum of the angular momentum 
vectors of the N electromagnetic waves in the $\pi^0$\,meson is 
zero, $\sum_ij(s_i)$ = 0.  Either the 
sum of the spin vectors of the electromagnetic 
waves in the $\pi^0$\,meson is zero, or each electromagnetic wave in the 
$\pi^0$\,meson has zero spin which would mean that they are linearly 
polarized. Linearly polarized electromagnetic waves are not expected to 
have angular momentum. That this is actually so was proven by Allen [61]. 
 Since the longitudinal  
oscillations in the $\pi^0$ and $\eta$ mesons do not have 
angular momentum and since the sum of the spin vectors $s_i$ of the 
electromagnetic  waves is zero, the intrinsic angular 
momentum of the $\pi^0$ and $\eta$ mesons is zero, or
\begin{equation} j(\pi^0,\eta) =  \sum_{i}\,j(\nu_i) +\sum_{i} \,j(s_i) = 
0 \quad (0\,\le\,i\le \mathrm{N})\,.\end{equation}
In the standing wave model the $\pi^0$ and $\eta$ mesons 
do not have an intrinsic angular momentum or spin, as it must be.

  We now consider particles such as the $\Lambda$ baryon whose
 mass is m($\Lambda$) = 1.0190\,$\cdot$\,2m($\eta$). The $\Lambda$ 
baryon seems to consist of the superposition of two oscillations of equal
 amplitudes and of frequencies $\omega$ and $\mathrm{-}\,\omega$,
 $|\mathrm{-}\,\omega |$ = $\omega$, at each of the N points of the 
lattice. The oscillations of such particles must be coupled what we have 
marked in Tables 1,2 by the $\ast$ sign. The particles contain then 
N circular oscillations, each with its own frequency and each 
having an angular momentum of $\hbar$/2 as we will see.

  The superposition of two perpendicular linearly polarized traveling 
waves of equal amplitudes and frequencies shifted in phase by 
$\pi$/2 leads to a circular wave with the constant angular momentum 
j = $\hbar$. As is well-known, the total energy of a traveling wave is 
the sum of the potential and the kinetic energy. In a traveling wave the 
kinetic energy is always equal to the potential energy. From  

\begin {equation} \mathrm{E}_{pot} + \mathrm{E}_{kin} = 
\mathrm{E}_{tot} = \hbar\omega\,,\end {equation}
\noindent
follows \begin{equation} \mathrm{E}_{tot} = 2\mathrm{E}_{kin} = 
2\frac{\Theta\omega^2}{2}\, = \hbar\omega 
,\end{equation}
\noindent
with the moment of inertia $\Theta$. It follows that the angular momentum 
j is
 
\begin{equation} j = \Theta\omega = \hbar\,.\end{equation}
\noindent
   This applies to a traveling wave  and corresponds to spin s = 1, or
 to a circularly polarized photon.

  We now add to one monochromatic circular  
oscillation with frequency $\omega$ a  second circular oscillation 
with $\mathrm{-}\,\omega$  of the same absolute value as $\omega$ but 
 shifted in phase by $\pi$,  having the same amplitude, as we have done 
 in [56]. Negative frequencies are permitted solutions of the equations 
for 
the lattice oscillations, Eq.(7). In other words we consider the circular 
oscillations

\begin{equation} x(t) = exp[i\omega t] + exp[-\,i(\omega t + 
\pi)]\,,\end{equation}

\begin{equation} y(t) = exp[i(\omega t + \pi/2)] + exp[-\,i(\omega t + 
                 3\pi/2)]\,\,.
\end{equation}
\noindent
This can also be written as 
\begin{equation} x(t) = exp[i\omega t] - exp[-i\omega t]\,,\end{equation}

\begin{equation} y(t) = i\cdot(exp[i\omega t] + exp[-i\omega t] 
)\,.\end{equation}
If we replace \emph{i} in the Eqs. above by $-$\,\emph{i} we have a 
circular oscillation turning in opposite direction.
The energy of the superposition of the two oscillations is the sum of the 
energies of both individual oscillations, and since in circular 
oscillations
 E$_{kin}$ = E$_{pot}$  we have with Eq.(87) 

\begin{equation} 4\mathrm{E}_{kin} = 4\Theta\omega^2/2 = 
\mathrm{E}_{tot} = \hbar\omega\,,\end{equation}
\noindent
from which follows that the circular oscillation has an angular momentum

\begin{equation} j = \Theta\omega = \hbar/2\,. \end{equation}
\noindent
The superposition of two circular monochromatic  oscillations of equal
 amplitudes and frequencies $\omega$ and $\mathrm{-}\,\omega$ 
satisfies the necessary condition for spin \\s = 1/2 that the angular 
momentum is  j = $\hbar$/2.

   The standing wave model treats the $\Lambda$ 
baryon, which has spin s = 1/2 and a mass m($\Lambda$) = 
1.0190\,$\cdot$\,2m($\eta$), as the superposition of two  
particles  of the same type with N standing electromagnetic 
waves. The waves are circular because they are the 
superposition of two waves with the same absolute value 
of the frequency and the same amplitude.  The angular 
momentum  vectors  of all circular waves in the lattice cancel,   
except for \emph{the wave at the center of the crystal}. Each 
oscillation with frequency $\omega$ at $\phi$\,$>$\,0 has at its mirror 
position $\phi$\,$<$\,0 a wave with the frequency $-\,\omega$, 
which has a negative angular momentum, since j = 
mr$^2\omega$  and $\omega = \omega_0\phi$. 
Consequently the angular momentum vectors of both waves cancel. 
The center of the lattice oscillates, as all other lattice points do, but 
with the frequency $\nu$(0) which is determined by the longest 
possible wavelength, which is twice the sidelength d of the lattice, 
so $\nu$(0) = c/2d. As the other circular waves in the lattice the 
circular wave at the center has the angular momentum $\hbar$/2 
according to Eq.(94). The angular momentum of the center wave 
is the only angular momentum which is not canceled by an 
oscillation of opposite circulation. There are, as it must be, three
possible orientations of the axis of the circular oscillations.    
 
    The net angular momentum of the N
 circular oscillations in the lattice which are  superpositions 
of two oscillations reduces to the angular momentum
 of the center oscillation and is $\hbar$/2.  Since the 
circular oscillations in the $\Lambda$ baryon are the 
only possible contribution to an angular momentum the 
intrinsic angular momentum of the $\Lambda$ baryon is $\hbar$/2 or
\begin{equation} j(\Lambda) = \sum_i\,j(\omega_i) = j(\omega_0) =  
\hbar/2\,.\end{equation}             
 We have thus explained that the $\Lambda$ and likewise
 the $\Xi_c^0$ baryon satisfy the necessary condition 
that j = $\hbar$/2 for s = 1/2. The intrinsic angular momentum of the 
$\Lambda$ baryon is the consequence of the superposition of two 
circular oscillations of the same amplitude and the same absolute value
of the frequency. Spin 1/2 is caused by the composition of the
 particles, it is not a contributor to the mass of a particle.

   The other particles of the $\gamma$-branch, the 
$\Sigma^0$, $\Xi^0$, $\Lambda^+_c$, $\Sigma^0_c$ and 
$\Omega^0_c$ baryons are composites of a baryon with spin 1/2 
plus one or two $\pi$\,mesons
 which do not have spin. Consequently the spin of these particles is 
1/2. The spin of all particles of the $\gamma$-branch, exempting the 
spin of the $\Omega^-$\,baryon, has thus been explained.
 For an explanation of 
 s($\Sigma^{\pm,0}$)  = 1/2 and of s($\Xi^{-,0}$)  = 1/2, regardless
 whether the particles  are charged or neutral, we refer to [56].    

\section{The spin of the particles of the $\nu$-branch}

The characteristic particles of the neutrino-branch are the 
$\pi^\pm$\,mesons which have zero spin. At first glance it seems 
to be odd that the $\pi^\pm$\,mesons do not have spin, because
 it seems that the $\pi^\pm$\,mesons should have spin 1/2 from 
the spin of the charges e$^\pm$ in $\pi^\pm$. What happens to
 the spin of e$^\pm$ in $\pi^\pm$\,? The solution of
 this puzzle is in the composition of the $\pi^\pm$\,mesons which 
are, according to the standing wave model, made of a lattice of 
neutrinos and antineutrinos (Fig.\,2) each having spin 1/2, the lattice 
oscillations, and an elementary electric charge.

  The longitudinal oscillations in the neutrino lattice of the 
$\pi^\pm$\,mesons 
do not cause an angular momentum, $\sum_i\,j(\nu_i)$ = 0, as it was with 
the $\pi^0$\,meson. In the cubic lattice of N = O($10^9$) neutrinos and 
antineutrinos of the $\pi^\pm$\,me-sons the spin of 
 nearly all neutrinos and antineutrinos must cancel because conservation 
of angular momentum during the creation of the particle requires that the 
total angular momentum around a central axis is $\hbar$/2. In 
fact the spin vectors of all but the neutrino or antineutrino in the 
center of the lattice cancel. In order for this to be so
 the spin vector of any particular neutrino in the lattice 
has to be opposite to the spin vector of the neutrino at its 
mirror position. As is well-known only left-handed neutrinos and 
right-handed antineutrinos exist. From $\nu$ = $\nu_0\phi$ (Eq.13) follows 
that the direction of motion of the neutrinos in e.g.\,\,the upper right 
quadrant ($\phi$\,$>$\,0) is opposite to the direction of motion in the 
lower left quadrant ($\phi$\,$<$\,0). Consequently the spin vectors of all 
neutrinos or antineutrinos in opposite quadrants are opposite and cancel.
 The only angular 
momentum remaining from the spin of the neutrinos of the lattice is the 
angular momentum of the neutrino or antineutrino at the \emph{center of  
the lattice} which does not have a mirror particle. 
Consequently the electrically neutral neutrino lattice consisting of 
N$^\prime$/2 neutrinos and N$^\prime$/2 antineutrinos and the center
 particle, each with spin j($n_i$) = 1/2, has an intrinsic 
angular momentum j = $\sum_i\,j(n_i)$ = \emph{j(n$_0$}) = $\hbar$/2.

   But electrons or positrons added to the neutral neutrino 
lattice have likewise spin 1/2. If the spin of the electron or positron 
added to the neutrino lattice is opposite to the spin of the neutrino or
 antineutrino in the center of the lattice then the net spin of the
 $\pi^+$ or $\pi^-$ mesons is zero, or
 \begin{equation} j(\pi^\pm) = \sum_i\,j(\nu_i) +\sum_i\,j(n_i) +
 j(\mathrm{e}^\pm) = j(n_0) + j(\mathrm{e}^\pm) = 0
\,\quad(0\leq i \leq \mathrm{N})\,.\end{equation}
It is important for the understanding of the structure of 
the $\pi^\pm$\,mesons to realize that s($\pi^\pm$) = 0 can only be 
explained if the $\pi^\pm$\,mesons consist of a \emph{neutrino lattice} 
to which an electron or positron is added whose spin is opposite to the 
net spin of the neutrino lattice. \emph{Spin 1/2 of the elementary 
electric  
charge can only be canceled by something that has also spin 1/2, and 
the only conventional choice for that is a single neutrino}.

   \emph{The spin, the mass and the decay of $\pi^\pm$ require that 
the $\pi^\pm$\,mesons are made of a cubic neutrino lattice and an 
elementary electric charge} e$^\pm$.

   The spin of the K$^{\pm}$\,mesons is zero. With the spin of the
 K$^\pm$\,mesons we 
encounter the same oddity we have just observed with the spin of the 
$\pi^\pm$\,mesons, namely we have a particle which carries an  
elementary electric charge with spin 
1/2, and nevertheless the particle does not have spin. The explanation 
of s(K$^\pm$) = 0 follows the same lines as the explanation of 
the spin of the $\pi^\pm$\,mesons. In the standing wave model the
 K$^\pm$\,mesons are described by the state (2.)$\pi^\pm$ + $\pi^0$,
 that means by the second mode of the lattice oscillations 
of the $\pi^\pm$\,mesons plus a $\pi^0$\,meson. The second mode
 of the longitudinal oscillations of a neutral neutrino lattice does  
not have a net intrinsic angular momentum $\sum_i\,j(\nu_i)$ = 0. But
 the spin of the neutrinos contributes an angular momentum $\hbar$/2, 
which originates from the neutrino or antineutrino in the center of the 
lattice, just as it is with the neutrino lattice in the 
$\pi^\pm$\,mesons, so $\sum_i\,{ j(n_i)} = j(n_0) = \hbar/2$. Adding 
an elementary electric charge with a spin opposite to the net intrinsic  
angular momentum of the neutrino lattice creates the charged 
(2.)$\pi^\pm$ mode which has zero spin
\begin{equation}j((2.)\pi^\pm) = \sum_i\,{j(n_i)} + j(\mathrm{e}^\pm) 
= j(n_0) + j(\mathrm{e}^\pm) = 0\,.\end{equation}
 As discussed in Section 6 it is 
necessary to add a $\pi^0$\,meson to the second mode of the 
$\pi^\pm$\,mesons in 
order to obtain the correct mass and the correct decays of the 
K$^\pm$\,mesons. Since the $\pi^0$\,meson does not 
have spin the addition of the $\pi^0$\,meson does not add to the 
intrinsic angular momentum of the K$^\pm$\,mesons. So 
 s(K$^\pm$) = 0 as it must be.

  The explanation of s = 0 of the K$^0$ and 
$\overline{{\mathrm{K}}^0}$\,mesons described by 
the state (2.)$\pi^\pm$ + $\pi^\mp$  is different because there is 
now no electric charge whose spin canceled the spin of the neutrino 
lattice in $\pi^\pm$.  The longitudinal oscillations of the second mode
of the neutrino oscillations of $\pi^\pm$  in K$^0$ as well 
 as of the basic $\pi^\mp$ mode do not create an angular momentum, 
$\sum_i\,j(\nu_i)$ = 0. The sum of the spin vectors of the neutrinos in  
K$^0$ and $\overline{{\mathrm{K}}^0}$ is determined by the 
neutrinos in the second mode of the $\pi^\pm$\,mesons,
or the (2.)$\pi^\pm$ state, and the basic  $\pi^\mp$ mode, 
each have N$^\prime$/2 neutrinos and 
N$^\prime$/2 antineutrinos plus a center neutrino or antineutrino, so 
the number of all neutrinos and antineutrinos in the sum of both states, 
the K$^0$,$\overline{{\mathrm{K}}^0}$\,mesons, is 2N. Since the 
size of the lattice of the K$^\pm$\,mesons and the 
 K$^0$\,mesons is the same it follows that two neutrinos are at each 
lattice point of the K$^0$ or $\overline{{\mathrm{K}}^0}$\,mesons.  
We assume that Pauli's exclusion principle applies to neutrinos 
as well. Consequently each neutrino at each lattice point must 
share its location with an antineutrino. That means that the contribution 
of the spin of all neutrinos and antineutrinos to the intrinsic angular 
momentum of the K$^0$\,meson is zero or $\sum_i\,j(2n_i)$ = 0. 
The sum of the spin vectors of the two opposite charges in 
either the K$^0$  or the  $\overline{{\mathrm{K}}^0}$\,mesons, 
or in the (2.)$\pi^\pm$ + $\pi^\mp$ state, is also 
zero. Since neither the lattice oscillations nor the spin of the 
neutrinos and antineutrinos 
nor the  electric charges contribute an angular momentum 
\begin{equation} j(\mathrm{K}^0) = \sum_i\,j(\nu_i) + 
\sum_i\,j(2n_i) + j(\mathrm{e}^+ + \mathrm{e}^-) = 0\,. \end{equation}
 The intrinsic angular momentum
of the K$^0$ and $\overline{{\mathrm{K}}^0}$\,mesons
 is zero, or s(K$^0$,$\overline{{\mathrm{K}}^0}$) = 0, as it 
must be. In simple terms since, e.g., the structure of K$^0$ is 
(2.)$\pi^+$ + $\pi^-$, the spin of K$^0$ is the sum of the spin of  
(2.)$\pi^+$ and of $\pi^-$, both of which do not have spin. 
It does not seem possible to arrive at
 s(K$^0$,$\overline{{\mathrm{K}}^0}$) = 0  if both particles 
do not contain the N pairs of neutrinos and antineutrinos required by the
 (2.)$\pi^\pm$ + $\pi^\mp$ state which we have suggested in Section 6.  

   In the case of the neutron one 
must wonder how it comes about that a particle which seems to be the 
superposition of two particles without spin ends up with spin 1/2. The 
neutron,
 which has  a mass $\approx$ 2m(K$^\pm$) or 2m(K$^0$), is either  
the superposition of a K$^+$ and a K$^-$ meson or of a K$^0$ and a 
$\overline{\mathrm{K}^0}$ meson. The intrinsic angular 
momentum of the superposition of K$^+$ and K$^-$ is either 0 or $\hbar$, 
which means that the neutron cannot be the superposition of K$^+$ 
and K$^-$. For a proof of this statement we refer to [56].

   On the other hand the neutron can be the superposition of a 
K$^0$  and a $\overline{\mathrm{K}^0}$\,meson. A significant change 
in the lattice occurs when a K$^0$ and a  $\overline{\mathrm{K}^0}$ meson
 are superposed. Since each K$^0$\,meson contains N neutrinos 
and N antineutrinos, as we discussed 
in context with the spin of K$^0$, the number of all 
neutrinos and antineutrinos in superposed K$^0$  and 
$\overline{\mathrm{K}^0}$ lattices is 4N. Since the size of the 
lattice of the proton as well of the neutron is the same as the   
size of K$^0$ each of the N lattice 
points of the neutron now contains four neutrinos, a muon neutrino 
and an anti-muon neutrino as well as an 
electron neutrino and an anti-electron neutrino. The 
$\nu_\mu,\bar{\nu}_\mu,\nu_e,\bar{\nu}_e$ quadrupoles  
oscillate just like individual neutrinos do because we learned from 
Eq.(7) that the ratios of the oscillation frequencies are independent 
of the mass as well as of the interaction constant between the lattice 
points. In the neutrino quadrupoles the spin of the 
neutrinos and antineutrinos cancels, $\sum_i\,j(4n_i)$ = 0. The
 superposition of two circular neutrino lattice oscillations,  that means 
the 
circular oscillations of frequency $\omega_i$,  contribute as before  
the angular momentum of the center circular oscillation, so 
$\sum_i\,j(\omega_i) = j(\omega_0) = \hbar/2$. The spin 
 and charge of the four electrical charges e$^+$e$^-$e$^+$e$^-$ 
 hidden in the sum of the K$^0$ and $\overline{\mathrm{K}^0}$ 
mesons cancel, j(4e$^\pm$) = 0. It follows  
 that the intrinsic angular momentum of a neutron created by the 
superposition of  a K$^0$  and a $\overline{\mathrm{K}^0}$ meson 
comes from the circular neutrino lattice oscillations only and is 
\begin{equation} j(\mathrm{n}) = \sum_i\,j(\omega_i) + \sum_i\,j(4n_i) +
 j(4\mathrm{e}^\pm) =  \sum_i\,j(\omega_i) = j(\omega_0) = \hbar/2\,,  
\end{equation}
as it must be. In simple terms, the spin of the neutron originates from 
the superposition of two circular neutrino lattice 
oscillations  with the frequencies $\omega$ and $\mathrm{-}\,\omega$
 shifted in phase by $\pi$, which 
produces the angular momentum $\hbar$/2 at the center.

   The spin of the proton is 1/2 and is unambiguously defined by the decay 
of the neutron n $\rightarrow$ p + e$^-$ + $\bar{\nu}_e$. We have 
suggested in Section 7 that 3/4$\cdot$N$^\prime$ anti-electron neutrinos 
of the neutrino lattice of the neutron are removed in
 the $\beta$-decay of the neutron and that the other N$^\prime$/4
 anti-electron neutrinos leave with the emitted electron. The intrinsic 
angular momentum of the proton originates then from the spin of the 
central $\nu_\mu\bar{\nu}_\mu\nu_e$ triplet, from the spin of
 the e$^+$e$^-$e$^+$ triplet which is part of the remains of the neutron,
 and from the angular momentum of the center of the lattice oscillations 
with the superposition of two circular oscillations. The spin of the 
central $\nu_\mu\bar{\nu}_\mu\nu_e$ triplet
is canceled by the spin of the e$^+$e$^-$e$^+$ triplet. According to the
 standing wave model the intrinsic angular momentum of the proton is

\begin{equation}
 j(\mathrm{p}) = j(\nu_\mu\bar{\nu}_\mu\nu_e)_0\,  +  
j{(\mathrm{e}^+\mathrm{e}^-\mathrm{e}^+})
+ j(\omega_0) = j(\omega_0) = \hbar/2\,,  \end{equation}
\noindent as it must be.
                         
   The other mesons of the neutrino branch, the D$^{\pm,0}$ and 
D$_s^\pm$ mesons, both having zero spin, are 
superpositions of a proton and an antineutron of opposite spin, or of 
their antiparticles, or of a neutron and an antineutron of opposite spin 
in D$^0$.  The spin of D$^\pm$ and D$^0$ does therefore not 
pose a new problem. 

   For an explanation of the spin of $\mu^\pm$ we refer to [62]. Since  
all muon or anti-muon  
neutrinos have been removed from the $\pi^\pm$ lattice in the
$\pi^\pm$ decay it follows that a neutrino vacancy is at the center of 
the $\mu^\pm$ lattice (Fig.\,6). Without a neutrino in the center of
 the lattice the sum of the spin vectors of all neutrinos in the 
$\mu^\pm$ lattice is zero. However the $\mu^\pm$\,mesons consist of 
the neutrino lattice plus an electric charge e$^\pm$ whose spin is 1/2. 
The spin of  the $\mu^\pm$\,mesons originates from the spin of the 
elementary electric charge carried
by the $\mu^\pm$\,mesons and is consequently s($\mu^\pm$) = 1/2. 
The same considerations apply for the spin of $\tau^\pm$,
 s($\tau^\pm$) = 1/2.   

   An explanation of the spin of the mesons and baryons can only be 
valid if the same explanation also applies to the antiparticles of these 
particles whose spin is the $\emph{same}$ as that of the ordinary 
particles. The antiparticles of the $\gamma$-branch consist of 
electromagnetic waves whose frequencies differ from the frequencies 
of the ordinary particles only by their sign.
The angular momentum of the superposition of two circular oscillations 
with $\mathrm{-}\,\omega$ and $\omega$  has the same angular 
momentum as the superposition of two circular oscillations 
 with frequencies of opposite  sign, as in $\Lambda$. Consequently 
the spin of the antiparticles of the $\gamma$-branch is the same as 
the spin of the ordinary 
particles of the $\gamma$-branch. The same considerations apply 
to the circular neutrino lattice oscillations which cause the spin of the 
neutron, the only particle 
of the $\nu$-branch which has spin. In the standing wave model 
the spin of the neutron and the antineutron is the same.

   Let us summarize: The spin of the particles of the $\nu$-branch 
originates from the 
angular momentum vectors of the oscillations and the spin vectors 
of the neutrinos in the particles and the spin vector 
of the elementary electric charge or charges a particle carries. The 
contribution 
of all or all but one of the O($10^9$) oscillations and O($10^9)$ 
neutrinos to the intrinsic
angular momentum of the particles must cancel, otherwise the 
spin cannot be either 0 or 1/2. It requires the symmetry of a cubic 
lattice for this to happen. The center of the lattices alone determines 
the intrinsic angular momentum of the oscillations and neutrinos in the 
lattice.
Adding to that the spin vector of one (or more) elementary electric 
charges with spin 1/2 and we arrive at the total intrinsic angular 
momentum of a particle. The most illuminating case are the 
$\pi^\pm$\,mesons which do not have spin although they carry 
an elementary electric charge. Actually the neutrino lattice 
of the $\pi^\pm$\,mesons has spin 1/2 from its central neutrino, 
but this spin vector is canceled by the spin of the electric charge 
e$^\pm$, so s($\pi^\pm$) = 0.   

   From the foregoing we arrive also at an understanding of the reason for 
the astonishing fact that the intrinsic angular momentum or spin of the 
particles is independent of the mass of the particles,
as exemplified by the spin $\hbar$/2 of the electron being the
same as the spin $\hbar/2$ of the proton, notwithstanding the fact that
 the mass of the proton is 1836 times larger than the mass of the 
electron. However,  
in our model, the spin of the particles including the electron 
is determined solely by the angular momentum $\hbar$/2 at the
 center of the lattice, the other angular momentum vectors in the 
particles cancel. The spin is independent of the  
number of the lattice points in a cubic lattice. Hence the mass 
of the particles  in the other 
$10^9$  lattice points is inconsequential for the intrinsic angular
 momentum of the particles.  In our model the spin of 
the particles is independent of the mass of the particles, as it must be.

\section{The spin and magnetic moment \\ of the electron} 

    The model of the electron we have proposed in Section 9
 has, in order to be valid, to pass a crucial test; the model has 
to explain satisfactorily the spin and the magnetic moment of the
 electron. When Uhlenbeck and Goudsmit [50] (U\&G) discovered
 the existence of the spin of the electron they also proposed that the 
electron has a magnetic moment with a value equal to Bohr's 
magnetic moment $\mu_B$ = e$\hbar$/2m(e$^\pm$)c$_{\star}$. Bohr's 
magnetic moment results from the motion of an electron on a circular  
orbit around a proton. The magnetic moment of the electron postulated
by U\&G has been confirmed experimentally, but has been corrected by 
  about  0.11\% for the so-called anomalous magnetic moment. 
If one tries to explain the magnetic moment of the electron
  with an electric charge moving on a circular orbit around the 
particle center, analogous to the magnetic moment of hydrogen, one ends 
up with velocities larger than the velocity of light, which cannot be, as 
already noted by U\&G. It remains to be explained how the magnetic
 moment of the electron comes about.

   We will have to explain the spin of the electron first. The spin, or 
the intrinsic angular momentum of a particle is, of course, the sum of the 
angular momentum vectors of all components of the particle. In the  
 electron these are the neutrinos and the electric oscillations. Each 
neutrino has spin 1/2 and in order for the electron to have 
s = 1/2 all, or all but one, of the spin vectors of the neutrinos in their 
lattice must cancel.
If the neutrinos are in a simple cubic lattice as in Fig.\,7 and the
 center particle of the lattice is not  a neutrino, as in Fig.\,7, the 
spin vectors of all neutrinos cancel, $\Sigma\,j(n_i)$ = 0, provided 
that the spin vectors of the 
 neutrinos of the lattice point in opposite direction at their mirror
 points in the lattice. Otherwise the spin vectors of the neutrinos 
would add up and make a very large angular momentum. We 
follow here the procedure we used in [56] to explain the spin
of the muons. The spin vectors of all electron neutrinos in the electron 
cancel, just as the spin vectors of all muon and electron neutrinos
 in the muons cancel, because there is
 a neutrino vacancy at the center of their lattices, Figs.\,6,7.

   We will now see whether the electric oscillations in the electron 
contribute to its angular 
 momentum. As we said in context with Eq.(51) there must be two 
times as many electric oscillations in the electron lattice than there are 
neutrinos. The oscillations can either be standing waves consisting 
of two linear oscillations or two circular oscillations  with the 
frequencies $\omega$ and $-\,\omega$.  Both the standing 
 waves and the circular oscillations are non-progressive and can be 
part of the \emph{rest mass} of a particle. We will now assume that  
the electric oscillation are circular.  Circular oscillations have an 
angular
momentum $\vec{j} = m\,\vec{r}\times\vec{v}$. And, as in the case 
of the spin vectors of the neutrinos, all or all but one of the O$(10^9)$ 
angular momentum vectors of the electric oscillations must cancel in 
order for the electron to have spin 1/2. We describe the 
superposition of the two oscillations by

\begin{equation} x(t) = exp[i\omega t] + exp[-\,i(\omega t + 
\pi)]\,,\end{equation}

\begin{equation} y(t) = exp[i(\omega t + \pi/2)] + exp[-\,i(\omega t + 
                 3\pi/2)]\,\,,
\end{equation}

\noindent
that means by the superposition of an oscillation with the 
frequency $\omega$ and a second oscillation with the frequency
 $\mathrm{-}\,\omega$ as in Eqs.(89,90). The oscillation with 
$\mathrm{-}\,\omega$ is shifted in phase by $\pi$. 
 Negative frequencies are permitted solutions of the equations 
of motion in a cubic lattice, Eqs.(7,13). As is well-known 
 oscillating electric charges should emit radiation. However,
 this rule does already not hold in the hydrogen atom, so we
will assume that the rule does not hold within the electron either.

   In circular oscillations the kinetic energy is always equal to the  
potential energy  and the sum of both is the total energy. From
\begin{equation} \mathrm{E}_{pot} + \mathrm{E}_{kin} =
 2\,\mathrm{E}_{kin} = \mathrm{E}_{tot} \end{equation}
follows, as discussed in Eqs.(93,94), that the angular
 momentum of the superposition of the two circular oscillations is
\begin{equation} j = \Theta\,\omega = \hbar/2\,. \end{equation} 
That means that each of the O$(10^9$) superposed circular  
electric oscillations has an angular momentum  $\hbar$/2.

   The circulation of the circular oscillations in Eqs.(101,102) is 
opposite  
for all $\phi$ of opposite sign. It follows from the equation for  
the displacements u$_n$  of the lattice points Eq.(5)
\begin{equation} u_n = Ae^{i(\omega\,t\, +\, n\phi)}\,,
\end{equation}
\noindent
 that the velocities of the lattice points are given by
\begin{equation} v_n = \dot{u}_n = i\,\omega_n\,u_n\,. \end{equation}
The sign of $\omega_n$ changes with the sign of $\phi$ because 
the frequencies are given by Eq.(13), that means by  
\begin{equation} \omega_n = \pm\,\omega_0(\,\phi_n + \phi_0\,)\,.
\end{equation}

Consequently the circulation of the electric oscillations is opposite  
to the circulation at the mirror
 points in the lattice and the angular momentum vectors cancel,
but for the angular momentum vector of the electric oscillation at 
the\emph{ center of the lattice}. The center circular oscillation has, as 
all 
other electric oscillations, the angular momentum  $\hbar$/2 as Eq.(104) 
says. The angular momentum of the entire electron lattice is therefore
\begin{equation} j(\mathrm{e}^\pm) = \sum_i\,j(n_i) + \sum_i\,j(el_i) =
 j(el_0) =\hbar/2\,,
\end{equation}
as it must be for spin s = 1/2. The explanation of the spin of the 
electron 
given here follows the explanation of the spin of the baryons, as well
as the explanation of the absence of spin in the mesons. A valid 
explanation of the spin must be applicable to all particles, in particular 
to the electron, the prototype of a particle with spin.

   We will now turn to the magnetic moment of the electron which is
 known with extraordinary accuracy, $\mu_e$  = 
1.001\,159\,652\,186\,$\mu_B$, according to the Review of Particle 
Physics [2], $\mu_B$ being Bohr's magneton. The decimals after 
1.00\,$\mu_B$  are caused by the anomalous magnetic moment which 
we will not consider. As is well-known the magnetic dipole moment of 
a particle with spin is, in Gaussian units, given by
\begin{equation} \vec{\mu} = g\,\frac{e\hbar}{2mc_{\star}}\,\vec{s}\,,
 \end{equation}
where g is the dimensionless Land\'{e} factor, m the rest mass of  
the particle that carries the charge e and $\vec{s}$  the spin vector. 
The g-factor has been introduced in order to bring the magnetic moment
 of the electron into agreement with the experimental facts. As U\&G
 postulated and as has been confirmed experimentally the 
g-factor of the electron is 2. With the spin s = 1/2  and g = 2  
the magnetic dipole moment of the electron is then 
\begin{equation} \mu_e = e\hbar/2m(e^\pm)c_{\star}\,,
 \end{equation}
or one Bohr magneton $\mu_B$  in agreement with the experiments, 
neglecting the anomalous moment. For a structureless point particle
Dirac [51] has explained why g = 2 for the electron. However 
we consider here an electron with a finite size and which is at rest. 
 When it  is at rest the electron has still its
 magnetic moment. Dirac's theory does therefore not apply here.

  The only part of Eq.(110) that can be changed in order to explain the 
g-factor of an electron with structure is the ratio e/m which deals with 
the 
spatial distribution of charge and mass. If part of the mass of the 
electron is
non-electromagnetic and the non-electromagnetic part of the mass does
not contribute to the magnetic moment of the electron, which to all 
that we know is true for neutrinos, then the ratio e/m in Eq.(109) is not
 e/m(e$^\pm$)  in the case of the electron. The elementary charge e 
certainly remains unchanged, but  e/m depends on what fraction of the 
mass of the electron is of electromagnetic origin and what fraction of the 
mass is  non-electromagnetic.  Only the current, not the mass of a 
current loop, determines the magnetic moment of a loop. From the very 
accurately known values of $\alpha_f$, m($\pi^\pm$)c$^2_\star$ 
and m(e$^\pm)$c$^2_\star$ and from Eq.(53)
for the energy in the electric oscillations in the electron 
$\mathrm{E}_\nu$(e$^\pm$) = 
$\alpha_f$/2\,$\cdot$\,m($\pi^\pm$)c$^2_\star$/2 = 
0.996570\,m(e$^\pm$)c$^2_\star$/2 follows
that  very nearly one half of the mass of the electron is of electric 
origin, the other half of m(e$^\pm$)  is made of neutrinos, Eq.(47),
and neutrinos do not contribute to the magnetic moment. That means     
that in the electron the mass that carries the charge e  is approximately
m(e$^\pm$)/2. The magnetic moment of the electron is then 
\begin{equation}
\vec{\mu}_e = g \frac{e\hbar}{2m(e^\pm)
/2\cdot c_{\star}}\vec{s}\,, \end{equation}
and with s = 1/2 we have $\mu_e$ = g\,$e\hbar$/$2m(e^\pm)c_{\star}$. 
Because of Eq.(110) the g-factor must be equal to one and is
 unnecessary. 

   In other words, if exactly one half of the mass of the 
electron consists of neutrinos, then it follows automatically that the 
electron has the correct magnetic moment 
$\mu_e$ = $e\hbar/2m(e^\pm)c_{\star}$ without the artificial 
g-factor.       

\section*{Conclusions}

   We conclude that  the standing wave model solves a number of  
problems for which an answer heretofore has been hard to come by. 
From the creation of the $\pi^0$\,meson and its decay into  
$\gamma\gamma$  after 10$^{-16}$ sec follows that the $\pi^0$\,meson   
and the other particles of the $\gamma$-branch must consist of standing 
electromagnetic waves. The rest masses of the particles of the 
$\gamma$-branch obey the integer multiple rule.
From the creation of the $\pi^\pm$\,mesons and their decay 
follows that the $\pi^\pm$\,mesons and the other particles of 
the $\nu$-branch must consist of a lattice of 
$\nu_\mu,\bar{\nu}_\mu,\nu_e,\bar{\nu}_e$ neutrinos, their 
oscillation energies and one or more elementary electric charges.
 From the explanation of the $\pi^\pm$\,mesons follows
that the rest mass of the $\mu^\pm$\,mesons is $\cong$ 3/4 
of the rest mass of the $\pi^\pm$\,mesons and it also follows 
that 1/2 of the rest mass of the electron or
positron must consist of electron neutrinos or anti-electron neutrinos.
The other half of the rest mass of the electron is in the energy of 
electric oscillations. The elementary electric charge of the electron is
the sum of the charges carried by the individual electric oscillations.
The rest mass of the $\mu^\pm$\,mesons is so much larger than
the rest mass of the electron or positron because the rest mass of
the muon neutrinos is so much larger than the rest mass of the 
electron or anti-electron neutrinos. We have verified the validity of
our explanation of m(e$^\pm$), m($\mu^\pm$) and m($\pi^\pm$)
and of the relation m($\nu_e$) = $\alpha_f$\,$\cdot$\,m($\nu_\mu$)
by showing that the calculated ratios m($\mu^\pm$)/m(e$^\pm$)
and m($\pi^\pm$)/m(e$^\pm$) agree within 1\% with the 
experimental values of these mass ratios.  
Only photons, neutrinos, charge and the weak nuclear force are 
needed to explain the  masses of the electron, the muon and of the 
stable mesons and baryons and their antiparticles. We can explain 
the  spin of the baryons and the absence of spin in the mesons, and
the spin of the electron and muon  as well; without making
 additional assumptions. We have also determined the rest masses 
of the e, $\mu$ and $\tau$ neutrinos and found that the mass of the 
electron neutrino is equal to the mass of the muon neutrino times the
fine structure constant.  

\vspace{1cm}

   {\bfseries Acknowledgments}. I gratefully acknowledge the contributions
 
of 
Dr. T. Koschmieder to this study. I thank Professor A. Gsponer \\ 
\indent for information about the history of the integer 
multiple rule.

\section*{References}

\noindent
[1] Gell-Mann,M. 1964. Phys.Lett.B {\bfseries111},1.

\smallskip
\noindent
[2] Yao,W.-M. et al.  2006. J.Phys.G {\bfseries 33},1.

\smallskip
\noindent
[3] El Naschie,M.S. 2002. Chaos,Sol.Frac. {\bfseries 14},649.

\smallskip
\noindent
[4] El Naschie,M.S.  2002. Chaos,Sol.Frac. {\bfseries 14},369.

\smallskip
\noindent
[5] El Naschie,M.S. 2003. Chaos,Sol.Frac. {\bfseries 16},353.

\smallskip
\noindent
[6] El Naschie,M.S. 2003. Chaos,Sol.Frac. {\bfseries 16},637.

\smallskip
\noindent
[7] Feynman,R.P. 1985. \emph{QED. The Strange Theory of Light and 
Matter}.\\
\indent\,Princeton University Press, p.152.

\smallskip
\noindent
[8] Koschmieder,E.L. 2003. Chaos,Sol.Frac. {\bfseries18},1129.
 
\smallskip
\noindent
[9] Nambu,Y. 1952. Prog.Th.Phys. {\bfseries 7},595.

\smallskip
\noindent
[10] Fr\"ohlich,H. 1958. Nucl.Phys. {\bfseries 7},148. 

\smallskip
\noindent
[11] Koschmieder,E.L. and Koschmieder,T.H. 1999.\\
\indent \,\,Bull.Acad.Roy.Belgique {\bfseries X},289,\\
\indent \,\, http://arXiv.org/hep-lat/0002016.

\smallskip 
\noindent
[12] Wilson,K.G. 1974. Phys.Rev.D {\bfseries10},2445.

\smallskip
\noindent
[13] Born,M. and v.\,Karman,Th. 1912. Phys.Z. {\bfseries13},297.

\smallskip
\noindent
[14] Blackman,M. 1935. Proc.Roy.Soc.A {\bfseries148},365;384.

\smallskip
\noindent
[15] Blackman,M. 1955. Handbuch der Physik VII/1, Sec.12.

\smallskip
\noindent
[16] Born,M. and Huang,K. 1954. \emph{Dynamical Theory of Crystal 
Lattices}.\\ 
\indent \,\,(Oxford).

\smallskip
\noindent
[17] Maradudin,A. et al. 1971. \emph{Theory of Lattice Dynamics in the 
Harmonic\\ 
\indent \,\, Approximation}. Academic Press, 2nd edition.

\smallskip
\noindent
[18] Ghatak,A.K. and Khotari,L.S. 1972. \emph{An introduction to Lattice\\ 
\indent \,\, Dynamics}. Addison-Wesley.

\smallskip
\noindent
[19] Rekalo,M.P., Tomasi-Gustafson,E. and Arvieux,J. 2002.\\
 \indent\,\, Ann.Phys.{\bfseries295},1.

\smallskip
\noindent
[20] Schwinger,J. 1962. Phys.Rev. {\bfseries128},2425.

\smallskip
\noindent
[21] Sidharth,B.J. 2001. \emph{The Chaotic Universe}, p.49.\\
\indent\,\,Nova Science Publishers, New York.

\smallskip
\noindent
[22] Perkins,D.H. 1982. \emph{Introduction to High-Energy Physics}.\\
\indent \,\,Addison Wesley.

\smallskip
\noindent
[23] Rosenfelder,R. 2000. Phys.Lett.B {\bfseries479},381.

\smallskip
\noindent
[24] Born,M. 1940. Proc.Camb.Phil.Soc. {\bfseries36},160.

\smallskip
\noindent
[25] Koschmieder,E.L. 2000. http://arXiv.org/hep-lat/0005027.

\smallskip
\noindent
[26] Melnikov,K. and van Ritbergen,T. 2000. Phys.Rev.Lett. 
{\bfseries84},1673.

\smallskip
\noindent
[27] Liesenfeld,A. et al. 1999. Phys.Lett.B {\bfseries468},20.

\smallskip
\noindent
[28] Bernard,V., Kaiser,N. and Meissner,U-G. 2000. \\
\indent  \,\,Phys.Rev.C {\bfseries62},028021.

\smallskip
\noindent
[29] Eschrich,I. et al. 2001. Phys.Lett.B {\bfseries522},233.

\smallskip
\noindent 
[30] Sommerfeld,A. 1952. \emph{Vorlesungen \"{u}ber Theoretische Physik}.\\
\indent  \,\,Bd.V, p.56.

\smallskip
\noindent
[31] Debye,P. 1912. Ann.d.Phys. {\bfseries39},789.

\smallskip
\noindent
[32] Bose,S. 1924. Z.f.Phys. {\bfseries26},178.

\smallskip
\noindent
[33] Bethe,H. 1986. Phys.Rev.Lett. {\bfseries58},2722.

\smallskip
\noindent
[34] Bahcall,J.N. 1987. Rev.Mod.Phys. {\bfseries59},505.

\smallskip
\noindent
[35] Fukuda,Y. et al. 1998. Phys.Rev.Lett. {\bfseries81},1562, 2003. 
{\bfseries90},171302.

\smallskip
\noindent
[36] Ahmad,Q.R. et al. 2001. Phys.Rev.Lett. {\bfseries87},071301.

\smallskip
\noindent
[37] Lai,A. et al. 2003. Eur.Phys.J.C {\bfseries30},33.

\smallskip
\noindent
[38] Koschmieder,E.L. 2003. http://arXiv.org/physics/0309025.

\smallskip
\noindent
[39] Barut,A.O. 1979. Phys.Rev.Lett. {\bfseries 42},1251.

\smallskip
\noindent
[40] Gsponer,A. and Hurni,J-P. 1996. Hadr.J. {\bfseries 19},367.

\smallskip
\noindent
[41] Barut,A.O. 1978. Phys.Lett.B {\bfseries73},310.

\smallskip
\noindent
[42] Born,M. and Stern,O. 1919. Sitzungsber.Preuss.Akad.Wiss.\\ 
\indent  \,\,{\bfseries33},901.

\noindent
\smallskip
\noindent
[43] Thomson, J.J. 1897. Phil.Mag. {\bfseries44},293.

\smallskip
\noindent
[44] Lorentz, H.A. 1903. \emph{Enzykl.Math.Wiss.} Vol.{\bfseries5},188.

\smallskip
\noindent
[45] Poincar\'{e}, H. 1905. Compt.Rend. {\bfseries 140},1504. Translated 
in:\\
\indent
 \,\,Logunov, A.A. 2001. \emph{On The Articles by Henri Poincare\\ 
\indent
``On The Dynamics of the Electron"}. Dubna JINR.

\smallskip
\noindent
[46] Ehrenfest, P. 1907. Ann.Phys. {\bfseries 23},204.

\smallskip
\noindent
[47] Einstein, A. 1919. Sitzungsber.Preuss.Akad.Wiss. {\bfseries20},349.

\smallskip
\noindent
[48] Pauli, W. 1921. \emph{Relativit\"{a}tstheorie}. B.G. Teubner. 
Translated in:\\
\indent  \,\,1958. Theory of Relativity. Pergamon Press.

\smallskip
\noindent
[49] Poincar\'{e}, H. 1906. Rend.Circ.Mat.Palermo {\bfseries21},129.

\smallskip
\noindent
[50] Uhlenbeck, G.E. and Goudsmit, S. 1925. Naturwiss. {\bfseries13},953 .

\smallskip
\noindent
[51] Dirac, P.A.M. 1928. Proc.Roy.Soc.London A{\bfseries117},610.

\smallskip
\noindent
[52] Gottfried, K. and Weisskopf, V.F. 1984. \emph{Concepts of Particle 
Physics}.\\
\indent
 \,\,Vol.1,\,p.38. Oxford University Press.

\smallskip
\noindent
[53] Schr\"{o}dinger, E. 1930. Sitzungsber.Preuss.Akad.Wiss. 
{\bfseries24},418. 

\smallskip
\noindent
[54] Mac Gregor, M.H. 1992. \emph{The Enigmatic Electron}. Kluwer.

\smallskip
\noindent
[55] Koschmieder,E.L. 2005. http://arXiv.org/physics/0503206.

\smallskip
\noindent
[56] Koschmieder,E.L. 2003. Hadr.J. {\bfseries26},471,\\
\indent\,\,2003. http://arXiv.org/physics/0301060.

\smallskip
\noindent
[57] Jaffe,R.L. 2001. Phil.Trans.Roy.Soc.Lond.A {\bfseries359},391.

\smallskip
\noindent
[58] G\"ockeler,M. et al. 2002. Phys.Lett.B {\bfseries545},112.

\smallskip
\noindent
[59] Rivas,M. 2001. \emph{Kinematic Theory of Spinning Particles}. 
Kluwer.  

\smallskip
\noindent
[60] Poynting,J.H. 1909. Proc.Roy.Soc.A {\bfseries82},560.

\smallskip
\noindent
[61] Allen,J.P. 1966. Am.J.Phys. {\bfseries34},1185.

\smallskip
\noindent
[62] Koschmieder,E.L. 2003. http://arXiv.org/physics/0308069,\\
\indent\,\,2005. \emph{Muons: New Research}. Nova.\\

\section*{Appendix}

\bigskip
\noindent
When the integer multiple rule was first proposed (Koschmieder, E.L. Bull.
Acad.Roy.Belgique, {\bfseries X},281,1999; arXiv.org/hep-ph/0002179;\,2000)
we relied on the data available at that time, Barnett et al. Rev.Mod.Phys.
 {\bfseries68},611,1996. Since then the data on the particle masses have 
multiplied and become more precise. It seems now that particles with 
masses even 70\,$\times$\,m($\pi^0$) obey the integer multiple rule. 
A list of the particles which follow the integer multiple rule  
 is given in Table 3.
 
\begin{table}\caption{The particles following the integer multiple rule}
\begin{tabular}{lcll|lcll}\\
\hline\hline\\
&J&m/m($\pi^0$) & multiples & &J&m/m($\pi^0$) & multiples\\
[0.5ex]\hline\\

$\pi^0$ & 0&1.0000 & 1.0000\,$\cdot$\,1$\pi^0$ &  $\eta_c$(1S) & 0 & 
22.0809 & 1.0037\,$\cdot$\,22$\pi^0$\\

$\eta$ & 0 & 4.0559 & 1.0140\,$\cdot$\,4$\pi^0$ & J/$\psi$ & 1 & 22.9441 & 
0.9976\,$\cdot$23$\pi^0$\\

$\eta^\prime$(958) & 0 &  7.0959 &1.0137\,$\cdot$\,7$\pi^0$ & 
$\chi_{c0}$(1P) & 0 & 25.2989 & 1.0119\,$\cdot$\,25$\pi^0$\\

$\eta$(1295) & 0 & 9.5868 & 0.9587\,$\cdot$\,10$\pi^0$ & $\chi_{c1}$(1P) & 
1 & 26.0094 & 1.0004\,$\cdot$\,26$\pi^0$\\ 

$\eta$(1405) & 0 & 10.445 & 1.0445\,$\cdot$\,10$\pi^0$ & $\eta_c$(2S) & 0 
& 26.9528 & 0.9983\,$\cdot$\,27$\pi^0$\\

$\eta$(1475) & 0 & 10.9352 & 0.9941\,$\cdot$\,11$\pi^0$ & $\psi$(2S) & 1 & 
27.3091 & 1.0115\,$\cdot$\,27$\pi^0$\\

& & & &  $\psi$(3770) & 1 & 27.9389 & 0.9978\,$\cdot$\,28$\pi^0$\\

$\Lambda$ & 1/2 & 8.2658 & 1.0332\,$\cdot$\,8$\pi^0$ & $\psi$(4040) & 1 & 
29.9237 & 0.9975\,$\cdot$\,30$\pi^0$\\ 

$\Lambda$(1405) & 1/2 & 10.4166 & 1.0417\,$\cdot$\,10$\pi^0$ & 
$\psi$(4191) & 1 & 31.0543 & 1.00175\,$\cdot$\,31$\pi^0$\\

$\Lambda$(1670) & 1/2 & 12.3725 & 1.0310\,$\cdot$\,12$\pi^0$ & 
$\psi$(4415) & 1 & 32.7538 & 0.9925\,$\cdot$\,33$\pi^0$\\

$\Lambda$(1800) & 1/2 & 13.335 & 1.0258\,$\cdot$\,13$\pi^0$ & & & &\\  

$\Sigma^0$ & 1/2 & 8.8359 & 0.9818\,$\cdot$\,9$\pi^0$ & $\Upsilon$(1S) & 1 
& 70.0884 & 1.0013\,$\cdot$\,70$\pi^0$\\

$\Sigma$(1660) & 1/2 & 12.2984 & 1.0249\,$\cdot$\,12$\pi^0$ & 
$\chi_{b0}$(1P) & 0 & 73.0455 & 1.0006\,$\cdot$\,73$\pi^0$\\
 
$\Sigma$(1750) & 1/2 & 12.9652 & 0.9973\,$\cdot$\,13$\pi^0$ & 
$\chi_{b1}$(1P) & 1 & 73.2926 & 1.0040\,$\cdot$\,73$\pi^0$\\

$\Xi^0$ & 1/2 & 9.7412 & 0.9741\,$\cdot$\,10$\pi^0$ & $\Upsilon$(2S) & 1 & 
75.8102 &
0.9975\,$\cdot$\,76$\pi^0$\\

$\Omega^-$ & 3/2 & 12.3907 & 1.0326\,$\cdot$\,12$\pi^0$ & $\chi_{b0}$(2P) 
& 0 & 75.8094 & 0.9975\,$\cdot$\,76$\pi^0$\\

$\Lambda_c^+$ & 1/2 & 16.9397 & 0.9965\,$\cdot$\,17$\pi^0$ & 
$\chi_{b1}$(2P) &1 & 75.9795 & 0.9997\,$\cdot$\,76$\pi^0$\\

$\Lambda_c$(2593)$^+$ & 1/2 & 19.2285 & 1.0120\,$\cdot$\,19$\pi^0$ & 
$\Upsilon$(3S) & 1 & 76.7185 & 0.9963\,$\cdot$\,77$\pi^0$\\

$\Sigma_c$(2455)$^0$ & 1/2 & 18.1792 & 1.00995\,$\cdot$\,18$\pi^0$ & $\Upsilon$(4S) 
& 1 & 78.3795 & 1.0049\,$\cdot$\,78$\pi^0$\\ 

$\Xi_c^0$ & 1/2 & 18.3069 & 1.01705\,$\cdot$\,18$\pi^0$ & $\Upsilon$(10860) & 1 & 
80.4954 & 1.0062\,$\cdot$\,80$\pi^0$\\
 
$\Xi^{\prime0}_c$ & 1/2 & 19.0996 & 1.0052\,$\cdot$19$\pi^0$ &  $\Upsilon$(11020) 
& 1 & 81.6364 & 0.9956\,$\cdot$\,82$\pi^0$\\

$\Xi_c$(2790) & 1/2 & 20.6843 & 0.9850\,$\cdot$\,21$\pi^0$\\

$\Omega_c^0$ & 1/2 & 19.9849 & 0.9992\,$\cdot$\,20$\pi^0$\\
[0.3cm]\hline\hline
\end{tabular}
\end{table}

  The masses are taken from the Review of Particle 
Physics [2]. Only particles, stable or unstable, with J\,$\le$\,1  are listed. 
To be on the safe side, we use only particles which the Particle Data 
Group considers to be ``established". The $\Omega^-$ baryon 
with J = 3/2 is also given for comparison, but is not 
included in the least square analysis.
 
   The correlation of the masses of the 
different $\eta$ mesons has also been investigated by Palazzi,
$<$http://particlez.org/p3a/abstract/2004-001.html$>$, using a mass unit of 
33.86\,MeV/c$^2$, which is 1.0034\,m($\pi^0$)/4. In his Fig.\,2a he shows 
that the masses of the $\eta$ mesons are integer multiples of his mass unit,
having a nearly perfect correlation coefficient. For the mass of $\psi$(4160) 
in Table 3 we have used a recent measurement of Ablikim et al. 
(arXiv:0705.4500) which places m($\psi$(4160)) at 4191.6\,MeV/c$^2$, 
which is 1.00175\,$\cdot$\,31m($\pi^0$). In all there are 41 particles which 
follow the integer multiple rule. The data of Table 3 are plotted in Fig.\,1,
 the line on Fig.\,1 is determined by Eq.(2).

\end{document}